\documentclass[11pt,a4paper]{article}

\usepackage[left=25mm,right=25mm,top=30mm,bottom=30mm]{geometry}
\usepackage{amsmath,amssymb,amsfonts}
\usepackage[linktocpage]{hyperref}
\usepackage{xcolor}
\usepackage{threeparttable}
\usepackage{enumerate}
\usepackage{bbm}
\usepackage{comment}
\usepackage{cite}

\newcommand{\ket}[1]{{\left|#1\right\rangle}}
\newcommand{\tr}{{\mathrm{tr}}}
\newcommand{\pdv}[1]{{\frac{\partial}{\partial {#1}}}}
\newcommand{\vb}[1]{\mbox{\boldmath $#1$}}

\newcommand{\deltasu}[2]{{\delta_{#1}}^{#2}}

\newcommand{\uu}[2]{{u_{#1}}^{#2}}

\newcommand{\UU}[2]{{\vb{u}_{#1}}^{ {#2}}}

\newcommand{\HU}[2]{{\hat{\vb{u}}_{#1}}^{~ {#2}}}
\newcommand{\hu}[2]{{\widehat{u}_{#1}}^{~#2}}
\newcommand{\TB}[1]{\tilde{\bar{#1}}}

\newcommand{\BB}{{\mathcal{B}}}
\newcommand{\CC}{{\mathcal{C}}}

\newcommand{\OO}{{\mathcal{O}}}

\newcommand{\II}{{\mathcal{I}}}
\newcommand{\JJ}{{\mathcal{J}}}

\newcommand{\LL}{{\mathcal{L}}}

\newcommand{\QQ}{{\mathcal{Q}}}
\newcommand{\CS}{{\mathcal{S}}}
\newcommand{\RR}{{\mathcal{R}}}
\newcommand{\NN}{{\mathcal{N}}}

\newcommand{\BZ}{{\boldsymbol {Z}}}
\newcommand{\bj}{{\bar \jmath}}

\newcommand{\schurB}{{\widehat{\BB}}}
\newcommand{\gschurB}{{\schurB_{R}}}
\newcommand{\schurD}{{\mathcal{D}}}
\newcommand{\gschurD}{{\schurD_{R(0,\bj)}}}
\newcommand{\schurDD}{{\bar{\mathcal{D}}}}
\newcommand{\gschurDD}{{\schurDD_{R(j,0)}}}
\newcommand{\schurC}{{\widehat{\CC}}}
\newcommand{\gschurC}{{\schurC_{R(j,\bj)}}}

\newcommand{\BTh}{\bar{\Theta}}
\newcommand{\TRep}[3]{{T_{#1}^{~{#2}}\left[#3\right]}}
\newcommand{\Imu}{\mathrm{i}}

\makeatletter

\@addtoreset{equation}{section}
\makeatother

\DeclareFontShape{OT1}{cmr}{mx}{n}{<->cmr10}{}

\begin{document}

\begin{titlepage}

\begin{flushright}
 OCU-PHYS 494
\end{flushright}

\vskip2cm

\begin{center}

{\huge\fontseries{mx}\selectfont
OPE Selection Rules for Schur Multiplets in 4D $\NN=2$ Superconformal Field
 Theories \par
}

\vskip1.5cm

{\large Kazuki Kiyoshige $^{\diamondsuit,1}$
and Takahiro Nishinaka$^{\heartsuit,2}$}

\vskip1cm

{\it
 $^1$Department of Physics, Graduate School of Science, Osaka City University\\
 3-3-138 Sugimoto, Sumiyoshi, Osaka 558-8585, Japan \\[2mm]
$^2$Department of Physical Sciences, College of Science and Engineering\\
 Ritsumeikan University, Shiga 525-8577, Japan
}

\end{center}

\vskip1.5cm

\begin{abstract}
We compute general expressions for two types of three-point functions of (semi-)short multiplets in four-dimensional $\mathcal{N}=2$ superconformal field theories.
These (semi-)short multiplets are called ``Schur multiplets'' and play an important role in the study of associated chiral algebras.
The first type of the three-point functions we compute involves two half-BPS Schur multiplets and an arbitrary Schur multiplet, while the second type involves one stress tensor multiplet and two arbitrary Schur multiplets. 
From these three-point functions, we read off the corresponding OPE selection rules for the Schur multiplets.
Our results particularly imply that there are non-trivial selection rules
on the quantum numbers of Schur operators in these multiplets.
We also give a conjecture on the selection rules for general Schur multiplets.

\end{abstract}

\renewcommand{\thefootnote}{\fnsymbol{footnote}}
\footnotetext{
	$^{\diamondsuit}$kiyoshig@sci.osaka-cu.ac.jp, 
	$^{\heartsuit}$nishinak@fc.ritsumei.ac.jp}
\renewcommand{\thefootnote}{\arabic{footnote}}

\end{titlepage}

\tableofcontents

\section{Introduction}\label{intro}

The space of four-dimensional $\NN=2$ superconformal field theories (SCFTs) has a rich structure.
The best known $\mathcal{N}=2$ SCFTs are the $SU(N_{c})$ gauge theories with $N_f$ fundamental matter hypermultiplets where $N_{f}=2N_{c}$ is satisfied so that the beta function vanishes.
While these theories are well-described by Lagrangian, there are many $\mathcal{N}=2$ SCFTs whose Lagrangian description is not known, such as Argyres-Douglas SCFTs \cite{Argyres:1995jj,Argyres:1995xn, Eguchi:1996vu},\footnote{For recent discussions on $\mathcal{N}=1$ Lagrangians that flow to the Argyres-Douglas SCFTs, see \cite{Maruyoshi:2016tqk, Maruyoshi:2016aim, Agarwal:2016pjo, Maruyoshi:2018nod, Agarwal:2017roi, Benvenuti:2017kud, Benvenuti:2017bpg, Giacomelli:2017ckh, Giacomelli:2018ziv, Carta:2018qke}.} Minahan-Nemeschanskey theories \cite{Minahan:1996fg, Minahan:1996cj}, and an infinite series of non-Lagrangian SCFTs of class $\mathcal{S}$ \cite{Gaiotto:2009we}.
To study general $\mathcal{N}=2$ SCFTs including these non-Lagrangian theories, we need a technique that relies only on the symmetry and unitarity of SCFTs.

Recently there was important progress in this direction.
The authors of \cite{Beem:2013sza} showed that
the operator product expansions (OPEs) of a special class of %the 
BPS local operators
%, in every four-dimensional $\mathcal{N}=2$ SCFT, 
are naturally encoded in a two-dimensional chiral algebra. %, which 
%is often dubbed the 4d/2d correspondence in the following.
%every four-dimensional $\mathcal{N}=2$ SCFT contains a special set of BPS local operators whose operator product expansions (OPEs) are naturally encoded in a two-dimensional chiral algebra.
These BPS operators are called ``Schur operators'' since they contribute to the Schur limit of the superconformal index \cite{Gadde:2011ik, Gadde:2011uv, Beem:2013sza}.
We call the superconformal multiplets including a Schur operator ``Schur multiplets.''

The existence of the associated chiral algebra implies, along with the four-dimensional unitarity and superconformal symmetry, that the ``$c$ central charge'' of any interacting four-dimensional $\mathcal{N}=2$ SCFTs is constrained by $c\geq \frac{11}{30}$ \cite{Liendo:2015ofa}, which is saturated by the minimal Argyres-Douglas SCFT.\footnote{Our normalization of the four-dimensional central charge is such that a free hypermultiplet has $c = \frac{1}{12}$ and $a=\frac{1}{24}$.}
Moreover, a similar analysis for $\mathcal{N}=2$ SCFTs with a flavor symmetry leads to a universal bound involving $c$ and the flavor central charge \cite{Beem:2013sza,Lemos:2015orc}.
For more recent works on the associated chiral algebras, see
\cite{Beem:2014rza, Lemos:2014lua, Beem:2014zpa, Buican:2015ina,
Cordova:2015nma, Cecotti:2015lab, Beem:2016cbd, Nishinaka:2016hbw,
Buican:2016arp, Xie:2016evu, Cordova:2016uwk, Lemos:2016xke,
Beem:2016wfs, Bonetti:2016nma, Song:2016yfd, Fredrickson:2017yka,
Cordova:2017mhb, Song:2017oew, Buican:2017fiq, Ito:2017ypt,
Beem:2017ooy, Neitzke:2017cxz, Pan:2017zie, Fluder:2017oxm,
Choi:2017nur, Buican:2017rya, Ito:2018wgj, Wang:2018gvb, Nishinaka:2018zwq}.

Let us 
briefly  %shortly 
sketch how these bounds on the central charges were
derived from 
the chiral algebra analysis.
%the four-dimensional unitarity and superconformal symmetry.
% along with the chiral algebra.
In four-dimensional $\NN=2$ SCFTs, the superconformal 
symmetry %symmetries
and the unitarity
impose strong constraints on superconformal multiplets appearing in %the
OPEs,
% of two Schur multiplets, 
which we call ``selection rules'' in the following.
In particular, %Among other,
 the selection rules 
\begin{align}
\schurB_{1}\times\schurB_{1}
&\sim
\schurB_{1}+\schurB_{2}+\sum_{\ell=0}^\infty\schurC_{0(\frac{\ell}{2},\frac{\ell}{2})}+\sum_{\ell=0}^\infty\schurC_{1(\frac{\ell}{2},\frac{\ell}{2})}+\cdots~,
\nonumber\\
\schurC_{0(0,0)}\times\schurC_{0(0,0)}
&\sim
\sum_{\ell=0}^\infty\schurC_{0(\ell,\ell)}+\sum_{\ell=0}^\infty\schurC_{1(\ell+\frac{1}{2},\ell+\frac{1}{2})}+\cdots~,
\nonumber\\
\schurC_{0(0,0)}\times\schurB_{1}
&\sim
\schurB_{1}+\sum_{\ell=0}^\infty\schurC_{1(\frac{\ell}{2},\frac{\ell}{2})}+\schurC_{0(\frac{\ell+1}{2},\frac{\ell+1}{2})}+\cdots~,
\label{eq:fusion0}
\end{align}
were used to derive the central charge bounds mentioned above. Here,
$\schurB_R$ and $\schurC_{R(j,\bj)}$ are two types of Schur multiplets
labeled by the $SU(2)_R$ charge $R$ and the spin $(j,\bj)$ of the
superconformal primary field, and the ellipses stand for non-Schur
multiplets.\footnote{For the precise definitions of
$\hat{\mathcal{B}}_R$ and $\widehat{\mathcal{C}}_{R(j,\bj)}$, see section \ref{secrev}. }
In particular, the $\schurB_1$ multiplet is a Schur multiplet including a flavor current, and the $\schurC_{0(0,0)}$ multiplet is the stress tensor multiplet.
The above %se 
selection rules (and unitarity) are crucial in deriving the central charge bounds.
For example, the bound $c\geq \frac{11}{30}$ was derived by
%the reality condition
interpreting the reality 
of the OPE coefficients for the second selection rule in \eqref{eq:fusion0}
in terms of the two-dimensional chiral algebra.\footnote{In this interpretation, it was assumed that $\schurC_{0(\frac{\ell}{2},\frac{\ell}{2})}$ for $\ell>0$ are absent in interacting $\mathcal{N}=2$ SCFTs since they involve higher spin currents.} 
This implies that identifying the selection rules for Schur multiplets provides a powerful tool to reveal universal constraints on general $\mathcal{N}=2$ SCFTs.

Moreover, the selection rules are also important in recovering four-dimensional OPEs from the two-dimensional chiral algebra. 
 Indeed, the 4d/2d correspondence of \cite{Beem:2013sza} implies that Schur operators with different quantum numbers could correspond to two-dimensional operators with the same quantum numbers. 
 Therefore, it is generically non-trivial to recover four-dimensional OPEs from two-dimensional OPEs. The selection rules, however, strongly constrain Schur multiplets appearing in the four-dimensional OPEs and therefore will be useful for reconstructing the four-dimensional OPEs from the associated chiral algebra.

In this paper, we study %two types of 
the selection rules 
for %of 
%OPEs of the form %Schur multiplets,
\begin{align}
%\schurB_{R_1} \times \schurB_{R_2}\,,\qquad
 \schurC_{0(0,0)} \times \OO^\mathrm{Schur}\,,
\label{eq:fusion1}
\end{align}
up to non-Schur multiplets, where $\OO^\mathrm{Schur}$ is an arbitrary Schur multiplet. Since the
Schur operator in the
stress tensor multiplet $\widehat{\mathcal{C}}_{0(0,0)}$ maps to the
Virasoro stress tensor in the associated chiral algebra, the selection
rules for \eqref{eq:fusion1} are particularly important in the study
of the 4d/2d correspondence.
%These selection rules are particularly important in the study of the 4d/2d correspondence. Indeed, $\schurB_{R}$ is a Schur multiplet whose bottom component is a Schur operator of dimension $2R$. 
%This Schur operator is called a ``Higgs branch operator'' because its
%vacuum expectation value often parameterizes the Higgs branch moduli
%space of vacua. Since a Higgs branch operator is always mapped to a
%generator of the chiral algebra \cite{Beem:2013sza} **************************, the selection rule for $\schurB_{R_1}\times \schurB_{R_2}$ strongly constrains the possible algebraic structure of associated chiral algebras.\footnote{By a generator, we mean an operator which is {\it not} realized as the normal ordered product or derivative of other operators.}
%On the other hand, the Schur operator in $\schurC_{0(0,0)}$ is mapped to the Virasoro stress tensor \cite{Beem:2013sza}. Therefore, the selection rules for $\schurC_{0(0,0)}\times
%\OO^\mathrm{Schur}$ 
In particular, they reveal how the four-dimensional operator 
associated with %of
a Virasoro primary is related to 
those %that 
of the Virasoro descendants. Indeed we find
%(see for example
%\eqref{C0Bsele})
 that, when four-dimensional Schur operators $\mathcal{O}$ and $\mathcal{O}'$
 respectively correspond to a Virasoro primary and its descendant in the
 associated chiral algebra, the $SU(2)_R$ charge of $\mathcal{O}$ is
 always smaller than or equal to that of $\mathcal{O}'$
(see for example \eqref{C0Bsele}). 
%the four-dimensional Schur operator of a
% two-dimensional primary operator is always smaller than or equal to
% those of its Virasoro descendants 
Note that the selection rules for $\widehat{\mathcal{C}}_{0(0,0)}
\times \widehat{\mathcal{C}}_{0(0,0)}$ and
$\widehat{\mathcal{C}}_{0(0,0)}\times \widehat{\mathcal{B}}_1$ were
already identified respectively in
\cite{Liendo:2015ofa} and \cite{Ramirez:2016lyk}, which we generalize to
\eqref{eq:fusion1} for all Schur multiplets
$\mathcal{O}^\mathrm{Schur}$ in this paper.

To derive the above selection rules, we study three-point functions of
the form %$\langle\schurB_{R_{1}}\schurB_{R_{2}}\OO \rangle$ and 
$\langle\schurC_{0(0,0)}{\OO_{1}}{\OO_{2}}\rangle$, where %$\OO,\,
$\OO_1$ and $\OO_2$ are arbitrary Schur multiplets.
Our strategy is to write down the most general ansatz for the
three-point functions and then impose the (semi-)shortening conditions
corresponding to the Schur multiplets. The same strategy was employed in
\cite{Kuzenko:1999pi,Liendo:2015ofa, Ramirez:2016lyk} to compute
several three-point functions.
%We restrict our studies to these two types, since for Schur multiplets in a general  higher-dimensional representation of both Lorentz group and $SU(2)_{R}$ symmetry the calculations are very complicated.
We stress that, since our analysis relies only on the (semi-)shortening conditions which purely follow from the superconformal algebra, our results are applicable to any four-dimensional $\NN=2$ SCFT.

Before studying the selection rules for
$\widehat{\mathcal{C}}_{0(0,0)}\times \mathcal{O}^\mathrm{Schur}$, we first
apply our strategy to the selection rules for 
\begin{align}
 \widehat{\mathcal{B}}_{R_1} \times \widehat{\mathcal{B}}_{R_2}\,,
\end{align}
 as a warm-up. While these rules were already
identified in \cite{Nirschl:2004pa},\footnote{See Eq.~(3.44) in
particular.} we believe it is worth showing an explicit derivation of
the rules. Moreover, we evaluate the most general expressions
for the three-point functions $\langle
\widehat{\mathcal{B}}_{R_1}\widehat{\mathcal{B}}_{R_2}
\mathcal{O}\rangle$ with $\mathcal{O}$ being an arbitrary Schur
multiplet, which contain more information than the selection rules.

Let us here make 
an observation %several observations 
on our OPE selection rules. %in the OPEs.
For some of the OPEs we study in this paper, the three-point function of the corresponding superconformal primary fields turns out to vanish even though three-point functions involving their descendants do not. 
This reflects the fact that the sum of the $U(1)_r$ charges of the superconformal primary fields in three-point functions is non-vanishing. 
On the other hand, we find that the sum of the $U(1)_r$ charges of {\it Schur operators} in these multiplets always vanishes, which suggests that the Schur operators play a central role in Schur multiplets. With this observation, we give a conjecture on the OPE selection rules for general Schur multiplets in section \ref{conc}.

The outline of this paper is as follow.
In section \ref{secrev}, we review the four-dimensional $\NN=2$ superconformal algebra and the (semi-)shortening conditions for the Schur multiplets, and also introduce a useful formalism \cite{Osborn:1998qu,Park:1999pd,Kuzenko:1999pi} to analyze the superconformal three-point functions.
In sections \ref{secBBfus} and \ref{secCOfus}, we derive the two types of three-point functions $\langle\schurB_{R_{1}}\schurB_{R_{2}}\OO^{\II} \rangle$ and $\langle\schurC_{0(0,0)} {\OO_{1}^{\II_{1}}}{\OO_{2}^{\II_{2}}} \rangle $ respectively. 
From these correlation functions, we present the $\schurB_{R_{1}}\times \schurB_{R_{2}}$ selection rule and the $\schurC_{0(0,0)} \times \OO^{\text{Schur}}$ selection rules.
section \ref{conc} is devoted to conclusions and discussions, where we conjecture
 more general selection rules between Schur multiplets as a natural generalization of our results.
In appendix \ref{appFierz}, we summarize the nilpotent structure of the Grassmann variables what we call Fierz identities, and appendices \ref{appCBO},\ref{appCDO}, and \ref{appCCC} are the details of our calculations.

\section{(Semi-)shortening conditions and three-point functions}\label{secrev}
In this section, we review the four-dimensional $\NN=2$ superconformal algebra and the short multiplets following \cite{Dolan:2002zh} and introduce a useful formalism constructed in \cite{Osborn:1998qu,Park:1999pd,Kuzenko:1999pi} for
the computations of correlation functions of SCFTs. 
We follow the convention of \cite{Beem:2013sza} unless otherwise stated.

\subsection{Superconformal shortening and semi-shortening conditions}

The four-dimensional $\NN=2$ superconformal algebra is the superalgebra $\mathfrak{su}(2,2|2)$, whose generators are the dilatation, translations, special conformal transformations, Lorentz transformations, Poincar\'e supercharges $Q^i_\alpha$ and $\widetilde{Q}_{\dot\alpha i}$, conformal supercharges $S_i^\alpha$ and $\widetilde{S}^{\dot\alpha i}$, and $SU(2)_{R}\times U(1)_{r}$ charges $\RR^{i}_{~j}$ and $r$.
Here $\alpha=\pm$ and $\dot{\alpha}=\dot{\pm}$ are the Weyl spinor indices and $ i,j=1,2$ are $SU(2)_R$ indices.

A general long multiplet of the four-dimensional $\NN=2$ superconformal algebra is labeled by five eigenvalues of the Cartan subalgebra for the primary state, namely, the conformal dimension $\Delta$, the Lorentz spin $(j,\bj)$, the irreducible representation $R$ of $SU(2)_{R}$, and the $U(1)_{r}$ charge $r$.\footnote{We take $R$ so that the Dynkin label for the irreducible $SU(2)_R$ representation is $2R$.}
Here, the superconformal primary field is defined as a state annihilated by all conformal supercharges, $S_i^\alpha$ and $\widetilde{S}^{\dot\alpha i}$.
We denote the superconformal primary field by $ \ket{\Delta,r}^{(i_{1}\cdots i_{2R})}_{(\alpha_{1}\cdots\alpha_{2j})(\dot{\alpha}_{1}\cdots \dot{\alpha}_{2\bj}) }$, where the parentheses in the scripts such as $\scriptsize{(i_{1}\cdots i_{2R_{1}})}$ denote the total symmetrization of the indices. 
Acting $Q^i_\alpha$ and $\widetilde{Q}_{\dot\alpha i}$ on the primary, we can generate $256(2R+1)(2j+1)(2\bj+1)$ components of the long multiplet.
These long multiplets satisfy unitarity bounds,
\begin{alignat}{2}
\Delta &\geq E_{i}\,, &\quad \text{if}\quad j_{i}\neq 0\,,&\label{bounds1}
\\
\Delta &= E_{i} -2 \quad \text{or} \quad \Delta \geq E_{i}\,, \quad&\text{if}\quad j_{i}= 0\,,&\label{bounds2}
\end{alignat}
where $E_{i}$ and $j_i$ are defined by
\begin{align}
E_{1}
:=
2R+2+2j_{1}+r \,,\qquad E_{2}
:=
 2R+2+2j_{2} -r\,, \qquad j_1 := j\,,\qquad j_2 := \bj\,.
\end{align}

An $\NN=2$ Poincar\'e supersymmetric field theory has 
a unitarity bound called the BPS bound, and 
if a long multiplet saturates the bound it becomes a short multiplet 
whose number of components is half the original one. 
The above unitarity bounds for $\mathcal{N}=2$ SCFTs play a similar role; 
when the eigenvalues $(\Delta,R,r, j,\bj)$ saturate the unitarity bounds \eqref{bounds1} or \eqref{bounds2}, a long multiplet is shortened.
If we consider the case of $ j=0, \Delta = E_{1} -2 $, 
the superconformal primary field 
$ \ket{\Delta}^{(i_{1}\cdots i_{2R})}_{(\dot{\alpha}_{1}\cdots\dot{\alpha}_{2\bj})}$
satisfies the condition
\begin{align}
\BB^1 :~ 
Q^{(i}_\alpha\ket{\Delta}^{i_{1}\cdots i_{2R})}_{(\dot{\alpha}_{1}\cdots\dot{\alpha}_{2\bj})} =0
\,, \quad \text{for}\quad \alpha=\pm.
\label{condB}
\end{align}
Similarly, for $\bj=0,\Delta =E_{2} -2 $, the superconformal primary field satisfies the condition 
\begin{align}
\bar{\BB}^{2} &:
\widetilde{Q}^{(i}_{\dot\alpha}\ket{\Delta}^{i_{1}\cdots i_{2R})}_{({\alpha}_{1}\cdots{\alpha}_{2j})} =0
\,, \quad \text{for}\quad \dot{\alpha}=\dot{\pm}.
\label{condBB}
\end{align}
These two conditions $\BB^1$ and $ \bar{\BB}^{2} $ are called shortening conditions.
On the other hand, for $\Delta = E_{1}$, the following 
 condition is possible:
\begin{align}
\CC^{1}:
\begin{cases}
\epsilon^{\alpha\beta}
Q^{(i}_\alpha\ket{\Delta}^{i_{1}\cdots i_{2R})}_{(\beta {\alpha}_{2}\cdots{\alpha}_{2j})}=0\,,
&\text{for}\quad j>0\,,
\\
\epsilon^{\alpha\beta}
Q^{(i}_\alpha Q^{i'}_{\beta}\ket{\Delta}^{i_{1}\cdots i_{2R})}=0\,,
&\text{for}\quad j=0\,.
\end{cases}\label{condC}
\end{align}
Similarly for $\Delta =E_{2}$, we can impose
\begin{align}
\bar{\CC}^{2}:&
\begin{cases}
\epsilon^{\dot{\alpha}\dot{\beta}}
\widetilde{Q}^{(i}_{\dot\alpha}
\ket{\Delta}^{i_{1}\cdots i_{2R})}_{(\dot{\beta} \dot{\alpha}_{2}\cdots\dot{\alpha}_{2\bj})} =0\,,
&\text{for}\quad \bj>0\,,
\\
\epsilon^{\dot{\alpha}\dot{\beta}}
\widetilde{Q}^{(i}_{\dot\alpha}\widetilde{Q}^{i'}_{\dot\beta}
\ket{\Delta}^{i_{1}\cdots i_{2R})}=0\,,
&\text{for}\quad \bj=0.
\end{cases}\label{condCC}
\end{align}
These two conditions $\CC^{1}$ and $\bar{\CC}^{2}$ are called semi-shortening conditions.

 The Schur multiplets are defined as multiplets satisfying a shortening condition or a semi-shortening condition for each of the chiralities.
There are four types of Schur multiplet, which are denoted as $\gschurB$, $\gschurD$, $\gschurDD$, and $\gschurC$ 
in the notation of \cite{Dolan:2002zh}.
 See table \ref{schurmult} for the definition of 
the four types and the relations among the quantum numbers
of their superconformal primary states.
 \begin{table}
 	\centering \renewcommand{\arraystretch}{1.5}
 	\begin{tabular}{|p{0.11596\textwidth}|p{0.11697\textwidth}|p{0.6\textwidth}|}
 		\hline
 		Multiplet &Condition& Conformal dimension and $U(1)_{r}
		 $ charge of the primary\\
 		\hline
 		$\gschurB$ &$\BB^{1}\cap\bar{\BB}^{2}$ & $\Delta=2R,\qquad\qquad\qquad r=0$,\\
 		\hline
 		$\gschurD$ & $\BB^{1}\cap\bar{\CC}^{2}$ & $\Delta=2R+\bj+1,\quad~~\quad r=\bj+1$,\\
 		\hline
 		$\gschurDD$ &$\CC^{1}\cap\bar{\BB}^{2}$ & $\Delta=2R+j+1,\quad~~\quad r=-j-1$,\\
 		\hline
 		$\gschurC$ &$\CC^{1}\cap\bar{\CC}^{2}$ & $\Delta=2R+j+\bj+2,~~~r=\bj-j$,\\
 		\hline
 	\end{tabular}
 	\caption{{
		The list of Schur multiplets. Their shortening 	conditions are shown in the middle column. 
	 The rightmost column denotes the conformal dimension and the $U(1)_{r}$
	 charge of the superconformal primary field. Here $(j,\bj)$ is the Lorentz spin of
	 the superconformal primary field field.}}
 		\label{schurmult}
 \end{table}

Some of these Schur multiplets contain important operators.
For example, the $\schurC_{0(0,0)}$ multiplet contains the stress-tensor operator and the $SU(2)_{R} \times U(1)_{r}$ conserved current operator.
We can regard the semi-shortening conditions as the conservation equations of these operators.
In this paper, we assume that the theory we are considering has a unique stress-tensor multiplet $\schurC_{0(0,0)}$. 
This assumption leads to a constraint 
on the three-point functions involving two stress tensor multiplets, as discussed later.
The $\schurC_{0(j,\bj)}$ multiplet is a higher-spin generalization of the stress-tensor multiplet $\schurC_{0(0,0)}$
and contains a higher spin current operator. 
It is expected that interacting (S)CFTs containing such a higher spin current have a decoupled free sector
\cite{Maldacena:2011jn,Alba:2013yda, Alba:2015upa}.
On the other hand, the $\schurB_{R}$ multiplets are half-BPS multiplets
whose superconformal primary field is annihilated by both 
$Q_\alpha{}^1$ and $\widetilde{Q}_{\dot\alpha 2}$.
As mentioned in section \ref{intro}, the $\schurB_1$ multiplet particularly contains a
 conserved flavor current.
 Finally, the $\schurD_{0(0,0)} \oplus\, \schurDD_{0(0,0)} $ multiplet is an $\NN=2$ free vector multiplet,
 whose superconformal primary field has the conformal dimension 1, and 
its (semi-)shortening conditions 
imply the massless equation of motion.

\subsection{Superspace formalism}
In this subsection, we review a useful superspace formalism 
following \cite{Osborn:1998qu,Park:1999pd,Kuzenko:1999pi}. We denote by
$z:= (x^\mu,\theta_i^{\alpha},\bar\theta^{\dot{\alpha} i})$ the
coordinate of the $\mathcal{N}=2$ superspace.
We then define 
chiral/anti-chiral variables $x_{\pm}^{\mu}$ and 
derivatives ${D^i_{\alpha}},{\bar{D}_{\dot{\alpha}i}}$ as
\footnote{Following the notation of \cite{Wess:1992cp},
	we define spinorial variables
	$x_{\alpha\dot{\alpha}}=x^{\mu} (\sigma_{\mu})_{\alpha\dot{\alpha}}\,,
	\tilde{x}^{\dot{\alpha}\alpha}=x^{\mu} (\bar{\sigma}_{\mu})^{\dot{\alpha}\alpha}$
	and the short-handed notations 
	$x^2=x^{\mu}x_{\mu}=-\frac{1}{2}\tr(x \tilde{x})\,,
	(\tilde{x}^{-1})_{\alpha\dot{\alpha}}=-\frac{1}{x^2}x_{\alpha\dot{\alpha}}$. We
	also use $\theta = (\theta^\alpha),\,\tilde\theta =
	(\theta_\alpha)=\epsilon_{\alpha \beta}\theta^{\beta},\, \bar \theta = (\bar\theta^{\dot\alpha})$ and 
	$\tilde{\bar\theta} = (\bar\theta_{\dot\alpha})=\epsilon^{\dot{\alpha} \dot{\beta}}\bar{\theta}_{\dot{\beta}}$. Complex conjugate of Grassmann variable is $(\theta^{\alpha}_{i})^{*}=\bar{\theta}^{\dot{\alpha}i}.$
}
\begin{align}
\begin{split}
x_{\pm}^{\mu}
&:= x^{\mu}\pm \Imu \theta_{i}^{\alpha}\sigma^{\mu}_{\alpha\dot{\alpha}}\bar{\theta}^{\dot{\alpha}i}\,,
\nonumber\\
{D^i_{\alpha}}
&:= \pdv{\theta^{\alpha }_i}
+\Imu (\sigma^{\mu}\bar{\theta}^i )_\alpha\pdv{x^\mu}
\,,\qquad
{\bar{D}_{\dot{\alpha}i}}
:=-\pdv{\bar{\theta}^{\dot{\alpha}i} }
-\Imu(\theta_i\sigma^{\mu})_{\dot{\alpha}}\pdv{x^\mu}\,.
\end{split}
\end{align}
In the superspace formalism, the (semi-)shortening conditions reviewed above
are expressed in terms of the covariant derivatives $D_\alpha^i$ and
$\bar{D}_{\dot\alpha i}$.
 For example, let $\LL_{(i_{1}\cdots i_{2R})}(z) $ be a superfield
 for the $\gschurB$ multiplet.
Then the shortening conditions for $\schurB_{R}$ imply
that any correlation function involving $\LL_{(i_{1}\cdots
i_{2R})}(z_{1})$ satisfies
 \begin{align}
 D^{\alpha}{}_{(i}\left\langle \LL_{i_{1}\cdots i_{2R})}(z) \OO_{1}(z_{1})\OO_{2}(z_{2})\cdots \right\rangle = 0 \,, 
 \label{exampB1}
 \nonumber \\
 \bar{D}^{\dot{\alpha}}{}_{(i} \left\langle \LL_{i_{1}\cdots i_{2R})}(z) \OO_{1}(z_{1})\OO_{2}(z_{2})\cdots \right\rangle=0\,,
 \end{align}
where $\OO_{1}(z_1),\,\OO_2(z_2),\,\cdots$ are superfields for general superconformal multiplets. 
 In the rest of this section, we review
 basic techniques to solve differential equations of this form.
We use them 
in sections \ref{secBBfus} and \ref{secCOfus}
to derive selection rules for the Schur multiplets.

To that end, we first introduce the following chiral and anti-chiral
 variables for given two points $z_{1}$, $z_{2}$ in the superspace:
\begin{align}
\begin{split}
x^\mu_{\bar{1}2}&
:=
 -x^\mu_{2\bar{1}}
:= 
x^\mu_{1-}-x^\mu_{2+}+2\Imu\theta_{1i}\sigma^{\mu}\bar{\theta}^i_{2}\,,
\nonumber\\
\theta_{12}&
:= \theta_{1}-\theta_{2}\,,\qquad \bar{\theta}_{12}
:= \bar{\theta}_{1}-\bar{\theta}_{2}\,,
\end{split}
\end{align}
where $x^\mu_{\bar{1}2}$ is anti-chiral for $z_{1}$ and chiral for $z_{2}$.
Next, we introduce superconformal covariant $U(2)_{R}$ and $SU(2)_{R}$ matrices respectively as
\begin{align}
\uu{i}{j}(z_{12})
:=
\delta_{i}^{j}+4\Imu
\theta_{12i}\tilde{x}_{\bar{1}2}^{-1}\bar{\theta}^j_{12}\,,\qquad
\hu{i}{j}(z_{12})
:= \left(\frac{x_{\bar{2}1}^2}{x^2_{\bar{1}2}}\right)^{\frac{1}{2}}\!\!\uu{i}{j}(z_{12})
\,, \qquad \det\hu{i}{j}(z_{12})=1\,,
\label{Rrep}
\end{align}
where $z_{12} := (x_{\bar 12}^\mu,\theta_{12_{\,} i}^{\alpha}
,\bar\theta_{12}^{{}_{\,}\dot\alpha i})$.
These matrices satisfy the relations
\begin{align}
\uu{i}{j}(z_{12})^{\dagger}\uu{j}{k}(z_{12})&=\uu{i}{j}(z_{21})\uu{j}{k}(z_{12})=
\delta_{i}^{k}\,,
\label{uu1}
\\
\hat{u}_{ij}(z_{12})&=-\hat{u}_{ji}(z_{21})\,,
\label{uminusu}
\end{align} 
where $\hat{u}_{ij}(z_{12}):=\hu{i}{k}(z_{12})\epsilon_{kj}$ \footnote{
	Except these $\hat{u}_{ij}(z_{12})$, all other $SU(2)_{R}$ indices are raised and lowered as 
	$C^{i}=\epsilon^{ij}C_{j}$, $C_{i}=\epsilon_{ij}C^{j}$.
}
with $\epsilon_{kj}$ being the $SU(2)_R$ invariant tensor.
We also define a spin $SL(2,\mathbb{C})$ covariant matrix
\begin{align}
I_{\alpha\dot{\alpha}}(x_{1\bar{2}})
:= \Imu \frac{x_{1\bar{2}\alpha\dot{\alpha}}}{\sqrt{x_{1\bar{2}}^2}}.
\label{spinrep}
\end{align}
These matrices \eqref{Rrep} and \eqref{spinrep} are building blocks of 
superconformal two-point functions.

Next, for given different three
points $z_{1}$, $z_{2}$, and $z_{3}$ in superspace,
 we introduce the superconformally covariant 
variable
$\BZ_{1}=(X_{1
\alpha\dot{\alpha}},\Theta^{i\alpha}_{1},\bar{\Theta}^{\dot{\alpha}}_{1i})$
by
\begin{align}
X_1
&:= 
\tilde{x}^{-1}_{1\bar{2}}\tilde{x}_{\bar{2}3}\tilde{x}^{-1}_{3\bar{1}}
\,,\qquad 
\bar{X}_1
:=
X_1^{\dagger}=-\tilde{x}_{1\bar{3}}^{-1}\tilde{x}_{\bar{3}2}\tilde{x}_{2\bar{1}}^{-1}\,,
\\
\tilde{\Theta}^i_{1\alpha}
&:=
\Imu\left(
(\tilde{x}^{-1}_{1\bar{3}})_{\alpha\dot{\alpha}}\bar{\theta}_{13}^{\dot{\alpha}i}
-(\tilde{x}^{-1}_{1\bar{2}})_{\alpha\dot{\alpha}}\bar{\theta}_{12}^{\dot{\alpha}i}
\right)\,,
\\
\TB{\Theta}_{1\dot{\alpha}i}
&:=
\Imu\left(\theta_{12i}^{\alpha}(\tilde{x}^{-1}_{\bar{1}2})_{\alpha\dot{\alpha}}
-\theta_{13i}^{\alpha}(\tilde{x}^{-1}_{\bar{1}3})_{\alpha\dot{\alpha}}
\right)\,.
\end{align}
The variable $\BZ_1$ transforms similarly to $z_{1}$.
In particular, the following identity will be important in our calculations below:
\begin{align}
X_{\alpha\dot{\alpha}}-\bar{X}_{\alpha\dot{\alpha}}
&=4\Imu \tilde{\Theta}^i_{\alpha}\TB{\Theta}_{\dot{\alpha}i}\,.
\label{barXtoX}
\end{align} 
Similar variables $\BZ_{2} $ and $\BZ_{3}$ are defined as
 cyclically permuting 
$z_1, z_2$ and $z_3$ in the above definition.\footnote{
The conformal dimension of  $(X_{\alpha\dot{\alpha}},\Theta^{i\alpha},\bar{\Theta}^{\dot{\alpha}}_{i})$
 is $(1,\frac{1}{2},\frac{1}{2})$, and the $U(1)_{r}$ charge is $(0,\frac{1}{2},-\frac{1}{2})$, respectively.} These $\BZ$ variables play central role in expressing three-point functions. 
We also define $SU(2)_{R}$ matrices 
using $\BZ_1$ as
\begin{align}
\UU{i}{j}(\BZ_{1})
&:=
\uu{i}{k}(z_{12})\uu{k}{l}(z_{23})\uu{l}{j}(z_{31})
=\delta_{i}^{j}-4\Imu \TB{\Theta}_{1i}X_1^{-1}\tilde{\Theta}^j_1\,,
\\[1mm]
\UU{i}{j} {}^{\dagger}(\BZ_{1})
&=\vb{u}^{-1}(\BZ_{1})=
\delta_{i}^{j}+4\Imu \TB{\Theta}_{1i}\bar{X}_1^{-1}\tilde{\Theta}^j_1\,,
\\
\HU{i}{j}(\BZ_{1})
&:= 
\left(\frac{\bar{X}^2_1}{X^2_1}\right)^{1/2}\UU{i}{j}(\BZ_{1})\,.
\end{align}

\subsection{(Semi-)shortening conditions for three-point functions}
Let us now consider a three-point function of three quasi
superfields $\Phi_{\II_i}(z_{i})$
for $i=1,2,3$.
We denote their conformal dimension 
and $U(1)_{r}$ charge 
by
 $(q_i+\bar{q}_{i})$ and $(\bar{q}_{i}-q_i)$ respectively. 
The subscript $\II_i$ expresses $SU(2)_R$ and $SL(2, \mathbb{C})$ indices collectively.
A general three-point function $\langle 
\Phi_{\II_1}(z_{1})\Phi_{\II_2}(z_{2})\Phi_{\II_3}(z_{3})
\rangle $ is 
written as
\begin{align}
\langle 
\Phi_{\II_1}(z_{1})\Phi_{\II_2}(z_{2})\Phi_{\II_3}(z_{3})
\rangle=
\frac{T^{\JJ_1}_{\II_1}\left[\hat{u}(z_{13}),I(x_{1\bar{3}},x_{3\bar{1}})\right]
	T^{\JJ_2}_{\II_2}\left[\hat{u}(z_{23}), I(x_{2\bar{3}},x_{3\bar{2}})\right]}
{ \left(x^2_{1\bar{3}}\right)^{q_1} \left(x^2_{\bar{1}3}\right)^{\bar{q}_1}
	\left(x^2_{2\bar{3}}\right)^{q_2} \left(x^2_{\bar{2}3}\right)^{\bar{q}_2}}
{\Large H}_{\JJ_1\JJ_2\II_3}(\BZ_3)\,,
 \label{3ptfunc_inH}
\end{align}
where $ T^{\JJ_1}_{\II_1}$ and $T^{\JJ_2}_{\II_2}$ are some functions 
composed of \eqref{Rrep} and \eqref{spinrep}
in the representation of $SU(2)_R\times SL(2,\mathbb{C})$ specified by 
$\II_1$ and $\II_2$.
On the other hand, the function $H(\BZ_{3})$ satisfies following homogeneity property 
\begin{align}
&{\Large H}_{\JJ_1\JJ_2\II_3}(\lambda\bar{\lambda}X,\lambda \Theta,\bar{\lambda}\bar{\Theta})
=\lambda^{2a}\bar{\lambda}^{2\bar{a}}{\Large H}_{\JJ_1\JJ_2\II_3}(X, \Theta,\bar{\Theta})\,,
\end{align}
where $a$ and $\bar a$ are fixed by
\begin{align}
a-2\bar{a}=\bar{q}_{1}+\bar{q}_{2}-{q}_{3}\,,\qquad 
\bar{a}-2a={q}_{1}+{q}_{2}-\bar{q}_{3}\,.
\end{align}
Therefore $H(\BZ_{3})$ has the conformal dimension $q_{3}+\bar{q}_{3}-(q_{1}+\bar{q}_{1})-(q_{2}+\bar{q}_{2})$ 
and the $U(1)_{r}$ charge $(\bar{q}_{3}-q_{3}+\bar{q}_{1}-q_{1}+\bar{q}_{2}-q_{2})$.
The function $H(\BZ_{3})$ is not fully determined by the global
superconformal symmetry.
When some of the $\Phi_{\mathcal{I}_i}(z_i)$ correspond to a
(semi-)short multiplet, their shortening conditions restrict the form of $H(\BZ_3)$.
For example in \cite{Kuzenko:1999pi,Liendo:2015ofa, Ramirez:2016lyk},
$H(\BZ_{3})$ is determined in such cases, up to an overall constant.

The formalism \eqref{3ptfunc_inH} is very useful when we consider the (semi-)shortening conditions such as \eqref{exampB1}.
Since the prefactor in \eqref{3ptfunc_inH}
\begin{align}
\frac{T^{\JJ_1}_{\II_1}\left[\hat{u}(z_{13}),I(x_{1\bar{3}},x_{3\bar{1}})\right]
	T^{\JJ_2}_{\II_2}\left[\hat{u}(z_{23}), I(x_{2\bar{3}},x_{3\bar{2}})\right]}
{\left(x^2_{1\bar{3}}\right)^{q_1} \left(x^2_{\bar{1}3}\right)^{\bar{q}_1}
	\left(x^2_{2\bar{3}}\right)^{q_2} \left(x^2_{\bar{2}3}\right)^{\bar{q}_2}}\,,
\end{align}
can be factorized into two-point functions
$\langle\Phi_{\II_1}(z_{1})\bar{\Phi}^{\JJ_1}(z_{3})\rangle$ and
$\langle\Phi_{\II_2}(z_{2})\bar{\Phi}^{\JJ_2}(z_{3})\rangle$, it 
trivially satisfies the (semi-)shortening conditions.
This implies that the (semi-)shortening conditions only constrain the function $H(\BZ_3)$.
Therefore, hereafter, we focus on $H(\BZ_{3})$. 
It is easy to find that 
 $D_{1}$, $\bar{D}_{1}$, $D_{2}$ and $\bar{D}_{2}$ act on $H(\BZ_{3})$ as
\begin{align}
\begin{split}
&D^{i}_{1 \alpha}H(\BZ_{3})
=-\Imu (\tilde{x}^{-1}_{\bar{3}1})_{\alpha\dot{\beta}}
\uu{j}{i}(z_{31})\bar{\mathcal{D}}^{\dot{\beta}j}H(\BZ_{3})\,,
\\
&\bar{D}_{1\dot{\alpha}i}H(\BZ_{3})
=\Imu(\tilde{x}^{-1}_{\bar{1}3})_{\beta\dot{\alpha}}
\uu{i}{j}(z_{13})\mathcal{D}^{\beta}_{j}H(\BZ_{3})\,,
\\
&D^{i}_{2\alpha}{\Large H}(\BZ_3)
=\Imu(\tilde{x}^{-1}_{\bar{3}2})_{\alpha\dot{\beta}}
\uu{j}{i}(z_{32})\bar{\QQ}^{\dot{\beta}j}H(\BZ_{3})\,,
\\
&\bar{D}_{2\dot{\alpha}i}{\Large H}(\BZ_3)
=-\Imu(\tilde{x}^{-1}_{\bar{2}3})_{\beta\dot{\alpha}}
\uu{i}{j}(z_{23})\QQ^{\beta}_{j}H(\BZ_{3})\,,
\end{split}
\end{align}
where derivatives 
$ \bar{\mathcal{D}}^{j\dot{\beta}},\mathcal{D}^{\beta}_{j},\QQ^{\alpha}_i$
and $\bar{\QQ}^{\dot{\alpha}i} $ are defined respectively as 
\begin{alignat}{2}
\mathcal{\bar{D}}^{\dot{\alpha}i}
&:=\pdv{\TB{\Theta}_{3\dot{\alpha}i}}
\,,\quad 
\mathcal{D}^{\alpha}_{i}
& := &
\pdv{\tilde{\Theta}^i_{3 \alpha}}
+4\Imu \TB{\Theta}_{3 i\dot{\alpha}}\pdv{X_{3\alpha\dot{\alpha}}}\,,
\nonumber\\	
\QQ^{\alpha}_{i}
& :=
\pdv{\tilde{\Theta}^i_{3 \alpha}}
\,,\quad 
\bar{\QQ}^{\dot{\alpha} i}
& := &
\pdv{\TB{\Theta}_{3 \dot{\alpha} i}}
-4\Imu \tilde{\Theta}^i_{3 \alpha }\pdv{X_{3\alpha\dot{\alpha}}}.
\end{alignat}
Moreover, quadratic derivatives such as $D^{(i}_{1
 \alpha}D^{i')\alpha}_{1 }H(\BZ_{3})$
are also concisely written in terms of
$\bar{\mathcal{D}}^{j\dot{\beta}},\mathcal{D}^{\beta}_{j},\QQ^{\alpha}_i$
and $\bar{\QQ}^{\dot{\alpha}i}$. For instance,
\begin{align}
D^{(i}_{1 \alpha}D^{i')\alpha}_{1}H(\BZ_{3})
&=- \frac{\uu{j}{i}(z_{31})\uu{j'}{i'}(z_{31})}{{x}^{2}_{\bar{3}1}}
\bar{\mathcal{D}}^{~j'}_{\dot{\beta}}\bar{\mathcal{D}}^{\dot{\beta}j}H(\BZ_{3})\,.
\end{align}
Therefore, the (semi-)shortening conditions are now translated into 
partial differential equations of $H(\BZ_{3})$
with respect to $\BZ_{3}$.

While the (semi-)shortening conditions of the first and second superfields,
$\Phi_{\II_1}(z_{1})$ and $\Phi_{\II_2}(z_{2})$, are easily expressed
 as partial differential equations for $H(\BZ_3)$,
it is not straightforward to translate the conditions for the third superfield $\Phi_{\II_3}(z_{3})$ 
into a similar equation for $H(\BZ_3)$.
 To consider the (semi-)shortening conditions of the third superfield, we change the variable 
from $\BZ_3$ to $\BZ_2$.\footnote{Here, we can also use $\BZ_1$ instead of $\BZ_2$.}
Indeed, using the cyclicity of $z_1,z_2$ and $z_3$, the correlation function \eqref{3ptfunc_inH} is also expressed as 
\begin{align}
\frac{T'^{\JJ_1}_{\II_1}\left[\hat{u}(z_{12}),I(x_{1\bar{2}},x_{2\bar{1}})\right]
	T'^{\JJ_3}_{\II_3}\left[\hat{u}(z_{32}), I(x_{3\bar{2}},x_{3\bar{2}})\right]}
{ \left(x^2_{1\bar{2}}\right)^{q_1} \left(x^2_{\bar{1}2}\right)^{\bar{q}_1}
	\left(x^2_{3\bar{2}}\right)^{q_3}\left(x^2_{\bar{3}2}\right)^{\bar{q}_3}}
{\Large G}_{\JJ_1\II_{2}\JJ_3}(\BZ_{2})\,,\label{defGfunc}
\end{align}
for some function $G(\BZ_2)$. 
The action of $D_{3\alpha}^{i}$ and $ \bar{D}_{3\dot{\alpha}i}$ on the $G(\BZ_{2})$ are given by 
\begin{align}
D_{3\alpha}^{i}G(\BZ_{2})
&=-\Imu(\tilde{x}_{\bar{2}3}^{-1})_{\alpha\dot{\beta}}\uu{j}{i}(z_{23})
\bar{\mathcal{S}}^{\dot{\beta}j}G(\BZ_{2})\,,
\label{barSderib}
\\
\bar{D}_{3\dot{\alpha}i}G(\BZ_{2})
&=\Imu(\tilde{x}_{\bar{3}2}^{-1})_{\beta\dot{\alpha}}
\uu{i}{j}(z_{32})\mathcal{S}_{j}^{\beta}G(\BZ_{2})\,,
\label{Sderib}
\end{align}
where the derivatives are now defined by
\begin{align}
\mathcal{S}^{\alpha}_i
:=
\pdv{\tilde{\Theta}^i_{2\,\alpha}}
+4\Imu \TB{\Theta}_{i\dot{\alpha}}\pdv{X_{2\,\alpha\dot{\alpha}}}
\,,\quad 
\mathcal{\bar{S}}^{\dot{\alpha}i}
:=
\pdv{\TB{\Theta}_{2\,\dot{\alpha}i}}
+4\Imu
 \tilde{\Theta}^i_{2\,\alpha}\pdv{\bar{X}_{2\,\alpha\dot{\alpha}}}\,.
\label{eq:S_and_barS}
\end{align}
As shown in \cite{Park:1999pd,Kuzenko:1999pi}, 
$\BZ_{3}$ and $\BZ_{2}$ are related as 
\begin{align}
\begin{split}
\tilde{x}_{\bar{2}3}X_3\tilde{x}_{\bar{3}2}=-(\bar{X}_2)^{-1}
\,,&\quad 
\tilde{x}_{\bar{2}3}\bar{X}_3\tilde{x}_{\bar{3}2}=-(X_2)^{-1}\,,
\\
\tilde{x}_{\bar{3}2}\tilde{\Theta}^i_{2}\uu{i}{j}(z_{23})=-X_3^{-1}\tilde{\Theta}^j_{3}
\,,&\quad 
\uu{i}{j}(z_{32})\TB{\Theta}_{2j}\tilde{x}_{\bar{2}3}=\TB{\Theta}_{3i}\bar{X}_{3}^{-1}\,.
\end{split}
\end{align}
Using these relations, 
we see that 
the function ${\Large G}_{\JJ_1\II_{2}\JJ_3}(\BZ_{2})$ 
is related to
 ${\Large H}_{\JJ_1\JJ_2\II_3}(\BZ_{3})$ 
by
\begin{align}\label{gneGtoH}
{\Large G}_{\JJ_1\II_{2}\JJ_3}({\BZ}_{2})
=\frac{
	\TRep{\JJ_1}{\LL}{\hat{\vb{u}}^{\dagger}({Z}_2),I(X_2, \bar{X}_2)}}
{\left(\bar{X}^2_2\right)^{\bar{q}_1}\left(X^2_2\right)^{q_1}}
{\Large H}_{\LL\II_{2}\JJ_3}
\left(\bar{X}_{2}^{-1},-\Imu\bar{X}^{-1}_{2}\tilde{\Theta}_{2},\Imu\TB{\Theta}_{2}X^{-1}_{2}\right).
\end{align}
It is important to consider the third superfield conditions since it is insufficient to fix the function $H(\BZ_{3})$ only 
considering the first and second superfields of the (semi-)shortening conditions in section \ref{secCDO}.

In the following sections, we will use the above formalism and
techniques to study the three-point functions of Schur multiplets.

\section{$\schurB_{R_{1}} \times \schurB_{R_{2}} $ Fusion} 
\label{secBBfus}

In this section,  we 
 %identify the selection rules for $\schurB_{R_1}\times \schurB_{R_2}$.
study the most general expressions for three-point functions of two half-BPS Schur multiplets $\gschurB$ and an arbitrary Schur multiplet $\OO^\II$.\footnote{
By definition, $SU(2)_{R}$ irreducible representation $R$ of
$\schurB_{R}$ must be $R\geq \frac{1}{2}$.} 
%For that purpose, we study three-point functions of two half-BPS Schur
%multiplets $\gschurB$ and an arbitrary Schur multiplet $\OO^\II$. 
Our result is particularly consistent with the fusion rules for
$\widehat{\mathcal{B}}_{R_1}\times \widehat{\mathcal{B}}_{R_2}$ which
were first
obtained in \cite{Nirschl:2004pa}.

The general expression for the three-point function $\langle\schurB_{R_{1}}\schurB_{R_{2}}\OO^{\II} \rangle$ is given by
	 \begin{align}
	 \begin{split}
	 \langle	\LL_{(i_{1}\cdots i_{2R_{1}})}(z_{1})\LL_{(j_{1}\cdots j_{2R_{2}})}(z_{2}) \OO^{\II}(z_{3}) \rangle
	 &=\frac{\hu{i_{1}}{l_{1}}(z_{13}) \cdots \hu{i_{2R_{1}}}{l_{2R_{1}}}(z_{13})
	 	\hu{j_{1}}{m_{1}}(z_{23})\cdots\hu{j_{2R_{2}}}{m_{2R_{2}}}(z_{23})}
	 {\left(x^2_{\bar{3}1} x^2_{\bar{1}3}\right)^{R_1} \left(x^2_{\bar{3}2}x^2_{\bar{2}3}\right)^{R_2}}
	 \\
	 &~\times {\Large H}_{(l_{1} \cdots l_{2R_{1}}) (m_{1}\cdots m_{2R_{2}})}^{\II}(\BZ_{3})\,,
	 \end{split}
	 \label{BBOgen}
\end{align}
where $\LL_{(i_{1}\cdots i_{2R})}(z)$ is the superfield of $\schurB_R$ multiplet, and the parentheses denote the total symmetrization of the indices.
Hereafter, we will often omit the parentheses with the understanding that the indices associated with the same Latin and Greek alphabet letters are always totally symmetrized.

Each of the $\schurB_{R_{1}}$ and $\schurB_{R_{2}}$ multiplets satisfies two shortening conditions as shown in table \ref{schurmult}.
As we have mentioned in the previous section, the shortening conditions
 are translated into differential equations for $H(\BZ_{3})$. 
 For the two $\schurB_R$ multiplets, the differential equations are written as
	 \begin{align}
	 \mathcal{D}^{\alpha}_{(l}{\Large H}_{l_{1} \cdots l_{2R_{1}}) (m_{1}\cdots m_{2R_{2}})}^{\II}(\BZ_{3})
	 &=0\,,
	 \label{BcondD} 
	 \\
		\bar{\mathcal{D}}^{\dot{\alpha}}_{ (l}{\Large H}_{l_{1} \cdots l_{2R_{1}}) (m_{1}\cdots m_{2R_{2}})}^{\II}(\BZ_{3})
		&=0\,,
		\label{BcondDbar}
		\\
		\QQ^{\alpha}_{ (m}{\Large H}_{ m_{1}\cdots m_{2R_{2}})(l_{1} \cdots l_{2R_{1}})}^{\II}(\BZ_{3})
		&=0\,,
		\label{BcondQ}
		\\
		\bar{\QQ}^{\dot{\alpha} }_{(m}{\Large H}_{m_{1}\cdots m_{2R_{2}})(l_{1} \cdots l_{2R_{1}})}^{\II}(\BZ_{3})&=0\,.
		\label{BcondQbar}
	 \end{align}

 It is easy to solve \eqref{BcondDbar} and \eqref{BcondQ}, since these are merely first-order linear equations for $\Theta $ or $\bar{\Theta}$. 
In contrast, \eqref{BcondD} and \eqref{BcondQbar} contain both $X$ and $\Theta$ (or $\bar{\Theta}$) derivations 
and therefore are more complicated.
However, if we use the $\bar\BZ_3 := (\bar{X}_{3},\Theta_{3}, \bar{\Theta}_{3})$ coordinate
 instead of $\BZ_3=(X_{3},\Theta_{3}, \bar{\Theta}_{3})$, the two equations \eqref{BcondD} and \eqref{BcondQbar}
 become simpler, because $\mathcal{D}^{\alpha}_i$ and $\mathcal{\bar{Q}}^{\dot{\alpha} i}$ are
 expressed in terms of $\bar\BZ_3$ as
 \begin{align}
 \mathcal{D}^{\alpha}_i=\pdv{\tilde{\Theta}^i_{3 \alpha}}
\,,\qquad 
 \mathcal{\bar{Q}}^{\dot{\alpha} i}=\pdv{\TB{\Theta}_{3 \dot{\alpha} i}}\,.
 \end{align}
Indeed, the most general solution to
\eqref{BcondDbar} and \eqref{BcondQ} is simply expressed in terms of $\BZ_3$ as %\cite{***}
\begin{align}
\begin{split}
H^{\II}_{(l_{1} \cdots l_{2R_{1}}) (m_{1}\cdots m_{2R_{2}})}({\BZ}_{3})
&=f^{\II}_{l_{1} \cdots l_{2R_{1}},m_{1}\cdots m_{2R_{2}}}(X_{3})
+\Theta^{\alpha}_{3m_{1}}\bar{\Theta}^{\dot{\alpha}}_{3l_{1}}
g^{\II}_{l_{2} \cdots l_{2R_{1}}, m_{2}\cdots m_{2R_{2}},\alpha \dot{\alpha}}(X_{3})
\\
&~+\Theta_{3m_{1}}\Theta_{3m_{2}}\bar{\Theta}_{3l_{1}}\bar{\Theta}_{3l_{2}}
h^{\II}_{l_{3} \cdots l_{2R_{1}}, m_{3}\cdots m_{2R_{2}}}(X_{3})\,,
\end{split}
\label{BBgeneral_ans}
\end{align} 
while that of \eqref{BcondD} and
\eqref{BcondQbar} is written in terms of $\bar{\BZ}_{3}$ as
\begin{align}
\begin{split}
H^{\II}_{(l_{1} \cdots l_{2R_{1}}) (m_{1}\cdots m_{2R_{2}})}(\bar{\BZ}_{3})
&=\bar{f}^{\II}_{l_{1} \cdots l_{2R_{1}},m_{1}\cdots m_{2R_{2}}}(\bar{X}_{3})
+\bar{\Theta}^{\dot{\alpha}}_{3 m_{1}}
\Theta^{\alpha}_{3 l_{1}}\bar{g}^{\II}_{l_{2} \cdots l_{2R_{1}}, m_{2}\cdots m_{2R_{2}}, \alpha \dot{\alpha}}(\bar{X}_{3})
\\
&~+\bar{\Theta}_{3m_{1}}\bar{\Theta}_{3m_{2}}\Theta_{3l_{1}}\Theta_{3l_{2}}
\bar{h}^{\II}_{l_{3} \cdots l_{2R_{1}}, m_{3}\cdots m_{2R_{2}}}(\bar{X}_{3})\,.
\end{split}
\label{BBgeneral_ans_bar}
\end{align}
Here, we use the short-hand notation
\begin{align}
\Theta_{3m_{1}}\Theta_{3m_{2}}
:=
\Theta_{3 m_{1}}^{\alpha}\epsilon_{\alpha\beta}\Theta_{3 m_{2}}^{\beta}
\,, \qquad 
\bar{\Theta}_{3m_{1}}\bar{\Theta}_{3m_{2}}
:=
\bar{\Theta}^{\dot{\alpha}}_{3m_{1}}\epsilon_{\dot{\alpha}\dot{\beta}}\bar{\Theta}^{\dot{\beta}}_{3m_{2}}\,.
\end{align}
The above two 
expressions, \eqref{BBgeneral_ans_bar} and \eqref{BBgeneral_ans},
must be equal 
under the relation
\eqref{barXtoX}. Therefore, our strategy is to rewrite
\eqref{BBgeneral_ans_bar} in terms of $\BZ_3$ by using \eqref{barXtoX} and restrict 
the parameters in the expression to be consistent with \eqref{BBgeneral_ans}.
This gives us the most general solution to the equations \eqref{BcondD}--\eqref{BcondQbar}.

After solving \eqref{BcondD}--\eqref{BcondQbar}, 
 we have to check it also satisfies the (semi-)shortening 
conditions of the third superfield $\OO^{\II}(z_{3})$.
 In this process, we relate $H(\BZ_3)$ to $G(\BZ_2)$ using
 \eqref{gneGtoH} and see if the $G(\BZ_2)$ satisfies the differential
 equations corresponding to the third set of (semi-) shortening conditions.

Below, we apply this strategy to evaluate the most general expression for the three-point
function $\langle\schurB_{R_{1}}\schurB_{R_{2}}\OO^{\II} \rangle$.

\subsection{$\langle{\schurB}_{R_{1}}{\schurB}_{R_{2}}{\schurB}_{R_3}\rangle$}

Let us first consider the case of $\OO^\II$ in the
$\schurB_{R_3}$ multiplet.
In this case, the function $H(\BZ_{3})$ has 
 dimension $-2R := 2R_{3}-2R_{1}-2R_{2}$ and 
vanishing $U(1)_r$ charge.
Note here that $R_{1}$, $R_{2}$, and $R_{3}$ are constrained by the inequalities $0\leq R_{1}+R_{3}-R_{2}$ and $0\leq R_{1}+R_{2}-R_{3}$.
In other words, $R_{3} $ must be such that
\begin{align}
|R_{1}-R_{2}|\leq R_{3}\leq R_{1}+R_{2}.
\end{align}
Moreover, since the primary of $\schurB_R$ is a scalar, so is the $H(\BZ_3)$. 
Therefore, the function $H(\BZ_{3})$ only carries $SU(2)_{R}$ indices.
The most general ansatz for $H(\BZ_3)$ is then written as
\footnote{$M_{ml}:= \Theta_{ m}^{~\alpha}	X_{ \alpha\dot{\alpha}}\bar{\Theta}^{\dot{\alpha}}_{~l} $. }
\begin{align}
H(\BZ_{3})&=\frac{1}{(X^{2}_{3})^{R}}\left(
	A' \epsilon_{l_{1}m_{1}} \epsilon_{l_{2}m_{2}}
	+B'\frac{M_{3m_{1}l_{1}}}{X^2_{3}} \epsilon_{l_{2}m_{2}}
	+C' \frac{M_{3m_{1}l_{1}}}{X^2_{3}} \frac{M_{3m_{2}l_{2}}}{X^2_{3}} 	 
	\right)\epsilon_{l_{3}m_{3}}\cdots \epsilon_{l_{R}m_{R}}
	\nonumber\\
	&~\times( \epsilon_{l_{R+1}k_{1}}\cdots \epsilon_{l_{2R_{1}}k_{2R_{1}-R}})
(\epsilon_{m_{R+1}k_{2R_{1}-R+1}}\cdots \epsilon_{m_{2R_{2}}k_{2R_{3}}})\,,
\end{align}
where $R\in \mathbb{Z}_{\geq0}$, and $k_1\cdots k_{2R_3}$ are the
$SU(2)_R$ indices associated with $\schurB_{R_3}$.
As mentioned above, indices associated with the same Latin and Greek alphabet letters,
 such as $l_{1} \cdots l_{2R_{1}}$, are totally symmetrized.
On the other hand, the same function should also be written in terms of
 $\bar \BZ_3$. The most general expression in terms of $\bar \BZ_3$
is given by
\begin{align}
H(\bar{\BZ}_{3})&=\frac{1}{(\bar{X}^{2}_{3})^{R}}\left(
A\epsilon_{l_{1}m_{1}}\epsilon_{l_{2}m_{2}}
+B\frac{\bar{M}_{3l_{1}m_{1}}}{\bar{X_{3}}^2} \epsilon_{l_{2}m_{2}}
+C \frac{\bar{M}_{3l_{1}m_{1}}}{\bar{X_{3}}^2} 
\frac{\bar{M}_{3l_{2}m_{2}}}{\bar{X_{3}}^2} 
\right)\epsilon_{l_{3}m_{3}}\cdots \epsilon_{l_{R}m_{R}}
\nonumber\\ 
&~\times(\epsilon_{l_{R+1}k_{1}}\cdots \epsilon_{l_{2R_{1}}k_{2R_{1}-R}})
(\epsilon_{m_{R+1}k_{2R_{1}-R+1}}\cdots \epsilon_{m_{2R_{2}}k_{2R_{3}}})\,.
\label{BBBbarX}
\end{align}
For the above two expressions to be consistent, the coefficients $A,B$
and $C$ have to satisfy some conditions. To identify the conditions,
we change the variables from $\bar \BZ_3$ to $\BZ_3$ in \eqref{BBBbarX}
by using
\eqref{barXtoX}.
Using the Fierz identities summarized in appendix \ref{appFierz}, 
we see that the conditions are
\begin{align}
B=4\Imu R A \,, \quad C=-16\frac{R(R-1)}{2} A\,,
 \end{align}
for an arbitrary constant $A$.
Up to an overall constant, the function $H(\BZ_3)$ is written as
 \begin{align}
H(\BZ_{3})
=\frac{\vb{u}(\BZ_{3})_{l_{1}m_{1}}\cdots\vb{u}(\BZ_{3})_{l_{R}m_{R}}}{(X^{2}_{3})^{R}}
( \epsilon_{l_{R+1}k_{1}}\cdots \epsilon_{l_{2R_{1}}k_{2R_{1}-R}})
(\epsilon_{m_{R+1}k_{2R_{1}-R+1}}\cdots \epsilon_{m_{2R_{2}}k_{2R_{3}}})\,. \label{eq:BBBans}
\end{align}
This is the most general expression for $H(\BZ_3)$ satisfying \eqref{BcondD}--\eqref{BcondQbar}.

Although it satisfies the shortening conditions of the ${\schurB}_{R_{1}}$ and ${\schurB}_{R_{2}}$ multiplets,
it is non-trivial whether the expression \eqref{eq:BBBans} satisfies the
shortening conditions of the third multiplet ${\schurB}_{R_3}$. To check
the shortening conditions for $\schurB_{R_3}$, let us
relate the $H(\BZ_3)$ to $G(\BZ_2)$ using \eqref{gneGtoH}. Indeed, as
reviewed above, the correlation function \eqref{BBOgen} is also written as 
\begin{align}
\frac{\hu{i_{1}}{l_{1}}(z_{12})\cdots\hu{i_{2R_{1}}}{l_{2R_{1}}}(z_{12})
	\hu{k_{1}}{n_{1}}(z_{32})\cdots\hu{k_{2R_{3}}}{n_{2R_{3}}}(z_{32})}{
	(x^2_{1\bar{2}}x^2_{2\bar{1}})^{R_{1}}(x^2_{3\bar{2}}x^2_{2\bar{3}})^{R_{3}}}
G_{(l_{1}\cdots l_{2R_{1}})(j_{1}\cdots j_{2R_{2}})(n_{1}\cdots n_{2R_{3}})}(\BZ_{2}).
\end{align}
The function $G(\BZ_2)$ is uniquely fixed by $H(\BZ_3)$ via \eqref{gneGtoH}, i.e.,
\begin{align}
G(\BZ_{2})&=
\frac{\vb{u}^{\dagger}_{l_{1}n_{1}}(\BZ_{2})\cdots \vb{u}^{\dagger}_{l_{R'}n_{R'}}(\BZ_{2})}{(\bar{X}_2^2)^{R'}}
(\epsilon_{l_{R'+1}j_{1}}\cdots \epsilon_{l_{2R_{1}}j_{R}})(\epsilon_{j_{R+1}n_{R'+1}}\cdots \epsilon_{j_{2R_{2}}n_{2R_{3}}})\,,
\label{eq:G_for_BBB}
\end{align}
where $R':= R_{1}+R_3-R_{2}$.
Now, the shortening conditions for $\schurB_{R_3}$ are
written as
\begin{align}
\mathcal{S}_{(n}^{\alpha}
G_{n_{1}\cdots n_{2R_{3}})}(\BZ_{2})=0
\,,\qquad 
\bar{\mathcal{S}}_{(n}^{\dot{\alpha}}G_{n_{1}\cdots
 n_{2R_{3}})}(\BZ_{2})=0\,,
\end{align}
with $\mathcal{S}^\alpha_n$ and $\bar{\mathcal{S}}^{\dot\alpha n}$
defined by \eqref{eq:S_and_barS}. Since \eqref{eq:BBBans} trivially
satisfies the above two equations, the expression \eqref{eq:G_for_BBB} also
satisfies the shortening conditions for the third Schur multiplet
$\schurB_{R_3}$.

In the rest of this paper, we omit the subscript $3$ of $X_3,\Theta_3$
and $\bar\Theta_3$ in the expression for $H(\BZ_3)$. Similarly, we omit
the subscript $2$ of $X_2,\Theta_2$ and $\bar \Theta_2$ in the
expression for $G(\BZ_2)$.

\subsection{$\langle{\schurB}_{R_{1}}{\schurB}_{R_{2}}{\schurDD_{R_{3}(j,0)}}\rangle$}
\label{subsec:BBD}
Let us turn to the case of $\OO^\II$ in the $\schurDD_{R(j,0)}$ multiplet.\footnote{Note here that the result for $\OO^\II$ in the $\schurD_{R(0,j)}$ multiplet follows from
	this case by CPT.}
In this case, the function $H(\BZ_{3})$ has dimension $-2R+j+1$ and $U(1)_{r}$ charge $-j-1$.
Since the highest
possible 
degree of $\bar{\Theta}$ in $H(\BZ_{3})$ is two (see \eqref{BBgeneral_ans}),
the only possible value of $j$ is $j=0$, and therefore the $U(1)_{r}$
charge of $H(\BZ_3)$
is indeed $-1$. 
From \eqref{BBgeneral_ans}, we see that the most general expression for such $H(\BZ_3)$ 
is given by
\begin{align}
H(\BZ_{3})&=A\frac{\bar{\Theta}_{l_{1}}\bar{\Theta}_{l_{2}}}{(X^2)^{R}}
(\epsilon_{l_{3}m_{1}}\cdots\epsilon_{l_{R+1}m_{R-1}})
(\epsilon_{l_{R+2}k_{1}}\cdots\epsilon_{l_{2R_{1}}k_{2R_{1}-R-1}})
(\epsilon_{m_{R}k_{2R_{1}-R}}\cdots\epsilon_{m_{2R_{2}}k_{2R_{3}}})\,,
\end{align} 
where $k_1\cdots k_{2R_3}$ are the $SU(2)_R$ indices for $\schurDD_{R_3(0,0)}$.
As mentioned at the end of the previous subsection,
$(X,\Theta,\bar\Theta)$ stands for $(X_3,\Theta_3,\bar\Theta_3)$ here.
However, when we change the variables from $\BZ_3$ to $\bar \BZ_3$, this
expression cannot be written in the form of
\eqref{BBgeneral_ans_bar}.\footnote{In particular,
	$\bar{\Theta}_{l_{1}}\bar{\Theta}_{l_{2}}$ cannot be mapped to
	$\bar{\Theta}_{m_{1}}\bar{\Theta}_{m_{2}}$.}
This means that there are no solutions to \eqref{BcondD}--\eqref{BcondQbar}.
Therefore, the $\schurDD_{R(j,0)}$ multiplet does not appear in the $\schurB_{R_{1}}\times \schurB_{R_{2}}$ selection rule.
Note that its conjugate implies that the $\schurD_{R(0,\bj)}$ multiplet also does not appear in the
$\schurB_{R_{1}}\times \schurB_{R_{2}}$ fusion.
By using the same argument, we can extend our results to that $H(\BZ_{3})$ for a non-vanishing correlation function $\langle{\schurB}_{R_{1}}{\schurB}_{R_{2}}\OO\rangle$ must be $U(1)_{r}$ neutral.

\subsection{$\langle{\schurB}_{R_{1}}{\schurB}_{R_{2}}{\schurC}_{R_3(j,\bj)}\rangle$}\label{secBBfusC}
Let us finally consider the case of $\OO^\II$ in the $\schurC_{R_3(j,\bj)}$.
By using the similar argument in previous section \ref{subsec:BBD}, $\bj=j$
is necessary for the three-point function to be non-vanishing. 
Therefore the function $H(\BZ_{3})$ has 
dimension $2-2R+2j$ and 
 $U(1)_{r}$ neutral, where we recall that $R:= R_1+R_2-R_3$.
The most general solution to \eqref{BcondDbar} and \eqref{BcondQ} is
written as
 \begin{align}\label{BBCX}
H(\BZ_{3})&=\frac{1}{X^{2(R-1)}}
\Bigg[\!\!
\left(A'\epsilon_{l_{1}m_{1}}\epsilon_{l_{2}m_{2}}
+B' \frac{M_{m_{1}l_{1}}}{X^2}\epsilon_{l_{2}m_{2}}
+C'\frac{M_{m_{1}l_{1}}}{X^2}\frac{M_{m_{2}l_{2}}}{X^2}\right)\!X_{\beta_{1}\dot{\beta}_{1}}\!
+D'\tilde{\Theta}_{m_{1}\beta_{1}}{\tilde{\bar{\Theta}}_{\dot{\beta}_{1}l_{1}}}
\epsilon_{l_{2}m_{2}}\Bigg]
\nonumber\\ 
&~\times\!
(\epsilon_{l_{3}m_{3}}\cdots\epsilon_{l_{R}m_{R}})
(\epsilon_{l_{R+1}k_{1}}\cdots \epsilon_{l_{2R_{1}}k_{2R_{1}-R}})
(\epsilon_{m_{R+1}k_{2R_{1}-R+1}}\cdots \epsilon_{m_{2R_{2}}k_{2R_{3}}})
\left(X_{\beta_{2}\dot{\beta}_{2}}\cdots X_{\beta_{2j}\dot{\beta}_{2j}}\right)\,,
\end{align}
where $k_i$ and $(\beta_i,\dot\beta_i)$ are the
$SU(2)_R$ and $SL(2,\mathbb{C})$ indices associated with
$\schurC_{R_3(j,j)}$.
On the other hand, the most general solution to \eqref{BcondD} and
\eqref{BcondQbar} is written as
\begin{align}
H(\bar{\BZ}_{3})&=\frac{1}{\bar{X}^{2(R-1)}}
\Bigg[\!\!
\left(
A\epsilon_{l_{1}m_{1}}\epsilon_{l_{2}m_{2}}
+B\frac{\bar{M}_{l_{1}m_{1}}}{\bar{X}^2}\epsilon_{l_{2}m_{2}}
+C\frac{\bar{M}_{ l_{1}m_{1}}}{\bar{X}^2}
\frac{\bar{M}_{ l_{2} m_{2} }}{\bar{X}^2}
\right)\!\bar{X}_{\beta_{1}\dot{\beta}_{1}}
+D\tilde{\Theta}_{l_{1}\beta_{1}}{\tilde{\bar{\Theta}}_{\dot{\beta}_{1}m_{1}}}
\epsilon_{l_{2}m_{2}}\Bigg]
\nonumber\\
&~\times\!
(\epsilon_{l_{3}m_{3}}\cdots\epsilon_{l_{R}m_{R}})
(\epsilon_{l_{R+1}k_{1}}\cdots \epsilon_{l_{2R_{1}}k_{2R_{1}-R}})
(\epsilon_{m_{R+1}k_{2R_{1}-R+1}}\cdots \epsilon_{m_{2R_{2}}k_{2R_{3}}})
\left(\bar{X}_{\beta_{2}\dot{\beta}_{2}}\cdots \bar{X}_{\beta_{2j}\dot{\beta}_{2j}} \right)\,.
\label{BBCbarX}
\end{align}
For the above two expressions to be consistent, the coefficients have to
satisfy some conditions. 
 Unless $R=0$, the conditions are
 \begin{align}
 B=4\Imu (R-1)A\,,\quad C=-16\frac{(R-1)(R-2-2j)}{2}A\,,\quad D=4\Imu (2j)A\,.
 \end{align}
On the other hand, for $R=0$ or equivalently $R_3=R_1+R_2$, all the
coefficients have to vanish. Therefore, the three-point function
$\langle
\schurB_{R_1}\schurB_{R_2}\schurC_{R_3(j,j)}\rangle$
vanishes if $R_3=R_1+R_2$.

Next, we consider the semi-shortening
conditions for $\schurC_{R_3(j,\bj)}$.
For that purpose, we relate $H(\BZ_3)$ to $G(\BZ_2)$ via
\eqref{gneGtoH}. Indeed, the three-point function \eqref{BBOgen} can be rewritten as
\begin{align}
&\frac{
	\left(\hu{i_{1}}{l_{1}}\cdots \hu{i_{2R_{1}}}{l_{2R_{1}}}(z_{12})\right)\!
	\left(\hu{k_{1}}{n_{1}}\cdots \hu{k_{2R_{3}}}{n_{2R_{3}}}(z_{32})\right)\!
	\left(I_{\delta_{1}\dot{\beta}_{1}}\cdots 	I_{\delta_{2j}\dot{\beta}_{2j}}(x_{3\bar{2}})\right)\!
	\left(	I_{\beta_{1}\dot{\delta}_{1}}\cdots	I_{\beta_{2j}\dot{\delta}_{2j}}(x_{2\bar{3}})\right)}
{(x^2_{1\bar{2}}x^2_{2\bar{1}})^{R_1}
	(x^2_{3\bar{2}}x^2_{2\bar{3}})^{R_{3}+j+1}}
\nonumber\\
&~\times G^{(\delta_{1}\cdots \delta_{2j})(\dot{\delta}_{1}\cdots \dot{\delta}_{2j})}_{
	(l_{1}\cdots l_{2R_{1}})(j_{1}\cdots j_{2R_{2}})(n_{1}\cdots n_{2R_{3}})}(\BZ_{2})\,.
\end{align}
and the explicit form of $G(\BZ_{2})$ becomes
\begin{align}
G(\BZ_{2})&=
\Bigg[
\Big(\vb{u}^{\dagger}_{l_{1}j_{1}}\vb{u}^{\dagger}_{l_{2}j_{2}}
+4\Imu (R-1)\frac{\bar{M}_{j_{1}l_{1}}}{\bar{X}^2}\vb{u}^{\dagger}_{l_{2}j_{2}}
-\frac{16(R-1)(R-2-2j)}{2}
\frac{\bar{M}_{j_{1}l_{1}}}{\bar{X}^2}\frac{\bar{M}_{j_{2}l_{2}}}{\bar{X}^2}
\Big) (\bar{X}^{-1})^{\dot{\delta}_{1}\delta_{1}}
\nonumber\\
&~~+4\Imu (2j)(\bar{X}^{-1}\tilde{\Theta}_{j_{1}})^{\dot{\delta}_{1}}
(\TB{\Theta}_{l_{1}}\bar{X}^{-1})^{\delta_{1}}\vb{u}^{\dagger}_{l_{2}j_{2}}
\Bigg](\epsilon_{l_{3}j_{3}}\cdots \epsilon_{l_{R}j_{R}})
(\bar{X}^{\scriptsize{-1}})^{\dot{\delta}_{2}\delta_{2}}\cdots (\bar{X}^{-1})^{\dot{\delta}_{2j}\delta_{2j}}
\nonumber\\
&~\times
\frac{\vb{u}^{\dagger}_{l_{R+1}n_{1}}\cdots \vb{u}^{\dagger}_{l_{2R_{1}}n_{2R_{1}-R}}(\BZ_{2})}{(\bar{X_2})^{2R_{1}-R+1}}
(\epsilon_{j_{R+1}n_{2R_{1}-R+1}}\cdots \epsilon_{j_{2R_{2}}n_{2R_{3}}})\,.
\label{BBCR}
\end{align}
In terms of $G(\BZ_2)$, the semi-shortening conditions for
$\schurC_{R_3(j,j)}$ are written as
\begin{align}
 \CS^{{\delta}_{1}}_{(n}{\Large G}_{n_{1}\cdots n_{2R_{3}})({\delta}_{1}\cdots\delta_{2j})(\dot{\delta}_{1}\cdots\dot{\delta}_{2j})}
=0\,,\qquad 
\bar\CS^{{\dot\delta}_{1}}_{(n}
{\Large G}_{n_{1}\cdots n_{2R_{3}})({\delta}_{1}\cdots\delta_{2j})	(\dot{\delta}_{1}\cdots\dot{\delta}_{2j})}
=0\,,
\label{thirdcondC1}
\end{align}
for $j>0$ and
\begin{align}
 \epsilon_{\alpha\beta}\CS^{\alpha}_{ (n}\CS^{\beta }_{n'}\,G_{n_{1}\cdots n_{2R_{3}})}
=0\,, \qquad \bar\CS_{\dot\alpha (n}\bar\CS^{\dot\alpha }_{n'}\,G_{n_{1}\cdots n_{2R_{3}})}
=0\,,
\label{thirdcondC2}
\end{align}
for $j=0$. 
%Here $n_i$ and $(\delta_i,\dot\delta_i)$ are respectively the $SU(2)_R$ and $SL(2,\mathbb{C})$ indices associated with $\schurC_{R_3(j,j)}$. 
It is straightforward to check \eqref{BBCR} satisfies these conditions.

In summary, the function $H(\BZ_{3})$ in $\langle{\schurB}_{R_{1}}{\schurB}_{R_{2}}{\schurC}_{R_3(j,j)}\rangle$ is given by, up to an overall constant
\begin{align}
H(\BZ_{3})&=\frac{1}{X^{2(R-1)}}
\Bigg(
\epsilon_{l_{1}m_{1}}\epsilon_{l_{2}m_{2}}
X_{ \beta_{1}\dot{\beta}_{1}}
+4\Imu (R-1) \frac{M_{m_{1}l_{1}}}{X^2}
\epsilon_{l_{2}m_{2}}X_{ \beta_{1}\dot{\beta}_{1}}
\nonumber\\
&~ -16\frac{(R-1)(R-2-2j)}{2}\!\frac{M_{m_{1}l_{1}}}{X^2}\frac{M_{m_{2}l_{2}}}{X^2}
X_{ \beta_{1}\dot{\beta}_{1}}
+4\Imu (2j)\tilde{\Theta}_{m_{1}\beta_{1}}\TB{\Theta}_{\dot{\beta}_{1}l_{1}}
\epsilon_{l_{2}m_{2}}
\Bigg)(\epsilon_{l_{3}m_{3}}\cdots \epsilon_{l_{R}m_{R}})
\nonumber\\ 
&~\times
( \epsilon_{l_{R+1}k_{1}}\cdots \epsilon_{l_{2R_{1}}k_{2R_{1}-R}})
(\epsilon_{m_{R+1}k_{2R_{1}-R+1}}\cdots \epsilon_{m_{2R_{2}}k_{2R_{3}}})\!
\left(X_{ \beta_{2}\dot{\beta}_{2}}\cdots X_{ \beta_{2j}\dot{\beta}_{2j}}\right)\,,
\label{BBCXans}
\end{align}
 for any integer or half-integer 
$j\geq 0$ and $R_3<R_1+R_2$.
 Therefore, the ${\schurC}_{R_3(j,j)}$ multiplet appears in
 $\schurB_{R_{1}}\times \schurB_{R_{2}}$ fusion for $|R_{1}-R_{2}|\leq R_{3}\leq R_{1}+R_{2}-1$.

\subsection{Selection rule}
\label{sec:BB}

In the above subsections, we have 
computed the most general expression for non-vanishing three-point
functions of the form $\langle{\schurB}_{R_{1}}{\schurB}_{R_{2}}\OO\rangle$.
From these results, we see that the selection rules for two
$\schurB_R$ multiplets are written as
\begin{align}
{\schurB}_{R_{1}}\times {\schurB}_{R_{2}}
\sim \sum^{R_{1}+R_{2}}_{R=|R_{1}-R_{2}|>0}{\schurB}_{R}
+ \sum^{R_{1}+R_{2}-1}_{R=|R_{1}-R_{2}|}\,\sum^{\infty}_{\ell=0}\schurC_{R(\frac{\ell}{2},\frac{\ell}{2})}\,,
\end{align}
up to non-Schur multiplets. This is particularly consistent with
Eq.~(3.44) of \cite{Nirschl:2004pa}.
%(In \cite{Nirschl:2017fkf}, they derived
%same selection rule from four-dimensional $\NN=4$ selection rule of BPS
%operators.)
%
Especially, for $R_{1} =R_{2}=1$, the selection rule is written as
 \begin{align}
 {\schurB}_{1}\times {\schurB}_{1}
 \sim {\schurB}_{1}+ {\schurB}_{2}
 + \sum^{\infty}_{\ell=0}\left[\schurC_{0(\frac{\ell}{2},\frac{\ell}{2})}+\schurC_{1(\frac{\ell}{2},\frac{\ell}{2})}\right],
 \end{align}
which is consistent with 
the harmonic superspace analysis in \cite{Arutyunov:2001qw}.

\section{$\schurC_{0(0,0)} \times \OO^{\text{Schur}} $ fusion}\label{secCOfus}

In this section, we 
turn to %study
 the selection rules for $\schurC_{0(0,0)}\times \OO^\mathrm{Schur}$ for an arbitrary Schur multiplet $\OO^\mathrm{Schur}$.
These selection rules are important in studying 
%the study of 
the corresponding two-dimensional chiral algebra, since the highest weight component of the $SU(2)_{R}$ current operator in the stress-tensor multiplet $\schurC_{0(0,0)}$ is mapped to the Virasoro stress-tensor operator in the chiral algebra \cite{Beem:2013sza}.
%For this purpose, 
To derive the selection rules, we compute the three-point
functions of the form $\langle
\schurC_{0(0,0)}\OO^{\II_1}\OO^{\II_2}\rangle$
for two Schur multiplets $\OO^{\II_1}$ and $\OO^{\II_2}$.

Recall here that the stress-tensor multiplet $\schurC_{0(0,0)}$ has two semi-shortening conditions 
\begin{align}
{\bar{\mathcal{D}}_{\dot{\alpha}}}^{~i}{\bar{\mathcal{D}}}^{\dot{\alpha}i'}H(\BZ_{3})
&=0\,,
\label{C00B}
\\ 
\epsilon_{\alpha\beta}{\mathcal{D}^{i\alpha}}\mathcal{D}^{i'\beta }H(\BZ_{3})
&=0\,.
\label{C00}
\end{align}
The most general solution to these two equations are respectively written as
\begin{align}
H(\BZ_{3})&=
f(X,\Theta)
+g^{k}_{\dot{\alpha}}(X,\Theta)\BTh^{\dot{\alpha}}_{{}_{\,}k}
+h_{(\dot{\alpha}\dot{\alpha}')}(X,\Theta)\BTh^{\dot{\alpha}\dot{\alpha}'}\,,
\label{ansC00}
\\
H(\bar{\BZ}_{3})&=
\tilde f(\bar{X},\BTh)
+\tilde g_{k,\alpha}(\bar{X},\BTh)\Theta^{k \alpha}
+\tilde h_{(\alpha\alpha')}(\bar{X},\BTh)\Theta^{\alpha\alpha'}\,
\label{ansC00Bar}
\end{align}
where
\begin{align}
\Theta^{\alpha\alpha'}
:=\Theta^{i\alpha}\epsilon_{ij}\Theta^{j\alpha'}
\,, \quad
\bar{\Theta}^{\dot{\alpha}\dot{\alpha}'}
:=\bar{\Theta}^{\dot{\alpha}}_{~\,i}\epsilon^{ij}\bar{\Theta}^{\dot{\alpha}'}_{~\,j}\,.
\end{align}
For the above two expressions to be consistent, the functions $f,g,h,\tilde f,\tilde g$, and $\tilde h$ have to satisfy some conditions. 
Moreover, they are also constrained by the (semi-)shortening conditions associated with $\OO^{\II_1}$ and $\OO^{\II_2}$. 
Below, we solve all these conditions to find general expressions for $H(\BZ_3)$. 
Since the concrete calculations are highly involved, we here write the results and, details of the computations are in appendices \ref{appCBO},\ref{appCDO}, and \ref{appCCC}.

\subsection{$\langle\schurC_{0(0,0)} \schurB_{R}{\OO^{\II}}\rangle$ }
\label{subsecCBO}

Let us first consider the three-point function $\langle\schurC_{0(0,0)}
\schurB_{R}{\OO^{\II}}\rangle$.
We denote by $\JJ(z)$ the superfield of the stress-tensor multiplet $\schurC_{0(0,0)}$.
The three-point function is then written as
\begin{align}
\langle	\JJ(z_{1})\LL_{(j_{1}\cdots j_{2R})}(z_{2}) \OO^{\II}(z_{3}) \rangle
&=\frac{\hu{j_{1}}{m_{1}}(z_{23})\cdots\hu{j_{2R}}{m_{2R}}(z_{23})}
{\left(x^2_{\bar{3}1} x^2_{\bar{1}3}\right) \left(x^2_{\bar{3}2}x^2_{\bar{2}3}\right)^{R}}
~{\Large H}_{(m_{1}\cdots m_{2R})}^{\II}(\BZ_{3})\,.
\end{align}
The (semi-)shortening conditions for $\schurC_{0(0,0)}$
and $\schurB_R$ are encoded in \eqref{BcondQ}, \eqref{BcondQbar}, \eqref{C00B}, and \eqref{C00}.
The three-point function is consistent with these four conditions only
 when $\OO^{\II}$ is $\schurB_{R}$, $\schurC_{R(j,j)}$, or
$\schurC_{R-1(j,j)}$.
Up to an overall constant, the expressions for $H(\BZ_{3})$ in these
 three cases are written as
\begin{align}
\schurB_{R}&:
\frac{\vb{u}_{{m_{1}}{k_{1}}}(\BZ_{3})}{X^2}
(\epsilon_{m_{2}k_{2}}\cdots \epsilon_{m_{2R}k_{2R}}),
\\
\schurC_{R(j,j)}&:
\left(
\epsilon_{m_{1}k_{1}}X_{\beta_{1}\dot{\beta}_{1}}
-4\Imu (2j)\tilde{\Theta}_{ m_{1}\beta_{1}}\tilde{\bar{\Theta}}_{\dot{\beta}_{1}k_{1}}
\right)
(\epsilon_{m_{2}k_{2}}\cdots \epsilon_{m_{2R}k_{2R}})
X_{\beta_{2}\dot{\beta}_{2}}\cdots X_{\beta_{2j}\dot{\beta}_{2j}}\,,
\\
\begin{split}
\schurC_{R-1(j,j)}&:
\frac{1}{X^2}\Bigg(
\frac{M_{m_{1}m_{2}}}{X^2}X_{ \beta_{1}\dot{\beta}_{1}}
-2j\Theta_{m_{1}\beta_{1}}\tilde{\bar{\Theta}}_{\dot{\beta}_{1}m_{2}}
+4\Imu j\frac{\Theta_{m_{1}}\Theta_{ m_{2}}}{X^2}
\tilde{\bar{\Theta}}_{\dot{\beta}_{1}i}
\big(X\tilde{\bar{\Theta}}^{i}\big)_{\beta_{1}}
\Bigg)
\\
&\qquad \times 
(\epsilon_{m_{3}k_{1}}\cdots \epsilon_{m_{2R}k_{2R-2}})
X_{ \beta_{2}\dot{\beta}_{2}}\cdots X_{ \beta_{2j}\dot{\beta}_{2j}}\,,
\end{split}
\label{C0BC}
\end{align}
where $k_i$ and $(\beta_i,\dot\beta_i)$ are respectively the $SU(2)_R$ and
$SL(2,\mathbb{C})$ indices associated with the third multiplet.
The derivations of these functions are given in appendix \ref{appCBO}.

We see that these are also consistent with the (semi-)shortening
conditions for the third Schur multiplet. Indeed, using \eqref{gneGtoH}, we see that
the function $G(\BZ_2)$ corresponding to the above $H(\BZ_3)$ is written as
\begin{align}
\schurB_{R}&:
\frac{\vb{u}_{j_{1}n_{1}}(\BZ_{2})}{X^2}
(\epsilon_{j_{2}n_{2}}\cdots \epsilon_{j_{2R}n_{2R}}),
\\
\schurC_{R(j,j)}&:
\Bigg(\frac{\epsilon_{j_{1}n_{1}}\tilde{\bar{X}}^{-1}_{\delta_{1}\dot{\delta}_{1}}}{X^{2}\bar{X}^{2}}
+\frac{4\Imu (2j)(\Theta_{j_{1}}\tilde{\bar{X}}^{-1})_{\dot{\delta}_{1}}
	(\tilde{\bar{X}}^{-1}\bar{\Theta}_{n_{1}})_{\delta_{1}}}{\bar{X}^{4}}\Bigg)
(\epsilon_{j_{2}n_{2}}\cdots \epsilon_{j_{2R}n_{2R}})
\tilde{\bar{X}}^{-1}_{\delta_{2}\dot{\delta}_{2}}\cdots 
\tilde{\bar{X}}^{-1}_{\delta_{2j}\dot{\delta}_{2j}}\,,
\\
\begin{split}
\schurC_{R-1(j,j)}&:
\frac{1}{\bar{X}^2}\Bigg(
\frac{\bar{M}_{j_{1}j_{2}}}{\bar{X}^2}\bar{X}_{\delta_{1}\dot{\delta}_{1}}
-2j\tilde{\Theta}_{j_{1}\delta_{1}}\tilde{\bar{\Theta}}_{\dot{\delta}_{1}j_{2}}
+4\Imu j\frac{\Theta_{j_{1}}\Theta_{j_{2}}}{\bar{X}^2}
\tilde{\bar{\Theta}}_{\dot{\delta}_{1}i}
\left(\bar{X}\bar{\Theta}^{i}\right)_{\delta_{1}}
\Bigg)
\\
&\qquad\times 
(\epsilon_{j_{3}n_{1}}\cdots \epsilon_{j_{2R}n_{2R-2}})
\tilde{\bar{X}}^{-1}_{\delta_{2}\dot{\delta}_{2}}\cdots 
\tilde{\bar{X}}^{-1}_{\delta_{2j}\dot{\delta}_{2j}}\,.
\end{split}
\label{CBCR-1}
\end{align}
Here $n_i$ and $(\delta_i,\dot\delta_i)$ are respectively the
$SU(2)_R$ and $SL(2,\mathbb{C})$ indices associated with the third multiplet $\OO^{\II}$.
These equations are all consistent with the (semi-)shortening conditions for the third multiplet.

Let us briefly comment on the case of $\OO^\II$ in $\schurC_{0(0,0)}$. 
When we assume
that there is only one stress tensor in the theory, the corresponding three-point function 
$\langle\schurC_{0(0,0)}\schurB_1\schurC_{0(0,0)}\rangle$
has to be symmetric under the action of $\mathbb{Z}_2$ exchanging the
first and the third multiplets. This $\mathbb{Z}_2$ symmetry implies
that the function $G(\BZ_2)$ is invariant under 
 $(X_{2},\Theta_{2},\bar{\Theta}_{2})\leftrightarrow
 (-\bar{X}_{2},-\Theta_{2},-\bar{\Theta}_{2})$. 
 However, the expression \eqref{CBCR-1} is not invariant under this $\mathbb{Z}_2$ action. 
Therefore, in an SCFT with unique stress tensor multiplet, the three-point function $\langle \schurC_{0(0,0)}\schurB_{1} \schurC_{0(0,0)} \rangle$ must vanish \cite{Kuzenko:1999pi}.

Before closing this subsection, let us also make a quick comment on the correlation function $\langle \schurC_{0(0,0)}\schurB_{R} \schurC_{R-1(j,j)} \rangle$.
In CFTs, any correlation function of {\it conformal} descendant fields is obtained by differentiating the correlation function of the conformal primary fields.
This particularly implies that, when a correlation function of conformal primary fields
vanishes, the corresponding descendant correlators also vanish.
This, however, is not the case for {\it superconformal} descendants in SCFTs. 
Indeed, when we set all Grassmann variables, $\theta_{1,2,3},\bar{\theta}_{1,2,3}$,
to zero in \eqref{C0BC}, the correlation function vanishes.
This shows that the correlator of the three superconformal primary fields in $\schurC_{0(0,0)}$, $\schurB_{R}$, and $\schurC_{R-1(j,j)} $ vanishes, while there are non-vanishing correlators involving superconformal descendants. 
This is a common feature of SCFTs \cite{Fortin:2011nq}.

\subsection{$\langle\schurC_{0(0,0)}\schurDD_{R(j,0)}{\OO^{\II}}\rangle$ }\label{secCDO}

Let us next consider $\langle\schurC_{0(0,0)}\schurDD_{R(j,0)}{\OO^{\II}}\rangle$.
We denote by $\bar{\NN}_{(j_{1}\cdots j_{2R})(\alpha_{1}\cdots\alpha_{2j})}(z)$
the superfield of a $\schurDD_{R(j,0)}$ multiplet. 
The three-point function is written as
\begin{align}
\langle \JJ(z_{1})\bar{\NN}_{(j_{1}\cdots j_{2R})(\alpha_{1}\cdots\alpha_{2j})}(z_{2})\OO^{\II}(z_{3})\rangle\!
=\frac{(\hu{j_{1}}{m_{1}}\cdots \hu{j_{2R}}{m_{2R}}(z_{23}))
	(I_{\alpha_{1}\dot{\gamma}_{1}}\cdots
	I_{\alpha_{2j}\dot{\gamma}_{2j}}(x_{2\bar{3}}))}
{x^2_{1\bar{3}}x^2_{3\bar{1}}(x^{2}_{2\bar{3}})^{R+j+1}(x^{2}_{3\bar{2}})^{R}}
{\Large H}_{(m_{1}\cdots m_{2R})}^{(\dot{\gamma}_{1}\cdots\dot{\gamma}_{2j})\,\II}
(\BZ_{3})\,.
\label{CDbOgen}
\end{align}
In this case, the function $H(\BZ_{3})$ has to satisfy the
semi-shortening conditions \eqref{C00B} and \eqref{C00} associated with the stress tensor multiplet $\schurC_{0(0,0)}$,
the shortening condition \eqref{BcondQ} of $\schurDD_{R(j,0)}$,
and the following semi-shortening condition of $\schurDD_{R(j,0)}$:
\begin{align}
\bar{\QQ}_{\dot{\gamma}(m}{\Large H}_{m_{1}\cdots m_{2R})}^{
	(\dot{\gamma}\dot{\gamma}_{2}\cdots\dot{\gamma}_{2j})\II}(\BZ_{3})=0\,,&\quad \text{for}\quad j>0\,,
\label{DcondQ}
\\
\bar{\QQ}_{\dot{\alpha}(m}\bar{\QQ}^{\dot{\alpha}}_{m'}
{\Large H}_{m_{1}\cdots m_{2R})}^{\II}(\BZ_{3})=0\,,&\quad \text{for}\quad j=0.
\label{DcondQ0}
\end{align}

We see that a non-vanishing solution to \eqref{C00B},
\eqref{C00}, \eqref{BcondQ} and \eqref{DcondQ}/\eqref{DcondQ0} exists if and only if
$\OO^{\II}$ is in the $\schurD_{R-\frac{1}{2}(0,j-\frac{1}{2})}$,
$\schurD_{R(0,j)}$, $\schurD_{R-1(0,j)}$,
$\schurC_{R-1(j_{1},j+j_{1}+1)}$,
$\schurC_{R+\frac{1}{2}(j_{1},j+j_{1}+\frac{1}{2})}$, or
$\schurC_{R-\frac{1}{2}(j_{1},j+j_{1}+\frac{1}{2})}$ multiplets. 
Moreover, considering the 
(semi-)shortening conditions for each of these
third multiplets, we find that the only possible third multiplets
$\OO^\II$ which can have a non-vanishing $H(\BZ_3)$ are
$\schurD_{R(0,j)}$, $\schurC_{R+\frac{1}{2}(j_{1},j+j_{1}+\frac{1}{2})}$,
 and $\schurC_{R-\frac{1}{2}(j_{1},j+j_{1}+\frac{1}{2})}$.
The explicit expressions for $H(\BZ_{3})$ for these three cases 
are written, up to an overall constant, as
\begin{equation}
\begin{split}
\schurD_{R(0,j)}&:
\frac{\vb{u}_{m_{1}k_{1}}(\BZ_{3})}{X^2}
(\epsilon_{m_{2}k_{2}}\cdots \epsilon_{m_{2R}k_{2R}})
(\epsilon_{\dot{\gamma}_{1}\dot{\beta}_{1}}\cdots\epsilon_{\dot{\gamma}_{2j}\dot{\beta}_{2j}})\,,\qquad
 \text{for}\qquad j\neq \frac{1}{2}\,,
\\
\schurC_{R+\frac{1}{2}(j_{1},j+j_{1}+\frac{1}{2})}&:
\bar{\Theta}_{\dot{\beta}_{1} k_{1}}
(\epsilon_{m_{1}k_{2}}\cdots \epsilon_{m_{2R}k_{2R+1}})
(\epsilon_{\dot{\gamma}_{1}\dot{\beta}_{2}}\cdots\epsilon_{\dot{\gamma}_{2j}\dot{\beta}_{2j+1}})
X_{\beta_{1}\dot{\beta}_{2j+2}}\cdots X_{\beta_{2j_{1}}\dot{\beta}_{2j+2j_{1}+1}}\,, 
\\
\schurC_{R-\frac{1}{2}(j_{1},j+j_{1}+\frac{1}{2})}&:
\frac{\tilde{\bar{\Theta}}_{\dot{\beta}_{1} k}
	\vb{u}^{k}_{~m_{1}}(\BZ_{3})}{X^2}
(\epsilon_{m_{2}k_{1}}\cdots \epsilon_{m_{2R}k_{2R-1}})
(\epsilon_{\dot{\gamma}_{1}\dot{\beta}_{2}}\cdots\epsilon_{\dot{\gamma}_{2j}\dot{\beta}_{2j+1}})
X_{\beta_{1}\dot{\beta}_{2j+2}}\cdots X_{\beta_{2j_{1}}\dot{\beta}_{2j+2j_{1}+1}}\,,
\end{split}
\label{CDO}
\end{equation}
where $ j_{1}\geq 0$ is an integer or a half-integer, and $k_i$ and
$(\beta_i,\dot\beta_i)$ are respectively the $SU(2)_R$ and
$SL(2,\mathbb{C})$ indices for the third multiplet.
Note that the first expression for the case of $\OO^\II$ in
the $\schurD_{R(0,j)}$ multiplet is only for $j \neq 1/2$. In the
case 
$j=1/2$, the function $H(\BZ_3)$ has two independent terms as
\begin{align}\label{CDC1/2}
H(\BZ_{3})=
\frac{1}{X^2}
\Bigg(A \vb{u}_{m_{1}k_{1}}(\BZ_{3})\epsilon_{\dot{\gamma}_{1}\dot{\beta}_{1}}
+\frac{B}{X^2}\left(\left(\Theta_{m_{1}}X\right)_{\dot{\gamma}_{1}}\tilde{\bar{\Theta}}_{\dot{\beta}_{1}k_{1}}
-M_{m_{1}k_{1}}\epsilon_{\dot{\gamma}_{1}\dot{\beta}_{1}}\right)\Bigg)
(\epsilon_{m_{2}k_{2}}\cdots \epsilon_{m_{2R}k_{2R}})\,,
\end{align}
where $A$ and $B$ are arbitrary constants.
For the detail of derivations of \eqref{CDO} and \eqref{CDC1/2}, see appendix \ref{appCDO}.

Note here that the second and third lines of \eqref{CDO} are proportional to $\bar\Theta$, which means that the three-point functions of the superconformal primaries vanish. 
This can also be seen from the fact that the sum of the $U(1)_r$ charges of the superconformal primaries does not vanish.

\subsection{$\langle\schurC_{0(0,0)} \schurC_{R(j,\bj)}{\OO^{\II}}\rangle$ }

Let us finally consider the correlation function 
$\langle\schurC_{0(0,0)} \schurC_{R(j,\bj)}{\OO^{\II}}\rangle$.
Since we have already studied the cases of $\OO^{\II}= \schurB_{R'}$,
$\schurD_{R'(0,\bj)}$, and $\schurDD_{R'(j,0)}$, the 
only remaining case we have to study here
 is $\OO^{\II}= \schurC_{R'(j_{2},\bj_{2})}$,
in which case the three-point function 
is written as
\begin{align}
\begin{split}
&\left\langle 
\JJ(z_1) \JJ_{\,\,(j_1\cdots j_{2R})}^{( \alpha_1\cdots \alpha_{2j_1}), (\dot{\alpha}_1 \cdots \dot{\alpha}_{2\bj_1})}(z_2)
\JJ_{\,\,(k_1\cdots k_{2R'})}^{(\beta_1\cdots \beta_{2j_2}),(\dot{\beta}_1\cdots \dot{\beta}_{2\bj_2})}(z_3)
\right \rangle \\
&\qquad=
\frac{\left(\hu{j_{1}}{m_{1}}\cdots \hu{j_{2R}}{m_{2R}}(z_{23})\right)\!\!
	(I_{\alpha_1 \dot{\gamma}_1} \cdots I_{\alpha_{2j_1} \dot{\gamma}_{2j_1}}(x_{2\bar{3}}))\,
	(I_{\gamma_1 \dot{\alpha}_1} \cdots I_{\gamma_{2\bj_1} \dot{\alpha}_{2\bj_1}}(x_{3\bar{2}}))}
{x^2_{1\bar{3}}x^2_{3\bar{1}}(x^2_{2\bar{3}})^{q_2}(x^2_{3\bar{2}})^{\bar{q}_2}}
\\
& ~~\qquad
\times {\Huge H}_{\,\,(m_1\cdots m_{2R})(k_1\cdots k_{2R'})}^{
	\,\,( \gamma_1\cdots \gamma_{2\bj_1}) (\dot{\gamma}_1 \cdots \dot{\gamma}_{2j_1})
	(\beta_1\cdots \beta_{2j_2})(\dot{\beta}_1\cdots \dot{\beta}_{2\bj_2})}(\BZ_{3})\,,
\end{split}
\label{genCCC}
\end{align}
where $\JJ_{\,\,(j_1\cdots j_{2R})}^{( \alpha_1\cdots \alpha_{2j}),
(\dot{\alpha}_1 \cdots \dot{\alpha}_{2\bj})}(z)$ is the superfield in
the $\schurC_{R(j,\bj)}$ multiplet.
In this case, the 
$H(\BZ_{3})$ has to satisfy the 
semi-shortening conditions \eqref{C00B} and \eqref{C00} associated with
$\schurC_{0(0,0)}$ and the semi-shortening conditions for the other two $\schurC_{R(j_{1},\bj_{1})}$
 multiplets.
The semi-shortening conditions for the second multiplet, $\schurC_{R(j_{1},\bj_{1})}$, are written as
\begin{subequations}
	\begin{alignat}{2}
	&\QQ_{\gamma(m}
	H^{(\gamma\gamma_2 \cdots \gamma_{2\bj_{1}})
		(\dot{\gamma}_1 \cdots \dot{\gamma}_{2j_1})\II}_{m_1 \cdots m_{2R})}
	=0&
&\quad \text{for}\quad \bj_{1}>0\,, \label{Cjj} 
	\\
	&\QQ_{\alpha (m}\QQ^{\alpha }_{m'}H^{\II}_{m_{1}\cdots m_{2R})}
	=0&
&\quad \text{for}\quad \bj_{1}=0\,, \label{Cjj0} 
	\end{alignat}
\end{subequations}
and
\begin{subequations}
	\begin{alignat}{2}
	&\bar{\QQ}_{\dot{\gamma} (m}
	H^{(\gamma_1 \cdots \gamma_{2\bj_{1}})
		(\dot{\gamma}\dot{\gamma}_2 \cdots \dot{\gamma}_{2j_1})\II}
	_{m_1 \cdots m_{2R})}
	=0&
&\quad \text{for}\quad j_{1}>0\,, \label{CjjB}
	\\
	&\bar{\QQ}_{\dot{\alpha}(m}\bar{\QQ}^{\dot{\alpha}}_{m'}
	H^{\II}_{m_{1}\cdots m_{2R})}
	=0&
&\quad \text{for}\quad j_{1}=0\,, \label{CjjB0}
	\end{alignat}
\end{subequations}
while the semi-shortening conditions for the third multiplet,
$\schurC_{R'(j_2,\bj_2)}$, are similarly expressed in
terms of $G(\BZ_2)$.

We have found that a non-vanishing $H(\BZ_3)$ in \eqref{genCCC} satisfying all these
semi-shortening conditions is possible only for the
following two types of correlator \footnote{
We will show other type correlation functions does not satisfy semi-shortening conditions in appendix \ref{appCCC}.}:
\begin{align}
\langle\schurC_{0(0,0)} \schurC_{R(j+\ell_{1},j)}\schurC_{R(j+\ell_{2},j+\ell_{1}+\ell_{2})}\rangle \label{CCCR}\,,
\\
\langle \schurC_{0(0,0)}\schurC_{R(j+\ell_{1},j)}\schurC_{R+1(j+\ell_{2},j+\ell_{1}+\ell_{2})} \rangle\label{CCCR-1}\,,
\end{align}
up to charge conjugation, where $\ell_1$ and $\ell_2$ are non-negative (half-)integers.
Note that the function $H(\BZ_3)$ is $U(1)_r$ neutral for all these cases.
Since general solutions for $H(\BZ_3)$ in these two cases are highly
involved, it is beyond the scope of this paper to identify the most
general expression for the allowed $H(\BZ_3)$. However, we find a
special solution for each of the above two types of
correlators,\footnote{
	Our method is as follows.
	We first solve 
all the semi-shortening conditions for smaller spins $j=1/2,1$ and so on
to find the explicit expression for $H(\BZ_3)$. 
We then guess an ansatz \eqref{CCRCR} and \eqref{CCRCR1} for general $ j$,
and check that the ansatz satisfies the (semi-)shortening conditions for
arbitrary $j$.}
which is sufficient to identify the selection rule.

Let us first focus on \eqref{CCCR}. Unless $j=\ell_1=0$, our special solution is written as
\begin{align}
H(\BZ_{3})&=\frac{1}{X^2}
(\epsilon_{m_2k_2}\cdots\epsilon_{m_{2R}k_{2R}})
(\tilde{X}^{\dot{\beta}_{2j+1}\beta_{2j+1}}\cdots\tilde{X}^{\dot{\beta}_{2j+2\ell_{2}}\beta_{2j+2\ell_{2}}})
(\epsilon^{\dot{\gamma}_{2j+1}\dot{\beta}_{2j+2\ell_{2}+1}}
\cdots
\epsilon^{\dot{\gamma}_{2j+2\ell_{1}}\dot{\beta}_{2j+2\ell_{1}+2\ell_{2}}})
\nonumber\\
&~\times \Biggl(
\sum_{k=0}^{2j}(-1)^k\frac{(2\ell_{2}+2)_{k}}{(2\ell_{1}+2)_{k}}
\begin{pmatrix}
2j\\k
\end{pmatrix}
\!
\Bigg[
\epsilon_{m_1k_1}
+4\Imu (2j+1)
\left(\epsilon_{m_1k_1}(\TB{\Theta}_{\dot{\gamma}m}\tilde{\Theta}^m_{\gamma})
+\frac{2R}{2R+1}(\TB{\Theta}_{\dot{\gamma}k_{1}}\tilde{\Theta}_{m_{1}\gamma})\right)
\frac{\tilde{X}^{\dot{\gamma}\gamma}}{X^2}
\nonumber\\
&
-4(2j+1)(2j+2)
\epsilon_{m_1k_1}
\frac{\Theta_{\gamma\gamma'}\bar{\Theta}_{\dot{\gamma}\dot{\gamma}'}
	\tilde{X}^{\dot{\gamma}\gamma}\tilde{X}^{\dot{\gamma}'\gamma'}}{X^4}
\Bigg]\!
\left(\frac{\tilde{X}^{\dot{\gamma}_1\gamma_1}\tilde{X}^{\dot{\beta}_1\beta_1}}{X^2}\right)
\cdots\left(\frac{\tilde{X}^{\dot{\gamma}_k\gamma_k}\tilde{X}^{\dot{\beta}_k\beta_k}}{X^2}\right)
\nonumber\\
&~\times
\left(\epsilon^{\gamma_{k+1}\beta_{k+1}}\epsilon^{\dot{\gamma}_{k+1}\dot{\beta}_{k+1}}
\cdots\epsilon^{\gamma_{2j}\beta_{2j}}\epsilon^{\dot{\gamma}_{2j}\dot{\beta}_{2j}}\right)
\Biggr)\,.
\label{CCRCR}
\end{align}
Here $(2\ell+2)_k:= \frac{(2\ell+2+k-1)!}{(2\ell+2-1)!}$ is the Pochhammer
symbol, and $\gamma$ (and $\gamma' $ when present) are contracted after totally symmetrized as
$(\gamma\gamma' \gamma_{1} \cdots)$\,.
Similarly, $m$ is also contracted after totally symmetrized as $(m m_{1}m_{2} \cdots)$.
On the other hand, in the case $j=\ell_1=0$, we find a solution with several free parameters:
\begin{align}
\begin{split}
H(\BZ_{3})&=\frac{\epsilon_{m_1k_1}\cdots\epsilon_{m_{2R}k_{2R}}}{X^2}
\Bigg[
\left(A-B\frac{m_{1}}{X^2}+C\frac{m_{1}^2-m_{2}}{X^4}\right)
\tilde{X}^{\dot{\beta}_{1}\beta_{1}}\tilde{X}^{\dot{\beta}_{2}\beta_{2}}
\\
&~
+2\ell_{2}\left((4\Imu A-B)\bar{\Theta}^{\dot{\beta}_{1}}\Theta^{\beta_{1}}
+ \left(4\Imu B-2C\right)\frac{\bar{\Theta}^{\dot{\beta}_{1}}_{~j}M^{j}_{~l}\Theta^{l\beta_{1}} 
	-\bar{\Theta}^{\dot{\beta}_{1}}\Theta^{\beta_{1}} m_{1}}{X^2}\right)
\tilde{X}^{\dot{\beta}_{2}\beta_{2}}
\\
&~
-\frac{2\ell_{2}(2\ell_{2}-1)}{2}(16A+8\Imu B-2C)\bar{\Theta}^{\dot{\beta}_{1}}
\Theta^{\beta_{1}}\bar{\Theta}^{\dot{\beta}_{2}}\Theta^{\beta_{2}}
\Bigg]
\left(\tilde{X}^{\dot{\beta}_{3}\beta_{3}}
\cdots\tilde{X}^{\dot{\beta}_{2\ell_{2}}\beta_{2\ell_{2}}}\right),
\end{split}
\label{CCC00R}
\end{align}
where $A,B$, and $C$ are arbitrary constants, and $m_{1}$ and $m_{2}$ are
combinations of variables defined in \eqref{mdef}.

Note that, for $j=\ell_1=R=0$, the three-point function $\langle
\schurC_{0(0,0)}\schurC_{0(0,0)}\schurC_{0{(\ell_{2},\ell_{2})}} \rangle$ has
a $\mathbb{Z}_{2}$-symmetry, under the assumption of uniqueness of the stress
tensor multiplet.
The $\mathbb{Z}_2$ symmetry implies that the function $H(\BZ_3)$ has to be invariant under 
$(X_{3},\Theta_{3},\bar{\Theta}_{3})\leftrightarrow (-\bar{X}_{3},-\Theta_{3},-\bar{\Theta}_{3})$.
This condition constrains \eqref{CCC00R} as follows.
\begin{itemize}
	\item 
If $\ell_{2}$ is a half-integer such that $\ell_2 \geq \frac{3}{2}$, 
the $\mathbb{Z}_2$ symmetry implies $A=B=C=0$.
Therefore, no $\schurC_{0{(\ell_{2},\ell_{2})}}$ for such $\ell_2$ appears in the $\schurC_{0(0,0)}\times\schurC_{0(0,0)}$ fusion.
	
\item	If $\ell_{2}$ is an integer 
such that $\ell_2\geq 1$, 
the $\mathbb{Z}_2$ symmetry implies
	$B=2\Imu A, C=0$, and therefore our solution \eqref{CCC00R} reduces to up to an over all constant
	\begin{align}
	\begin{split}
		H(\BZ_{3})&=\frac{A}{X^2}
	\Bigg[\left(1-2\Imu\frac{m_{1}}{X^2}\right)
	\tilde{X}^{\dot{\beta}_{1}\beta_{1}}
	+2\Imu(2\ell_{2})
	\Big(\bar{\Theta}^{\dot{\beta}_{1}}\Theta^{\beta_{1}}
	-4\Imu\frac{\bar{\Theta}^{\dot{\beta}_{1}}_{~j}M^{j}_{~l}\Theta^{l\beta_{1}} 
		-\bar{\Theta}^{\dot{\beta}_{1}}\Theta^{\beta_{1}} m_{1}}{X^2}\Big)\Bigg]
	\\
	&~\times
	(\tilde{X}^{\dot{\beta}_{2}\beta_{2}}	\cdots\tilde{X}^{\dot{\beta}_{2\ell_{2}}\beta_{2\ell_{2}}})\,.
	\end{split}
	\end{align}

	\item	If $\ell_{2}=\frac{1}{2}$, the $\mathbb{Z}_2$ symmetry implies $A=0, C=2\Imu B$, which reduces our solution \eqref{CCC00R} to
	\begin{align}
	H(\BZ_{3})&=\frac{A}{X^2}
	\left(
	\left(\frac{m_{1}}{X^2}-2\Imu \frac{m_{2}-m_{1}^{2}}{X^2}
	\right)\tilde{X}^{\dot{\beta}_{1}\beta_{1}}
	+\bar{\Theta}^{\dot{\beta}_{1}}\Theta^{\beta_{1}}
	\right)\,.
	\end{align}
		
\item If $\ell_{2}=0$, the $\mathbb{Z}_2$ symmetry implies $B=2\Imu A$. 
Therefore our solution \eqref{CCC00R} reduces to
	\begin{align}
	H(\BZ_{3})&=\frac{A}{X^2}
	\left(1-2\Imu \frac{m_{1}}{X^2}\right)
	+\frac{C}{X^2}
	\left(\frac{m_{1}^2-m_{2}}{X^4}\right)\,.
	\end{align}
This corresponds to the three-point functions of the stress tensor multiplet and is consistent with \cite{Kuzenko:1999pi} (see also \cite{Liendo:2015ofa} for its implication in the associated two-dimensional algebras).
As shown in the paper, the two constants $A$ and $C$ are related to the conformal anomalies, $a$ and $c$, of the four-dimensional SCFTs as
	\begin{align}
	A=\frac{3}{32\pi^6}(4a-c)\,,\quad C=\frac{1}{8\pi^6}(4a-5c)\,.
	\end{align}

\end{itemize}

\vspace{1mm}
Let us now turn to 
the second type of correlator, \eqref{CCCR-1}.
In this case, the function $H(\BZ_{3})$ has 
 dimension $2\ell_{2}$
 and vanishing $U(1)_{r}$ charge.
We see that there is no non-trivial solution for $\ell_2=0$. For $\ell_2>0$,
we find the following special solution:
\begin{align}
\begin{split}
H(\BZ_{3})&=
\bar{\Theta}^{\dot{\beta}_{1}}_{k_{1}}\Theta^{\beta_{1}}_{k_{2}}
\left(
\tilde{X}^{\dot{\beta}_{2}\beta_{2}}\cdots\tilde{X}^{\dot{\beta}_{2\ell_{2}}\beta_{2\ell_{2}}}
\right)
(\epsilon_{m_{1}k_{3}}\cdots\epsilon_{m_{2R}k_{2R+2}})
\\
&~\times(
\epsilon^{{\gamma}_{1}{\beta}_{2\ell_{2}+1}}
\cdots
\epsilon^{{\gamma}_{2j}{\beta}_{2j+2\ell_{2}}})
(\epsilon^{\dot{\gamma}_{1}\dot{\beta}_{2j+2\ell_{2}+1}}
\cdots
\epsilon^{\dot{\gamma}_{2j+2\ell_{1}}\dot{\beta}_{2j+2\ell_{1}+2\ell_{2}}})\,,
\end{split}
\label{CCRCR1}
\end{align}
up to a constant prefactor.\footnote{
It has very recently been shown in \cite{Agarwal:2018zqi} that the square
of the OPE coefficient of $\widehat{\mathcal{C}}_{0(0,0)}\times
\widehat{\mathcal{C}}_{n-1(\frac{n-1}{2},\frac{n-1}{2})} \supset
\widehat{\mathcal{C}}_{n(\frac{n}{2},\frac{n}{2})}$ for $n\in \mathbb{N}$ is proportional to
$\prod_{i=1}^n(c-c_i)$ with $c_n\equiv
\frac{n(6n+5)}{6(2n+3)}$. This implies that the constant prefactor
of $H(\BZ_3)$ for this channel vanishes when $c=c_i$ for $i=1,2,\cdots,n$.
}
Note here that, for $R=j=\ell_{1}=0$, 
the $\mathbb{Z}_2$ symmetry discussed above constrains
\eqref{CCRCR1} as
\begin{align}
	H(-\bar{X_{3}},-\Theta_{3},-\bar{\Theta}_{3})=(-1)^{2\ell_{2}-1}
	\left(\bar{\Theta}^{\dot{\beta}_{1}}_{k_{1}}\Theta^{\beta_{1}}_{k_{2}}
	\right)\!
	\left(
	\tilde{X}^{\dot{\beta}_{2}\beta_{2}}\cdots\tilde{X}^{\dot{\beta}_{2\ell_{2}}\beta_{2\ell_{2}}}
	\right).
	\label{CCRCR2}
\end{align}
It must be equal to $H(X_{3},\Theta_{3},\bar{\Theta}_{3})$, and therefore, $\ell_{2}$ must be a half-integer, otherwise correlation function should vanish.

\subsection{Selection rules }
\label{sec:CO}

We here write down the selection rules for
$\schurC_{0(0,0)}\times \OO^\mathrm{Schur}$ read
off from the three-point functions we computed above. Note that all the
following rules are only up to non-Schur multiplets.

\subsubsection{$\schurC_{0(0,0)} \times \schurB_{R}$ fusion}
The $\schurC_{0(0,0)} \times \schurB_{R}$ selection rules are as follows.
\begin{itemize}
	\item For $R>1$, the selection rule is
	\begin{align}
	\schurC_{0(0,0)} \times \schurB_{R}
	\sim
	\schurB_{R}
	+ \sum^{\infty}_{\ell=0}\left[
	\schurC_{R(\frac{\ell}{2},\frac{\ell}{2})}+\schurC_{R-1(\frac{\ell}{2},\frac{\ell}{2})}
	\right].\label{C0Bsele}
	\end{align}
	\item For $R=1$, because of the $\mathbb{Z}_{2}$-symmetry, the selection rule is
	\begin{align}
	\schurC_{0(0,0)} \times \schurB_{1}
	\sim
	\schurB_{1}
	+ \sum^{\infty}_{\ell=0}\left[
	\schurC_{1(\frac{\ell}{2},\frac{\ell}{2})}+\schurC_{0(\frac{\ell+1}{2},\frac{\ell+1}{2})}
	\right]\,,
	\end{align}
	 which is consistent with \cite{Ramirez:2016lyk}.
	\item For $R=\frac{1}{2}$, the rule is
	%
	 %\note{Takahiro}{By definition, $R$ of $\schurC_{R(j,\bj)}$ has to be non-negative,
	 %	 and therefore there is no $\schurC_{-1/2(j,\bj)}$.}
	 %
	\begin{align}
	\schurC_{0(0,0)} \times \schurB_{\frac{1}{2}}
	\sim
	\schurB_{\frac{1}{2}}
	+ \sum^{\infty}_{\ell=0}
	\schurC_{\frac{1}{2}(\frac{\ell}{2},\frac{\ell}{2})}.
	\end{align}
	
\end{itemize}
\subsubsection{$\schurC_{0(0,0)} \times \schurDD_{R(j,0)}$ fusion}
The selection rules for $\schurC_{0(0,0)} \times \schurDD_{R(j,0)}$ are
written 
as follows.
\begin{itemize}
	\item For $R>0$,
	\begin{align}
	\schurC_{0(0,0)} \times \schurDD_{R(j,0)}
	\sim\schurDD_{R(j,0)}
	+\sum_{\ell=0}^{\infty }\left[
	\schurC_{R+\frac{1}{2}(j+\frac{\ell}{2}+\frac{1}{2},\frac{\ell}{2})}
	+
	\schurC_{R-\frac{1}{2}(j+\frac{\ell}{2}+\frac{1}{2},\frac{\ell}{2})} 
	\right].
	\label{fusCD}
	\end{align}
	\item For $R=0$, 
	\begin{align}
	\schurC_{0(0,0)} \times \schurDD_{0(j,0)}
	\sim\schurDD_{0(j,0)}
	+\sum_{\ell=0}^{\infty }\left[
	\schurC_{\frac{1}{2}(j+\frac{\ell}{2}+\frac{1}{2},\frac{\ell}{2})}
	\right].
	\label{fusCD0}
	\end{align}
\end{itemize}
Note here that, for the $\schurC_{R,(j,\bj)}$ type multiplets on the
right-hand sides, the corresponding three-point functions of the
superconformal primaries vanishes. This reflects the fact that the
sum of the $U(1)_r$ charges of the primaries in three-point function is non-vanishing. 
However, the sum of the $U(1)_r$ charges of the {\it Schur operators} in the same
multiplets vanishes, which implies that the three-point functions of the Schur operators can be non-trivial.
Note also that the selection rules for $\schurC_{0(0,0)}\times \schurD_{R(0,j)}$
are obtained by taking the charge conjugate of \eqref{fusCD} and \eqref{fusCD0}.

% *************************
% In both cases, the superconformal primary does not conserve $U(1)_{r}$ charge.
% Left-hand side of them have the $U(1)_{r}$ charge $-j-1 $, in contrast right-hand side have $-j-1 $ or $-j-\frac{1}{2}$. 
% However, Schur operators summarized in table \ref{schurop} conserve the $U(1)_{r}$ charge.
% Schur operators in $\schurDD_{R(j,0)}$ and $\schurC_{R(j,\bj)}$ have $U(1)_{r}$ charge $ -j-\frac{1}{2}$ and $\bj-j$ respectively.
% Both sides of \eqref{fusCD} and \eqref{fusCD0} have $U(1)_{r}$ charge $ -j-\frac{1}{2}$.
% We can admit that the selection rules conserve the $U(1)_{r}$ charge as Schur operator.
% ************************

\subsubsection{$\schurC_{0(0,0)}\times \schurC_{R(j,\bj)}$ fusion }
The selection rules for $\schurC_{0(0,0)}\times \schurC_{R(j,\bj)}$ 
are written as follows.
\begin{itemize}
	\item For $ j-\bj=\ell_{1}\geq\frac{1}{2}$, $R>\frac{1}{2}$,
	\begin{align}
	\begin{split}
	\schurC_{0(0,0)}\times \schurC_{R(\bj+\ell_{1},\bj)}
	&\sim\schurDD_{R+\frac{1}{2}(\ell_{1}-\frac{1}{2},0)}+
	\schurDD_{R-\frac{1}{2}(\ell_{1}-\frac{1}{2},0)}
	\\
	&~+
	\sum_{\bj+\frac{\ell}{2}=0}^{\infty} \left[
	\schurC_{R(\bj+\frac{\ell}{2}+\ell_{1},\bj+\frac{\ell}{2})}
	\right]
	+\sum_{\ell=1}^{\infty} \left[
	\schurC_{R+1(\bj+\frac{\ell}{2}+\ell_{1},\bj+\frac{\ell}{2})}
	\right]
	+\sum_{\ell=1}^{2\bj} \left[
	\schurC_{R-1(\bj-\frac{\ell}{2}+\ell_{1},\bj-\frac{\ell}{2})}
	\right].
	\end{split}
	\label{CCjneq0fus1}
	\end{align}
	\item For $ j-\bj=\ell_{1}\geq\frac{1}{2}$, $R=\frac{1}{2}$,
	\begin{align}
	\schurC_{0(0,0)}\times \schurC_{\frac{1}{2}(\bj+\ell_{1},\bj)}
	\sim\schurDD_{1(\ell_{1}-\frac{1}{2},0)}+
	\schurDD_{0(\ell_{1}-\frac{1}{2},0)}
	+\sum_{\bj+\frac{\ell}{2}=0}^{\infty} 
	\left[
	\schurC_{\frac{1}{2}(\bj+\frac{\ell}{2}+\ell_{1},\bj+\frac{\ell}{2})}
	\right]
	+\sum_{\ell=1}^{\infty} 
	\left[
	\schurC_{\frac{3}{2}(\bj+\frac{\ell}{2}+\ell_{1},\bj+\frac{\ell}{2})}
	\right].
	\label{CCjneq0fus2}
	\end{align}
	\item For $ j-\bj=\ell_{1}\geq\frac{1}{2}$, $R=0$,
	\begin{align}
	\schurC_{0(0,0)}\times \schurC_{0(\bj+\ell_{1},\bj)}
	\sim\schurDD_{\frac{1}{2}(\ell_{1}-\frac{1}{2},0)}
	+
	\sum_{\bj+\frac{\ell}{2}=0}^{\infty} \left[
	\schurC_{0(\bj+\frac{\ell}{2}+\ell_{1},\bj+\frac{\ell}{2})}
	\right]
	+\sum_{\ell=1}^{\infty} \left[
	\schurC_{1(\bj+\frac{\ell}{2}+\ell_{1},\bj+\frac{\ell}{2})}
	\right].
	\label{CCjneq0fus3}
	\end{align}
	\item For $\bj=j>0$, $R>1$, 
	\begin{align}
	\schurC_{0(0,0)}\times \schurC_{R(j,j)}
	\sim\schurB_{R}+ \schurB_{R+1}
	+\sum_{j+\frac{\ell}{2}=0}^{\infty} \left[
	\schurC_{R(j+\frac{\ell}{2},j+\frac{\ell}{2})}\right]
	+\sum_{\ell=1}^{\infty} 
	\left[
	\schurC_{R+1(j+\frac{\ell}{2},j+\frac{\ell}{2})}
	\right]
	+\sum_{\ell=1}^{2j} 
	\left[
	\schurC_{R-1(j-\frac{\ell}{2},j-\frac{\ell}{2})}
	\right].
	\label{CCjeqbjfus}
	\end{align}
	\item For $\bj=j>0$, $R=1$,
	\begin{align}
	\schurC_{0(0,0)}\times \schurC_{1(j,j)}
	\sim\schurB_{1}+ \schurB_{2}
	+\sum_{j+\frac{\ell}{2}=0}^{\infty} \left[
	\schurC_{1(j+\frac{\ell}{2},j+\frac{\ell}{2})}\right]
	+\sum_{\ell=1}^{\infty} 
	\left[
	\schurC_{2(j+\frac{\ell}{2},j+\frac{\ell}{2})}
	\right]
	+\sum_{\ell=1}^{2j} 
	\left[
	\schurC_{0(j-\frac{\ell}{2},j-\frac{\ell}{2})}
	\right].
	\label{CCfus3}
	\end{align}
	When $j$ is an integer, the stress-tensor multiplet $\schurC_{0(0,0)}$ in the last term on the right-hand side % of \eqref{CCfus3} 
	must be excluded by the $\mathbb{Z}_{2}$-symmetry, under the assumption of uniqueness of the stress tensor. 
	\item For $\bj=j>0$, $R=\frac{1}{2}$,
	\begin{align}
	\schurC_{0(0,0)}\times \schurC_{\frac{1}{2}(j,j)}
	\sim\schurB_{\frac{1}{2}}+ \schurB_{\frac{3}{2}}
	+\sum_{j+\frac{\ell}{2}=0}^{\infty} \left[
	\schurC_{\frac{1}{2}(j+\frac{\ell}{2},j+\frac{\ell}{2})}\right]
	+\sum_{\ell=1}^{\infty} 
	\left[
	\schurC_{\frac{3}{2}(j+\frac{\ell}{2},j+\frac{\ell}{2})}
	\right].
	\end{align}
	\item For $\bj=j=0,R=0$,
	\begin{align}
	\schurC_{0(0,0)}\times \schurC_{0(0,0)}
	\sim
	\sum_{\ell=0}^{\infty} 
	\left[
	\schurC_{0(\ell,\ell)}
	+\schurC_{1(\ell+\frac{1}{2},\ell+\frac{1}{2})}
	\right].
	\label{CC0fus}
	\end{align}
\end{itemize}
The cases of $ j-\bj< 0$ are 
obtained by the charge conjugates of the above ones.

\section{Conclusions and discussions}
 \label{conc}
In this paper, we have computed the three-point functions of the form
$\langle
\widehat{\mathcal{B}}_{R_1}\widehat{\mathcal{B}}_{R_2}\mathcal{O}\rangle$
and $\langle
\widehat{\mathcal{C}}_{0(0,0)}\mathcal{O}_1\mathcal{O}_2\rangle$ for
arbitrary Schur multiplets $\mathcal{O}, \mathcal{O}_1$ and
$\mathcal{O}_2$. We have obtained the most general
expressions for these three-point functions, except for the two types
correlators in \eqref{CCCR} and \eqref{CCCR-1}. For the two correlators
in \eqref{CCCR} and \eqref{CCCR-1}, we have found special solutions to the
semi-shortening conditions. From these results, we have derived the OPE selection
rules for $ \widehat{\mathcal{C}}_{0(0,0)} \times
\mathcal{O}^\mathrm{Schur}$ up to non-Schur multiplets, where $\mathcal{O}^\mathrm{Schur}$ is an arbitrary Schur
multiplet. Our selection rules are listed in sub-section
\ref{sec:CO}. We have also shown in sub-section \ref{sec:BB} that our results on the three-point functions $\langle
\widehat{\mathcal{B}}_{R_1}\widehat{\mathcal{B}}_{R_2}\mathcal{O}\rangle$
are consistent with the selection rules for
$\widehat{\mathcal{B}}_{R_1}\times \widehat{\mathcal{B}}_{R_2}$ obtained
in \cite{Nirschl:2004pa}.
%Our results
%are summarized in sub-section \ref{sec:BB} and sub-section \ref{sec:CO}. 
%we have derived the superconformal selection rules for $\schurB_{R_{1}} \times \schurB_{R_{2}}$ and $\schurC_{0(0,0)} \times \OO^{\text{Schur}}$, up to non-Schur multiplets, by solving the (semi-)shortening conditions for the Schur multiplets.
 % Our results are summarized in section \ref{sec:BB} and section \ref{sec:CO}. 
 % To derive these results, we evaluated the most general expressions for the three-point functions of the form $\langle \schurB_{R_1}\schurB_{R_2}\OO\rangle$ and $\langle \schurC_{0(0,0)}\OO_1\OO_2\rangle$ for arbitrary Schur multiplets $\OO,\OO_1$, and $\OO_2$, except for the two types of correlators in \eqref{CCCR} and \eqref{CCCR-1}. 
 % For the two correlators in \eqref{CCCR} and \eqref{CCCR-1}, we find special solutions to the semi-shortening conditions. 
We emphasize that our analysis 
relies %rely 
only on the shortening conditions for Schur multiplets and therefore does not depend on any detail of four-dimensional $\mathcal{N}=2$ SCFTs.

Let us here discuss an interesting constraint appearing in the selection rules for $\schurC_{0(0,0)}\times \OO^\mathrm{Schur}$. 
Suppose that $\OO$ and $\OO'$ are two Schur multiplets so that $\OO'$ appears in the OPE of $\schurC_{0(0,0)}\times \OO$, i.e., $\schurC_{0(0,0)}\times \OO \supset
\OO'$. 
We denote by $R^{(s)}$ and $R'{}^{(s)}$ the $SU(2)_R$ charges of the Schur operators in $\OO$ and $\OO'$, respectively.
We also denote respectively by $h$ and $h'$ the holomorphic dimensions of the two-dimensional operators associated with the Schur operators in $\OO$ and $\OO'$. See table \ref{schurop} for the relation between the two-dimensional holomorphic dimension and four-dimensional quantum numbers of operators. 
Now, we see that our selection rules imply
\begin{align}
R^{\text{(s)}}<R'^{\text{(s)}} \; \Longrightarrow\; h< h'
 \,,\qquad R^{\text{(s)}}> R'^{\text{(s)}} \;\Longrightarrow\;
 h> h'\,.
\label{eq:const_Rh}
\end{align}
In the 4d/2d correspondence of \cite{Beem:2013sza}, this means that the
$SU(2)_R$ charge of the four-dimensional ancestor of a two-dimensional operator is always
smaller than or equal to those of its Virasoro descendants. Since the
$SU(2)_R$ symmetry is broken in the associated chiral algebra
\cite{Beem:2013sza}, this relation between the $SU(2)_R$ charge and
the holomorphic dimension is surprising.\footnote{
Note that, since $h$ is related to the $SU(2)_R$ charge $R^{(s)}$ and the spin $(j^{(s)},\bj^{(s)})$ of
the Schur operator by $h=R^{(s)} + j^{(s)} + \bj^{(s)}$, the
constraint \eqref{eq:const_Rh} can also be regarded as a constraint on the
$SU(2)_R$ charges and spins of the Schur operators in $\OO$ and $\OO'$.}
See also \cite{Bonetti:2018fqz} for a remarkable discussion on reconstructing 
the $SU(2)_R$-filtration of the chiral algebra.

Another interesting observation is that, in some of the OPE channels allowed by our selection rules, the
three-point function of the superconformal primaries vanishes even
though those of their descendants do not. As mentioned already, this is
a common feature of SCFTs \cite{Fortin:2011nq}. Indeed, the vanishing of the
three-point function of the primaries reflects the fact that the sum of
their $U(1)_r$ charges is non-vanishing. 
Therefore, our selection rules for Schur multiplets do not imply the non-vanishing of the three-point
function of the corresponding superconformal primaries.

On the other hand, when we focus on the {\it Schur operator} in each Schur multiplet, we see that the sum of their $U(1)_r$ charges vanishes whenever the corresponding three Schur multiplets are allowed by the selection rules.
This seems to suggest that the three-point functions of Schur {\it operators} are 
 always non-vanishing whenever the corresponding Schur {\it multiplets}
 have non-vanishing three-point functions.\footnote{ While this could in principle be checked by using the three-point functions we computed, it is not straightforward to extract the correlators of Schur components from our superfield correlators. We leave it to future work. } 
This observation leads us to a conjecture on BPS selection rules of general Schur multiplets. Suppose that a Schur multiplet $\OO_3^\mathrm{Schur}$ appears in the OPE of $\OO_1^\mathrm{Schur}$ and $\OO_{2}^\mathrm{Schur}$, i.e., $\OO^{\text{Schur}}_{1}\times \OO^{\text{Schur}}_{2} \supset \OO^{\text{Schur}}_{3}$.
Let us denote the $SU(2)_R$ and $U(1)_r$ charges of the Schur operators in the multiplets respectively by $R^{(s)}_i$ and $r^{(s)}_i$ for $i=1,2$, and $3$. Then we conjecture that the following two conditions are satisfied:
\begin{align}
 r^{(s)}_1 + r_2^{(s)} = r^{(s)}_3, \qquad |R_{1}^{\text{(s)}}-R_{2}^{\text{(s)}}| & \leq R^{\text{(s)}}_{3}\leq R_{1}^{\text{(s)}}+R_{2}^{\text{(s)}}\,.
\label{eq:const_ch}
\end{align}
Note that these conditions are necessary for the three-point functions of the Schur operators to be non-vanishing. 
Recognizing \eqref{eq:const_Rh} and \eqref{eq:const_ch} as principle of selection rule related to four-dimensional $\NN=2$ SCFT whose stress tensor is unique, we recover our all selection rules in section \ref{sec:BB} and \ref{sec:CO}. 
We leave the detailed study of this conjecture to future work.

 \begin{table}%[!h]
	\centering \renewcommand{\arraystretch}{1.5}
	\begin{tabular}{|p{0.11596\textwidth}|p{0.270\textwidth}|p{0.14596\textwidth}|p{0.08596\textwidth}|p{0.14596\textwidth}|}
		\hline
		Multiplet & Schur operator & $U(1)_{r}$ charge&$SU(2)_{R}$& $h$\\
		\hline
		$\gschurB$ &$\Phi^{(i_{1}\cdots i_{2R})}$ & $ r=0$&$R$&$R$\\
		\hline
		$\gschurD$ & 
		$\bar{\QQ}_{(\dot{\alpha}}^{(i}\Phi^{i_{1}\cdots i_{2R})}_{\dot{\alpha}_{1}\cdots\dot{\alpha}_{2\bj})}$ & 
		$r=\bj+\frac{1}{2}$&
		$R+\frac{1}{2}$&$R+j+1 $\\
		\hline
		$\gschurDD$ &${\QQ}_{({\alpha}}^{(i}\Phi^{i_{1}\cdots i_{2R})}
		_{{\alpha}_{1}\cdots{\alpha}_{2j})}$ & 
		$r=-j-\frac{1}{2}$&
		$R+\frac{1}{2}$&$R+\bj+1 $\\
		\hline
		$\gschurC$ &
		$\bar{\QQ}_{(\dot{\alpha}|}^{(i}\QQ_{(\alpha}^{i'}
		\Phi^{i_{1}\cdots i_{2R})}_{{\alpha}_{1}\cdots{\alpha}_{2j})|\dot{\alpha}_{1}\cdots\dot{\alpha}_{2\bj})}
		$ & $r=\bj-j$&$R+1$&$R+j+\bj+2$\\
		\hline
	\end{tabular}
	\caption{{Schur operators in Schur multiplets with their $U(1)_r$ and $SU(2)_R$ 	charges.
 Here $\Phi$ is the superconformal primary field of the multiplet. 
			The rightmost column shows the holomorphic dimension of the corresponding operator in the two-dimensional chiral algebra.}}
	\label{schurop}
\end{table}

\section*{Acknowledgements}

The authors are particularly grateful to Sanefumi Moriyama for collaboration at the early stage of
this work as well as for various useful discussions during this project. The authors
thank Matthew Buican for illuminating discussions. 
The authors are  grateful to Hiroshi Itoyama, Kazunobu Maruyoshi,  Yutaka Matsuo, and Satoshi Yamaguchi for discussions.
The authors are also
grateful to Nobuhito Maru for his encouragement. Most of our computations are done with the Mathematica package
``grassmann.m'' provided by M.~Headrick to whom the authors are
grateful.
The work of T.~N. is partially supported by JSPS Grant-in-Aid for Early-Career Scientists 18K13547.
The work of K.~K is supported by JSPS KAKENHI Grant Number JP18J22009.

\appendix
\section{Fierz identities}\label{appFierz}
In this appendix, we summarize 
useful identities for Grassmann variables $\Theta^{i\alpha}$ and
$\bar{\Theta}^{\dot{\alpha}}_{~i}$,
which we call 
Fierz identities.
We first introduce 
the following variables:
\begin{align}
M^{i}_{~j}
&\equiv \Theta^{i\alpha}X_{\alpha\dot{\alpha}} \bar{\Theta}^{\dot{\alpha}}_{~j}\,,
\label{Mdef}
\\
m_l &\equiv \tr M^l \,,
\label{mdef}
\\
H_{ij}^{\dot{\alpha}\alpha}
&\equiv{\bar{\Theta}^{\dot{\alpha}}}_{~j}(\epsilon\Theta)_{i}^{~\alpha}.
\end{align}
From the nilpotent structure of $\Theta$ and $\bar{\Theta}$, we see that the
following Fierz identities hold\footnote{$(\epsilon M)_{ij}=\epsilon_{ik}M^{k}_{~j}$}
\begin{align}
\begin{split}
&m_1m_2=0\,,\quad 
m_2^2=m_1^4\,,\quad 
(\epsilon M)_{(ii')}(m_2+2m_1^2)=0\,,
\\
&(\epsilon M)_{(ii')}(\epsilon M)_{(jj')}m_{1}=\frac{1}{12}(\epsilon_{ij}\epsilon_{i' j'}+\epsilon_{i j'}\epsilon_{i' j})m_1^3\,.
\end{split}
\label{mfirez}
\end{align}
Moreover, by using the variables $m_l$, we can expand powers of $\bar{X}^2$ as \cite{Liendo:2015ofa}
\begin{align}
\frac{1}{(\bar{X}^2)^{\Delta}}
=\frac{1}{(X^2)^{\Delta}}\left(
1-4\Imu \Delta\frac{m_1}{X^2}
+8\Delta\frac{m_2}{X^4}-8\Delta^2\frac{m_1^2}{X^4}
+\frac{32i}{3}\Delta(\Delta^2-1)\frac{m_1^3}{X^6}
+\frac{32}{3}\Delta^2(\Delta^2-1)\frac{m_1^4}{X^8}
\right)\,.
\label{X-bartoX}
\end{align}
We can also derive 
\footnote{
	$\bar{M}^{i}_{~j}:=\Theta^{i\alpha}\bar{X}_{\alpha\dot{\alpha}} \bar{\Theta}^{\dot{\alpha}}_{~j}
	\,,\bar{m}_l := \tr \bar{M}^l.$}
\begin{align}
\frac{M_{(ii')}}{X^4}=\frac{\bar{M}_{(ii')}}{\bar{X}^4}
\,, \qquad \frac{{m}^2_{1}-{m}_{2}}{{X^6}}=\frac{\bar{m}^2_{1}-\bar{m}_{2}}{\bar{X^6}}=
\frac{\Theta^{\alpha\alpha'}X_{\alpha\dot{\alpha}}X_{\alpha'\dot{\alpha}'} \bar{\Theta}^{\dot{\alpha}\dot{\alpha}'}}{X^6}\,.
\label{MXBMBX}
\end{align}
The following Fierz identities are derived by using the Mathematica
package \verb|grassmann.m| \cite{Grasmman}:
\begin{align}
({\Theta}^{i \alpha}{\Theta}^{j\beta}\epsilon_{\alpha\beta})\bar{\Theta}
\Theta^{\dot{\alpha}_1\alpha_1}\bar{\Theta}\Theta^{\dot{\alpha}_2\alpha_2}
&=0\,,
\\
(m_1^2+m_2)\bar{\Theta}\Theta^{\dot{\alpha}_1\alpha_1}\bar{\Theta}\Theta^{\dot{\alpha}_2\alpha_2}
&=0\,,\label{TTBm1m2}
\\
m_1(\epsilon M)_{(ij)}
\bar{\Theta}\Theta^{\dot{\alpha}_1\alpha_1}\bar{\Theta}\Theta^{\dot{\alpha}_2\alpha_2}
&=0\,,
\\
m_1	\bar{\Theta}\Theta^{\dot{\alpha}_1\alpha_1}\bar{\Theta}\Theta^{\dot{\alpha}_2\alpha_2}
&=\frac{1}{3X^2}(m_1^2+m_2)\bar{\Theta}\Theta^{\dot{\alpha}_1\alpha_1}{\tilde{X}}^{\dot{\alpha}_2\alpha_2}\,,
\\
(\epsilon M)_{(ij)}
\bar{\Theta}\Theta^{\dot{\alpha}_1\alpha_1}\bar{\Theta}\Theta^{\dot{\alpha}_2\alpha_2}
&=\frac{ m_1}{X^2}(\epsilon M)_{(ij)}\bar{\Theta}\Theta^{\dot{\alpha}_1\alpha_1}	
{\tilde{X}}^{\dot{\alpha}_2\alpha_2}\,,
\\
\label{MTTBm1}
(\epsilon M)_{(ij)}\bar{\Theta}\Theta^{\dot{\alpha}\alpha}m_1^2
&=(\epsilon M)_{(ij)}\bar{\Theta}\Theta^{\dot{\alpha}\alpha} {m_2}=0\,,
\\
(2m_1^2-m_2)\bar{\Theta}\Theta^{\dot{\alpha}\alpha}
&=\frac{m_1^3}{X^2}{\tilde{X}}^{\dot{\alpha}\alpha}
\label{TTBm1m1}
\,,
\\
m_1^3\bar{\Theta}\Theta^{\dot{\alpha}\alpha}
&=\frac{m_1^4}{2X^2}{\tilde{X}}^{\dot{\alpha}\alpha}\,,
\\
(\epsilon M)_{(ij')}(\epsilon M)_{(i'j')}
\bar{\Theta}\Theta^{\dot{\alpha}\alpha}
&=\frac{(\epsilon_{ij}\epsilon_{i'j'})}{12}\left((m_1^2+m_2)
\bar{\Theta}\Theta^{\dot{\alpha}\alpha}-\frac{m_1^3}{X^2}{\tilde{X}}^{\dot{\alpha}\alpha}\right),
\\
(\epsilon M)_{(ij)}(\epsilon M)_{(i'j')}
\bar{\Theta}\Theta^{\dot{\alpha}_1\alpha_1}\bar{\Theta}\Theta^{\dot{\alpha}_2\alpha_2}
&=-\frac{m_1^4(\epsilon_{ij}\epsilon_{i'j'})}{24(X^2)^2}
{\tilde{X}}^{\dot{\alpha}_1\alpha_1}	{\tilde{X}}^{\dot{\alpha}_2\alpha_2}\,,
\\
H_{(ij)}^{\dot{\alpha}_1\alpha_1}\bar{\Theta}\Theta^{\dot{\alpha}_2\alpha_2}&=0\,,
\label{HTTB}
\\
m_1^3H_{(ij)}^{\dot{\alpha}\alpha}
&=0\,,
\\
m_1 H_{(ij)}^{\dot{\alpha}\alpha}
+(\epsilon M)_{(ij)}\bar{\Theta}\Theta^{\dot{\alpha}\alpha}
&= \frac{m_1}{X^2}(\epsilon M)_{(ij)} {\tilde{X}}^{\dot{\alpha}\alpha}\,,
\label{Hm1}
\\
\frac{m_2}{X^2}H_{(ij)}^{\dot{\alpha}\alpha}
&=-\frac{m_1^2}{X^2}\frac{\epsilon
	M_{(ij)}}{X^2}{\tilde{X}}^{\dot{\alpha}\alpha}\,,
\\
\frac{	(m_1^2+m_2)}{X^2}H_{(ij)}^{\dot{\alpha}\alpha}
&=-\frac{m_1}{X^2}(\epsilon M)_{(ij)}\bar{\Theta}\Theta^{\dot{\alpha}\alpha}.
\end{align}

\section{$\langle{\schurC}_{0(0,0)}{\schurB}_{R}{\OO}^{\II}\rangle$}\label{appCBO}
In this appendix, we describe a derivation of the most general
expression for the correlation function $\langle{\schurC}_{0(0,0)}{\schurB}_{R}{\OO}^{\II}\rangle$ for various ${\OO}^{\II}$.
Since the case of ${\OO}^{\II}={\schurB}_{R'}$ has been analyzed in
section \ref{secBBfusC}, we here focus on 
${\OO}^{\II}={\schurDD}_{R'(j,0)}$, ${\schurD}_{R'(0,j)}$, and ${\schurC}_{R'(j,\bj)} $.

\subsection{$\langle{\schurC}_{0(0,0)}{\schurB}_{R}{\schurDD}_{R'(j,0)}\rangle$}\label{apCBD}

Let us first consider the case of $\mathcal{O}^\II = \bar{\mathcal{D}}_{R'(j,0)}$.
In this case, $H(\bar{\BZ}_{3})$ has dimension $2(R'-R) +j-1$ and 
$U(1)_{r}$ charge $-j-1$.
Since the absolute value of the $U(1)_{r}$ charge 
is at most one,
the spin 
$ j$ must be zero.
Then the only 
possibility for $R'$ is $R'=R-1$, which implies
\begin{align}
H(\bar{\BZ}_{3})=\frac{A}{\bar{X}^{4}}\bar{\Theta}_{m_{1}}\bar{\Theta}_{m_{2}}\epsilon_{m_{3}k_{1}}\cdots \epsilon_{m_{2R}k_{2R-2}}\,.
\end{align}
However, this is not consistent with \eqref{ansC00} unless $A=0$, and therefore the correlation function $\langle{\schurC}_{0(0,0)}{\schurB}_{R}{\schurDD}_{R'(j,0)}\rangle$ has to vanish.
From charge conjugation, we see that $\langle{\schurC}_{0(0,0)}{\schurB}_{R}{\schurD}_{R'(0,j)}\rangle$ also vanishes.

\subsection{$\langle{\schurC}_{0(0,0)}{\schurB}_{R}\, {\schurC}_{R'(j,\bj)}\rangle$}

Let us next turn to the case of ${\OO}^{\II} = {\schurC}_{R'(j,\bj)}$.
%and 
In this case, the function $H(\BZ_{3})$ has 
%the
dimension $2(R'-R) +j+\bj $ and 
%the 
$U(1)_{r}$ charge $\bj-j$. Recall that $|\bj-j|$ is at most one.
Moreover, since charge conjugation exchanges $j$ and $\bj$, we only need to
study the cases of $j\geq \bj$.
%For charge conjugation
%$({\schurC}_{R'(j,\bj)})^*={\schurC}_{R'(\bj,j)}$ and $|\bj-j|\leq1$, 
Therefore, we assume $-1 \leq\bj-j \leq 0$.
Then the possible 
%patterns 
combinations of $R'$ and $\bj - j$ are 
$(R',\bj-j)=(R,-1)$, $(R+\frac{1}{2},-\frac{1}{2})$,
$(R-\frac{1}{2},-\frac{1}{2})$, $(R+1,0)$, $(R,0)$, and $(R-1,0)$. We
study the most general expression for $H(\BZ_3)$ in each of these cases below. 

\subsubsection{$\bj=j-1\geq 0 \,, R'=R$}
In this case, the only possible candidate for $H(\BZ_3)$ is 
\begin{align}
H(\BZ_{3})=\frac{A}{X^2}\epsilon_{m_{1}k_{1}}\cdots \epsilon_{m_{2R}k_{2R}}
X_{\beta_{1}\dot{\beta}_{1}}\cdots X_{\beta_{2j}\dot{\beta}_{2j}}
\tilde{\bar{\Theta}}_{\dot{\beta}_{2j+1}\dot{\beta}_{2j+2}}.
\end{align}
%It does not fit 
This is not consistent with \eqref{ansC00Bar} unless $A=0$.
%by replacing $X \rightarrow \bar{X}$ and $A$ must be zero.

\subsubsection{$\bj=j-\frac{1}{2}\geq 0 \,, R'=R+\frac{1}{2}$}
In this case, the only possible candidate is
\begin{align}
H(\BZ_{3})=A\epsilon_{m_{1}k_{1}}\cdots \epsilon_{m_{2R}k_{2R}}
X_{\beta_{1}\dot{\beta}_{1}}\cdots X_{\beta_{2j-1}\dot{\beta}_{2j-1}}
X_{\beta_{2j}\dot{\alpha}}
\bar{\Theta}^{\dot{\alpha}}_{k_{2R+1}}\,,
\end{align}
which is not consistent with \eqref{BcondQbar} unless $A=0$.

\subsubsection{$\bj=j-\frac{1}{2}\geq0 \,, R'=R-\frac{1}{2}$}
In this case, there are three possible terms in $H(\BZ_{3})$; 
\begin{align}
H(\BZ_{3})
&=\frac{1}{X^2}\epsilon_{m_{1}k_{1}}\cdots \epsilon_{m_{2R-1}k_{2R-1}}
\Bigg(
A\bar{\Theta}^{\dot{\alpha}'}_{m_{2R}}
X_{\beta_{1}\dot{\beta}_{1}}
+
\Theta_{m_{2R}}^{\alpha}\bar{\Theta}^{\dot{\alpha}\dot{\alpha}'}
\left(B
\frac{X_{\alpha\dot{\alpha}}X_{\beta_{1}\dot{\beta}_{1}}}{X^2} 
+C \epsilon_{\dot{\beta}_{1}\dot{\alpha}}\,\epsilon_{\beta_{1}\alpha}
\right)\Bigg) 
\nonumber\\
&~\times
X_{\beta_{2}\dot{\beta}_{2}}\cdots X_{\beta_{2j-1}\dot{\beta}_{2j-1}}X_{\beta_{2j}\dot{\alpha}'}\,.
\end{align}
However, this is not consistent with \eqref{ansC00Bar} unless $A=B=C=0$.

\subsubsection{$ \bj=j\geq0 \,, R'=R$}
The only possible $H(Z_3)$ is of the form
\begin{align}
H(\BZ_{3})=\!
\Bigg(A\epsilon_{m_{1}k_{1}}X_{\beta_{1}\dot{\beta}_{1}}
+B X_{\beta_{1}\dot{\beta}_{1}}\frac{M_{m_{1}k_{1}}}{X^2}
+C\tilde{\Theta}_{m_{1}\beta_{1}}\tilde{\bar{\Theta}}_{\dot{\beta}_{1}k_{1}}
\Bigg)\epsilon_{m_{2}k_{2}}\cdots \epsilon_{m_{2R}k_{2R}}
X_{\beta_{2}\dot{\beta}_{2}}\cdots X_{\beta_{2j}\dot{\beta}_{2j}}\,.
\label{CBCRJ}
\end{align}
The constraint 
%of
\eqref{BcondQ} 
%gives 
is then expressed as
%, since 
\begin{align}
0=B\left(\epsilon_{m_{1}k_{1}}\cdots \epsilon_{m_{2R}k_{2R}}\right)
X_{\beta_{1}\dot{\beta}_{1}}\cdots X_{\beta_{2j}\dot{\beta}_{2j}}
\frac{\Theta_{m}^{\alpha}X_{\alpha\dot{\alpha}}}{X^2}\,,
\end{align}
which implies $B=0$. Moreover, for \eqref{CBCRJ} to be consistent with
\eqref{ansC00Bar}, we should set
%For setting 
$C=-4\Imu (2j)A$. Indeed, with this condition imposed, the function
\eqref{CBCRJ} can be rewritten as
%, we can rewrite $H(\BZ_{3})$ as $\bar{\BZ_{3}}$ function by using \eqref{HTTB}
\begin{align}
H(\bar{\BZ}_{3})&=A
(\epsilon_{m_{1}k_{1}}\bar{X}_{\beta_{1}\dot{\beta}_{1}}
-4\Imu (2j)\tilde{\Theta}_{k_{1}\beta_{1}}\tilde{\bar{\Theta}}_{\dot{\beta}_{1}m_{1}})
(\epsilon_{m_{2}k_{2}}\cdots \epsilon_{m_{2R}k_{2R}})
\bar{X}_{\beta_{2}\dot{\beta}_{2}}\cdots \bar{X}_{\beta_{2j}\dot{\beta}_{2j}}\,,
\end{align}
and therefore satisfies \eqref{C00}. We also see that this
expression satisfies \eqref{BcondQbar}.
%it fits \eqref{ansC00Bar}, namely 
%satisfies \eqref{BcondQbar} and \eqref{C00}.
%trivially. 

\subsubsection{$\bj-j=0 \,, R'=R-1$}
In this case, the possible $H(\BZ_{3})$ for $ j>0$ is of the form
\begin{align}
\begin{split}
H(\BZ_{3})&=\frac{1}{X^2}\Bigg(
\left(
AX_{\beta_{1}\dot{\beta}_{1}}
\frac{M_{m_{1}m_{2}}}{X^2}
+B\tilde{\Theta}_{m_{1}\beta_{1}}\tilde{\bar{\Theta}}_{\dot{\beta}_{1}m_{2}}
\right)
+\frac{C}{X^2}
\Theta_{m_{1}}\Theta_{m_{2}}\bar{\Theta}^{\dot{\alpha}\dot{\alpha}'}
X_{\beta_{1}\dot{\alpha}}\epsilon_{\dot{\beta}_{1}\dot{\alpha}'}\Bigg)
\\
&~\times\left(\epsilon_{m_{3}k_{1}}\cdots \epsilon_{m_{2R}k_{2R-2}}\right)
X_{\beta_{2}\dot{\beta}_{2}}\cdots X_{\beta_{2j}\dot{\beta}_{2j}}.
\end{split}
\end{align} 
%By setting $B=-2j A, ~C=-4\Imu j A$, it satisfies \eqref{BcondQbar} and
%\eqref{C00}.
For this to be consistent with \eqref{BcondQbar} and \eqref{C00}, we
must impose $B=-2j A$ and $C=-4\Imu j A$.
On the other hand, for $ j=0$, $H(\BZ_{3})$ is given by
\begin{align}
H(\BZ_{3})=
A\frac{M_{m_{1}m_{2}}}{X^4}
\left(\epsilon_{m_{3}k_{1}}\cdots \epsilon_{m_{2R}k_{2R-2}}\right).
\end{align}
%From
Using the Fierz identity \eqref{MXBMBX}, we see that this satisfies all the
(semi-)shortening conditions.
% trivially. 

\section{$\langle{\schurC}_{0(0,0)}{\schurDD}_{R'(j,0)}{\OO}^{\II}\rangle$}
\label{appCDO}
In this appendix, we describe the details of our computations of
$\langle{\schurC}_{0(0,0)}{\schurDD}_{R'(j,0)}{\OO}^{\II}\rangle$. We solve the equations \eqref{C00B},\eqref{C00}, 
\eqref{BcondQ}, and \eqref{DcondQ}/\eqref{DcondQ0} together with the
(semi-)shortening conditions associated with the third Schur multiplet ${\OO}^{\II}$.
We note here that the most general solution to
\eqref{DcondQ} or \eqref{DcondQ0} is written as
\begin{align}
\begin{split}
H&=
f_{1(m_1\cdots m_{2R})}^{(\dot{\gamma}_1\cdots \dot{\gamma}_{2j}),\,\II}
+f_{2(m_1\cdots m_{2R}m)}^{(\dot{\gamma}_1\cdots \dot{\gamma}_{2j}\dot{\gamma}),\,\II}
\tilde{\bar{\Theta}}^{~m}_{\dot{\gamma}}
+f_{3(m_1\cdots m_{2R-1}|}^{(\dot{\gamma}_1\cdots \dot{\gamma}_{2j}\dot{\gamma}),\,\II}
\tilde{\bar{\Theta}}_{\dot{\gamma}|m_{2R})}
+f_{4(m_1\cdots m_{2j-1}}^{\II (\dot{\gamma}_1\cdots \dot{\gamma}_{2j-1}}
{\bar{\Theta}}^{\dot{\gamma}_{2j})}_{m_{2R})}
\\
&~+
f_{5(m_1\cdots m_{2R})}^{(\dot{\gamma}_1\cdots \dot{\gamma}_{2j}\dot{\gamma}\dot{\gamma}'),\,\II}
\tilde{\bar{\Theta}}_{\dot{\gamma}\dot{\gamma}'}
+f_{6(m_1\cdots m_{2R-2} }^{(\dot{\gamma}_1\cdots \dot{\gamma}_{2j}),\,\II}\bar{\Theta}_{m_{2R-1}}\bar{\Theta}_{m_{2R})}
\\
&~+\left(2j\epsilon^{\dot{\beta}(\dot{\gamma}_1}
f_{7(m_1\cdots m_{2R})}^{\dot{\gamma}_2\cdots\dot{\gamma}_{2j})\dot{\beta}',\,\II}\,\tilde{\bar{\Theta}}_{\dot{\beta}\dot{\beta}'}
+(2j+2)\bar{\Theta}^{(k} \bar{\Theta}^{k')}\epsilon_{k(m_1}
f_{7|m_2\cdots m_{2R})k'}^{(\dot{\gamma}_1\cdots \dot{\gamma}_{2j}),\,\II}
\right)
\\
&~+
\bar{\Theta}^{(\dot{\gamma}|}_{~j}\tilde{\bar{\Theta}}^{j}_{\dot{\alpha}}\bar{\Theta}^{\dot{\alpha}}_{(m}
f_{8m_{1}\cdots m_{2R})}^{|\dot{\gamma}_{1}\cdots\dot{\gamma}_{2j}),\,\II}\,
+\tilde{\bar{\Theta}}_{\dot{\gamma}j}\tilde{\bar{\Theta}}^{j}_{\dot{\alpha}}\bar{\Theta}^{\dot{\alpha}}_{~(m_{1}}
f_{9m_{2}\cdots m_{2R})}^{(\dot{\gamma}_{1}\cdots\dot{\gamma}_{2j}\dot{\gamma}),\,\II}.
\end{split}
\label{semigsol}
\end{align}
Here $f_{i}$ are functions of $X$ and $\Theta$.
Note in particular that the superscripts $\dot\gamma_i$ in
$f_{7(m_1\cdots m_{2R})}^{(\dot\gamma_1\cdots \dot\gamma_{2j}),\,\II}$ 
are totally symmetric, which is implicit in the first term in
the bracket in \eqref{semigsol}.\footnote{
	For example in the case of $ j=1$, $f_7$ is defined with some function $f$ by
	$f_7^{(\gamma_1\gamma_{2})}=\frac{1}{2}\left(f^{\gamma_{1}\gamma_{2}}+f^{\gamma_{2}\gamma_{1}}\right) $
	while in \eqref{semigsol} we further symmetrize as in
	$\epsilon^{\beta(\gamma_{1}}f_{7}^{\gamma_{2})\beta'}=\frac{1}{2}\left(
	\epsilon^{\beta\gamma_{1}}f_{7}^{(\gamma_{2}\beta')}+\epsilon^{\beta\gamma_{2}}f_{7}^{(\gamma_{1}\beta')}
	\right)$
	using the function $f_7$ we have just defined, and $SU(2)_{R}$ indices are symmetrized similarly. }

The function $H(\BZ_{3})$ has dimension $\Delta_{3}-2R-j-3$ and $U(1)_{r}$ charge $r_{3}-j-1$, where $\Delta_{3}$ and $r_{3}$ are the dimension and the $U(1)_{r}$ charge of the third multiplet ${\OO}^{\II}$ respectively.
The case of ${\OO}^{\II}= \schurB_{R'}$ has already been studied in
section \ref{apCBD}.
%we can also use the same argument for 
Moreover, $H(\BZ_3)$ turns out to vanish for ${\OO}^{\II}= {\schurDD}_{R'(j',0)}$.
%In this case, 
Indeed, for ${\OO}^{\II}= {\schurDD}_{R'(j',0)}$, the $U(1)_{r}$ charge of $H(\BZ_3)$ is $-j-j'-2$, which is
not possible since the $U(1)_r$
charge is bounded from below by $-1$.
Therefore, ${\schurDD}_{R'(j',0)}$ dose not appear in the OPE
of ${\schurC}_{0(0,0)} \times {\schurDD}_{R(j,0)}$. In the rest of this appendix, we consider
the remaining cases ${\OO}^{\II}= \schurD_{R'(0,\bj)}$ and $\schurC_{R'(j',\bj')}$.

\subsection{$\langle{\schurC}_{0(0,0)}{\schurDD}_{R(j,0)}{\schurD}_{R'(0,\bj)}\rangle$}

For ${\OO}^{\II}= \schurD_{R'(0,\bj)}$, $H(\BZ_{3})$ has dimension $2(R'-R) +(\bj-j) -2$ and $U(1)_{r}$ charge $\bj-j$.
Up to charge conjugation, the possible values of the $U(1)_r$ charge is $0,\frac{1}{2}$, and $1$.
Therefore the possible combinations of $R'$ and $\bj$ are
$(R',\bj)=(R,j-1)$, $(R-\frac{1}{2},j-\frac{1}{2})$, $(R+\frac{1}{2},j-\frac{1}{2})$, $(R,j)$, and $(R-1,j)$.
The shortening condition of the third multiplet ${\schurD}_{R'(0,\bj)}$ is 
\begin{align}
\bar{\mathcal{S }}^{\dot{\alpha}}_{(n}G(\BZ_{2})_{n_{1}\cdots n_{2R'})}=0\,,
\label{thirdDshort}
\end{align}
and the semi-shortening condition is 
\begin{alignat}{2}
&\mathcal{S}^{\delta}_{(n}G_{n_{1}\cdots n_{2R'})(\delta\delta_{2}\cdots \delta_{2\bj})}(\BZ_{2})
=0\,,&\quad \text{for}\quad j > 0\,,&\label{thirdDsemishort}
\\
&\epsilon_{\alpha\beta}\mathcal{S}^{\alpha}_{(n}\mathcal{S}^{\beta}_{n'}G_{n_{1}\cdots n_{2R'})}(\BZ_{2})
=0\,,&\quad \text{for}\quad j = 0\,,&\label{thirdDsemishort0}
\end{alignat}
where indices $n_i,\,\delta_i$ and $\dot\delta_i$ are related with the ${\schurD}_{R'(0,\bj)}$ multiplet. 
Below, we solve these equations together with \eqref{C00B},\eqref{C00},\eqref{BcondQ}, and \eqref{DcondQ}/\eqref{DcondQ0}, for all possible values of $R'$ and $\bj$.

\subsubsection{$\bj=j-1\geq0 \,, R'=R$}
In this case, $H(\BZ_{3})$ is 
\begin{align}
H(\BZ_{3})=\frac{A}{X^4}(\epsilon_{m_{1}k_{1}}\cdots \epsilon_{m_{2R}k_{2R}})
{\bar{\Theta}}^{\dot{\gamma}_{1}\dot{\gamma}_{2}}
\epsilon^{\dot{\gamma}_{3}\dot{\beta}_{1}}\cdots\epsilon^{\dot{\gamma}_{2j}\dot{\beta}_{2j-2}}
\,,
\end{align}
which is inconsistent with \eqref{DcondQ} unless $A=0$.
\begin{comment}
\begin{align}
0=\frac{2A}{X^4}\left(\epsilon_{m_{1}k_{1}}\cdots \epsilon_{m_{2R}k_{2R}}\right)
\left(\epsilon^{\dot{\gamma}_{1}\dot{\beta}_{1}}\cdots\epsilon^{\dot{\gamma}_{2j-2}\dot{\beta}_{2j-2}}\right)
\epsilon^{\dot{\alpha}\dot{\gamma}_{2j-1}}
\bar{\Theta}^{\dot{\gamma}_{2j}i}.
\end{align}
\end{comment}

\subsubsection{$\bj=j-\frac{1}{2}\geq 0\,, R'=R-\frac{1}{2}$}
In this case, the only possible $H(\BZ_{3})$ is of the form\footnote{Recall here that
	$\bar{\Theta}^{\dot{\alpha}}_{~\dot{\gamma}_{2j}}=
	\bar{\Theta}^{\dot{\alpha}}_{~i}\epsilon^{ik}\tilde{\bar{\Theta}}_{\dot{\gamma}_{2j} k}$.}
\begin{align}
\begin{split}
H(\BZ_{3})
&=\frac{1}{X^4}
\Bigg(A\epsilon_{\dot{\gamma}_{1}\dot{\beta}_{1}}
\tilde{\bar{\Theta}}_{\dot{\gamma}_{2}m_{1}}
+B\frac{\epsilon_{\dot{\gamma}_{1}\dot{\beta}_{1}}}{X^2}
(\Theta_{m_{1}}X)_{\dot{\alpha}}\bar{\Theta}^{\dot{\alpha}}_{~\,\dot{\gamma}_{2}}
+\frac{C}{X^2}
(\Theta_{m_{1}}X)_{\dot{\beta}_{1}}
\tilde{\bar{\Theta}}_{\dot{\gamma}_{1}\dot{\gamma}_{2}}
\Bigg)
\\
&~\times
(\epsilon_{m_{2}k_{1}}\cdots \epsilon_{m_{2R}k_{2R-1}})
(\epsilon_{\dot{\gamma}_{3}\dot{\beta}_{2}}\cdots\epsilon_{\dot{\gamma}_{2j}\dot{\beta}_{2j-1}}).
\end{split}
\end{align}
For this to be consistent with \eqref{C00B} and \eqref{DcondQ}, we must
set $B=-\frac{4 \Imu }{j+1}A ~, C=-\frac{2\Imu (2j-1)}{j+1}A$.
Then the function $G(\BZ_{2})$ given by 
\eqref{gneGtoH} is now written as
\begin{align}
\begin{split}
G(\BZ_{2})&
=-\Imu A \Bigg(
(j+1)\epsilon_{\alpha_{1}\delta_{1}}
(\tilde{\bar{X}}^{-1}
\bar{\Theta}^{i}_{2})_{\alpha_{2}}\vb{u}^{\dagger}_{i j_{1}}(\BZ_{2})
-4\Imu \epsilon_{\alpha_{1}\delta_{1}}
(\Theta_{j_{1}}\tilde{\bar{X}}^{-1}\bar{\Theta}_{i})
(\tilde{\bar{X}}^{-1}\bar{\Theta}^{i})_{\alpha_{2}}
\\
&\qquad
-2\Imu (2j-1) \tilde{\Theta}_{j_{1}\delta_{1}}
(\tilde{\bar{X}}^{-1}\bar{\Theta}_{i})_{\alpha_{1}}
(\tilde{\bar{X}}^{-1}\bar{\Theta}^{i})_{\alpha_{2}}
\Bigg)(\epsilon_{j_{2}n_{1}}\cdots \epsilon_{j_{2R}n_{2R-1}})
(\epsilon_{\alpha_{3}\delta_{2}}\cdots\epsilon_{\alpha_{2j}\delta_{2j-1}})\,.
\end{split}
\end{align}
However, we see that this does not satisfy the shortening condition
\eqref{thirdDshort} unless $A=0$.
%\note{Takahiro}{I have added the prefactor $A$ in the above equation. Please
	%check if that is correct.}

\subsubsection{$\bj=j-\frac{1}{2} \geq 0 \,, R'=R+\frac{1}{2}$}
In this case, the possible solution is given by 
\begin{align}
H(\BZ_{3})=\frac{A}{X^2}\bar{\Theta}_{~k_{1}}^{\dot{\gamma}_{1}}
(\epsilon_{m_{1}k_{2}}\cdots \epsilon_{m_{2R}k_{2R+1}})
(\epsilon^{\dot{\gamma}_{2}\dot{\beta}_{1}}\cdots\epsilon^{\dot{\gamma}_{2j}\dot{\beta}_{2j-1}})\,,
\end{align}
which is not consistent with \eqref{DcondQ} unless $A=0$.
\begin{comment},
\begin{align}
0=\frac{A}{X^2}
\left(\epsilon_{mk_{1}}\epsilon_{m_{1}k_{2}}\cdots \epsilon_{m_{2R}k_{2R+1}}\right)
\left(\deltasu{\dot{\gamma}_{1}}{\dot{\gamma}_{1}}\epsilon^{\dot{\gamma}_{2}\dot{\beta}_{1}}\cdots\epsilon^{\dot{\gamma}_{2j}\dot{\beta}_{2j-1}}\right).
\end{align}
\end{comment}

\subsubsection{$\bj=j\geq0\,, R'=R$}
In this case, $H(\BZ_{3})$ is given by
\begin{align}
\begin{split}
H(\BZ_{3})&=\frac{1}{X^2}
\Bigg(
\epsilon^{\dot{\gamma}_{1}\dot{\beta}_{1}}
\left(A\epsilon_{m_{1}k_{1}}
+C\frac{M_{m_{1}k_{1}}}{X^2}\right)
+B\frac{(\Theta_{m_{1}}X)_{\dot{\gamma}_{1}}\tilde{\bar{\Theta}}_{\dot{\beta}_{1}k_{1}}}{X^2}
\Bigg)
\\
&~\times
(\epsilon_{m_{2}k_{2}}\cdots \epsilon_{m_{2R}k_{2R}})
(\epsilon^{\dot{\gamma}_{2}\dot{\beta}_{2}}\cdots\epsilon^{\dot{\gamma}_{2j}\dot{\beta}_{2j}})\,.
\end{split}
\end{align}
It is straightforward to show that this satisfies \eqref{C00}. Let us
next consider the semi-shortening condition for the second multiplet. For
$j>0$, the condition \eqref{DcondQ} reads
\begin{align}
0=4\Imu \Theta_{(m}^{\alpha}\epsilon_{m_{1})k_{1}}\frac{\left(
	B+	8\Imu jA +2j C
	\right)X_{\alpha\dot{\beta}}}{X^4}(\epsilon_{m_{2}k_{2}}\cdots)
(\epsilon^{\dot{\gamma}\dot{\beta}}\cdots)\,,
\label{eq:C-1-4}
\end{align}
which implies $B=-4\Imu (2j)A-2j C $.
On the other hand, for $ j=0$, the condition \eqref{DcondQ0} implies 
$C=-4\Imu A$. Note that the term proportional to $B$ in
\eqref{eq:C-1-4} does not exist for $j=0$.

To solve the conditions associated with the third multiplet, let us
relate the above $H(\BZ_3)$ to $G(\BZ_2)$ via \eqref{gneGtoH}. The
result is generally written as 
\begin{align}
\begin{split}
G(\BZ_{2})&=\frac{A}{X^2}
\Bigg(\!\!
\left(\epsilon_{j_{1}n_{1}}-(4\Imu +j C )\frac{M_{j_{1}n_{1}}}{X^2}\right)
\epsilon_{\alpha_{1}\delta_{1}}
-2j^2 C
\tilde{\Theta}_{j_{1}\alpha_{1}}(\tilde{X}^{-1}{\bar{\Theta}}_{n_{1}})_{\delta_{1}}
\Bigg)
\\
&~\times 
(\epsilon_{j_{2}n_{2}}\cdots \epsilon_{j_{2R}n_{2R}})
(\epsilon_{\alpha_{2}\delta_{2}}\cdots\epsilon_{\alpha_{2j}\delta_{2j}})\,,
\end{split}
\end{align}
where $C$ is a free parameter that is present only in the case of $j>0$. 
We see that this expression satisfies the shortening condition
\eqref{thirdDshort} for arbitrary
$C$. On the other hand, the semi-shortening condition implies $C=0$
unless $j=\frac{1}{2}$. For $j=\frac{1}{2}$, the semi-shortening condition does not
restrict the value of $C$.

\begin{comment}
In summary for $ j\neq \frac{1}{2}$ case 
\begin{align}
H(\BZ_{3})=\frac{A}{X^2}
\vb{u}_{m_{1}k_{1}}(\BZ_{3})
\left(\epsilon_{m_{2}k_{2}}\cdots \epsilon_{m_{2R}k_{2R}}\right)
\left(\epsilon_{\dot{\gamma}_{1}\dot{\beta}_{1}}\cdots\epsilon_{\dot{\gamma}_{2j}\dot{\beta}_{2j}}\right),
\end{align}
and $ j=\frac{1}{2}$ case
\begin{align}
H(\BZ_{3})=\frac{A}{X^2}
\Bigg(
\vb{u}_{m_{1}k_{1}}(\BZ_{3})
\epsilon_{\dot{\gamma}_{1}\dot{\beta}_{1}}
+\frac{C}{X^2}
\left(\left(\Theta_{m_{1}}X\right)_{\dot{\gamma}_{1}}\tilde{\bar{\Theta}}_{\dot{\beta}_{1}k_{1}}
-M_{m_{1}k_{1}}\right)
\Bigg)\left(\epsilon_{m_{2}k_{2}}\cdots \epsilon_{m_{2R}k_{2R}}\right).
\end{align}
\end{comment}

\subsubsection{$\bj=j \,, R'=R-1$}
In this case, the possible $H(\BZ_3)$ is of the form
\begin{align}
\begin{split}
H(\BZ_{3})&=\frac{1}{X^6}
\Bigg(
\Theta_{m_{1}}^{\alpha}\left(
AX_{\alpha\dot{\gamma}_{1}}\tilde{\bar{\Theta}}_{\dot{\beta}_{1}m_{2}}
+B X_{\alpha\dot{\beta}_{1}}\tilde{\bar{\Theta}}_{\dot{\gamma}_{1}m_{2}}
\right)+C
\epsilon_{\dot{\gamma}_{1}\dot{\beta}_{1}}
M_{m_{1}m_{2}}
+D
\Theta_{m_{1}}\Theta_{m_{2}}\tilde{\bar{\Theta}}_{\dot{\gamma}_{1}\dot{\beta}_{1}}\Bigg)
\\
&~\times
(\epsilon_{m_{3}k_{1}}\cdots \epsilon_{m_{2R}k_{2R-2}})
(\epsilon_{\dot{\gamma}_{2}\dot{\beta}_{1}}\cdots\epsilon_{\dot{\gamma}_{2j}\dot{\beta}_{2j}}).
\end{split}
\end{align}
Note that this expression vanishes if $j= 0$. Therefore there is no
non-trivial solution for $j=0$.

For $j>0$, the above expression is consistent with \eqref{C00} and \eqref{DcondQ} if and
only if $C=\frac{A-B}{2}$ and $D=\Imu(A+B)$. 
%only when we set these coefficients as $C=\frac{A-B}{2}$, $D=\Imu(A+B)$, $H(\BZ_{3})$ satisfies \eqref{C00} and \eqref{DcondQ}. 
%Meanwhile $ j=0$ case, trivially $A=B=D=0$ and it does not satisfy both \eqref{C00} and \eqref{DcondQ}.
%To check the third multiplet (semi-)shortening condition,
With these conditions imposed, the function $G(\BZ_{2})$ given by \eqref{gneGtoH} is written as
\begin{align}
\begin{split}
G(\BZ_{2})&=
\Bigg(
\left(
A \tilde{\Theta}_{j_{1}\alpha_{1}}(\tilde{\bar{X}}^{-1}\bar{\Theta}^{k})_{\delta_{1}}\vb{u}_{k j_{2}}^{\dagger}
+B \tilde{\Theta}_{j_{1}\delta_{1}} (\tilde{\bar{X}}^{-1}\bar{\Theta}^{k})_{\alpha_{1}}\vb{u}_{k j_{2}}^{\dagger}
\right)
\\
&~~+\frac{A-B}{2}\epsilon_{\alpha_{1}\delta_{1}}
\frac{\bar{M}_{j_{1}j_{2}}}{\bar{X}^{2}_{2}}
-\Imu (A+B) 
\Theta_{j_{1}}\Theta_{j_{2}}(\tilde{\bar{X}}^{-1}\bar{\Theta}_{i})_{\alpha_{1}}
(\tilde{\bar{X}}^{-1}\bar{\Theta}^{i})_{\delta_{1}}\Bigg)
\\
&~\times 
(\epsilon_{j_{3}n_{1}}\cdots \epsilon_{j_{2R}n_{2R-2}})
(\epsilon_{\alpha_{2}\delta_{2}}\cdots\epsilon_{\alpha_{2j}\delta_{2j}})\,.
\end{split}
\end{align}
It is then straightforward to see that $A=B=0$ is necessary for this to be consistent with \eqref{thirdDshort} and
\eqref{thirdDsemishort}. Therefore, no non-trivial solution exists in
this case.

\subsection{$\langle{\schurC}_{0(0,0)}{\schurDD}_{R(j,0)}{\schurC}_{R'(j_{1},\bj_{2})}\rangle$}

Let us now turn to the case of $\mathcal{O}^\II = \widehat{\mathcal{C}}_{R'(j_1,\bj_2)}$. 
The function $H(\BZ_{3})$ now has dimension $2(R'-R) +(j_{1}+\bj_{2}-j) -1$ and $U(1)_{r}$ charge $\bj_{2}-j_{1}-j-1$.
Since $|\bj_{2}-j_{1}-j-1|\leq 1$, the possible values
of $\bj_{2}-j_{1} $ are $ j$, $ j+\frac{1}{2}$, $ j+1$, $
j+\frac{3}{2}$, and $j+2$.
It is straightforward to see that no non-trivial $H(\BZ_3)$ is possible
for $\bj_{2}-j_{1}$=$ j+\frac{3}{2}$ and $ j+2$.
Therefore, the only possible values of $R'$ and $\bj_2-j_1$
%selections 
are $(R',\bj_{2}-j_{1})=
(R,j+1)$, $(R-1,j+1)$, $(R+\frac{1}{2},j+\frac{1}{2})$, $(R-\frac{1}{2},j+\frac{1}{2})$, and $(R,j)$.
The semi-shortening conditions for the third multiplet
%${\schurC}_{R'(j_{1},\bj_{2})}$ conditions 
are \eqref{thirdcondC1} and \eqref{thirdcondC2}. 
Below, we solve all the (semi-)shortening conditions for each of the
possible values of $\bj_2-j_1$ and $R'$.

\subsubsection{$\bj_{2}-j_{1}=j+1 \,, R'=R $}
In this case, the function $H(\BZ_{3})$ is of the form
\begin{align}
H(\BZ_{3})
&=A\frac{(\Theta_{m_{1}}X)_{\dot{\beta}_{1}}
	\tilde{\bar{\Theta}}_{\dot{\beta}_{2}k_{1}}}{X^2}
(\epsilon_{m_{2}k_{2}}\cdots \epsilon_{m_{2R}k_{2R}})
(\epsilon_{\dot{\gamma}_{1}\dot{\beta}_{3}}\cdots\epsilon_{\dot{\gamma}_{2j}\dot{\beta}_{2j}})
X_{\beta_{1}\dot{\beta}_{2j+3}}\cdots X_{\beta_{2j_{1}}\dot{\beta}_{2j+2j_{1}+2}}\,,
\end{align}
which is not consistent with the condition \eqref{C00} unless $A=0$.
\begin{comment}
\begin{align}
0=16A (2j_{1}+1)\bar{\Theta}_{(i}\bar{\Theta}_{j)}\frac{
X_{\beta_{1}\dot{\beta}_{2j+1}}\cdots X_{\beta_{2j_{1}}\dot{\beta}_{2j+2j_{1}}}X_{\alpha\dot{\beta}_{2j+2j_{1}+1}}}{X^4}
\Theta_{m_{1}}^{\alpha}\tilde{\bar{\Theta}}_{\dot{\beta}_{2j+2j_{1}+1}}^{k_{2R}}
\end{align}
\end{comment}

\subsubsection{$\bj_{2}-j_{1}=j+1 \,, R'=R-1\geq 0$}
In this case, the possible $H(\BZ_{3})$ is 
\begin{align}
\begin{split}
H(\BZ_{3})&=\frac{1}{X^4}
\left(A
(\Theta_{m_{1}}X)_{\dot{\beta}_{1}}\tilde{\bar{\Theta}}_{\dot{\beta}_{2} m_{2}}
+B \, \Theta_{m_{1}}\Theta_{m_{2}}\tilde{\bar{\Theta}}_{\dot{\beta}_{1}\dot{\beta}_{2}}
\right)
\left(\epsilon_{m_{3}k_{1}}\cdots \epsilon_{m_{2R}k_{2R-2}}\right)
\\
&
\times (\epsilon_{\dot{\gamma}_{1}\dot{\beta}_{3}}\cdots\epsilon_{\dot{\gamma}_{2j}\dot{\beta}_{2j+2}})
X_{\beta_{1}\dot{\beta}_{2j+3}}\cdots X_{\beta_{2j_{1}}\dot{\beta}_{2j_{1}+2j+2}}\,,
\end{split}
\end{align}
where $ j\geq 0$, $ j_{1}\geq 0$, $R\geq 1$.
We see that \eqref{C00} implies $B=-\Imu(2j_{1}) A$. Then the function $G(\BZ_2)$ is written as
\begin{align}
\begin{split}
G(\BZ_{2})&=
\frac{-A}{\bar{X}_{2}^{2}}
\left(
\tilde{\Theta}_{j_{1}\delta_{1}}
\left(\tilde{\bar{X}}^{-1}\bar{\Theta}^{k}\right)_{\delta_{2}}{ \vb{u}}_{kj_{2}}^{\dagger}(\BZ_{2})
-\Imu (2j_{1})  \Theta_{j_{1}}\Theta_{ j_{2}}
\left(\tilde{\bar{X}}^{-1}\bar{\Theta}_{k}\right)_{\delta_{1}}
\left(\tilde{\bar{X}}^{-1}\bar{\Theta}^{k}\right)_{\delta_{2}}
\right)
\\
&~\times
(\epsilon_{j_{3}n_{1}}\cdots \epsilon_{j_{2R}n_{2R-2}})
(\epsilon_{\alpha_{1}{\delta}_{3}}\cdots\epsilon_{\alpha_{2j}{\delta}_{2j+2}})
\tilde{\bar{X}}^{-1}_{\delta_{1}\dot{\delta}_{2j+3}}\cdots \tilde{\bar{X}}^{-1}_{\delta_{2j_{1}}\dot{\delta}_{2j+2j_{1}+2}}\,,
\end{split}
\end{align}
which does not satisfy the conditions \eqref{thirdcondC1} and
\eqref{thirdcondC2} unless $A=0$.

\subsubsection{$\bj_{2}-j_{1}=j+\frac{1}{2} \,, R'=R+\frac{1}{2}$ }

In this case, the only possible candidate for $H(\BZ_{3})$ is of the form
\begin{align}
H(\BZ_{3})=A \tilde{\bar{\Theta}}_{\dot{\beta}_{1} k_{1}}
X_{\beta_{1}\dot{\beta}_{2}}\cdots X_{\beta_{2j_{1}}\dot{\beta}_{2j_{1}+1}}
(\epsilon_{\dot{\gamma}_{1}\dot{\beta}_{2j_{1}+2}}\cdots\epsilon_{\dot{\gamma}_{2j}\dot{\beta}_{2j+2j_{1}+1}})
(\epsilon_{m_{1}k_{2}}\cdots \epsilon_{m_{2R}k_{2R+1}})\,.
\end{align}
By using \eqref{barXtoX}, it is rewritten as
\begin{align}
\begin{split}
H(\bar{\BZ}_{3})& = A\tilde{\bar{\Theta}}_{\dot{\beta}_{1} k_{1}}
\left(\bar{X}_{\beta_{1}\dot{\beta}_{2}}-4\Imu (2j_{1})\tilde{\bar{\Theta}}_{\dot{\beta}_{2}}\tilde{\Theta}_{\beta_{1}}\right)
\bar{X}_{\beta_{2}\dot{\beta}_{3}}\cdots \bar{X}_{\beta_{2j_{1}}\dot{\beta}_{2j_{1}+2}}
(\epsilon_{m_{1}k_{2}}\cdots \epsilon_{m_{2R}k_{2R+1}})
\\
& ~\times
(\epsilon_{\dot{\gamma}_{1}\dot{\beta}_{2j_{1}+2}}\cdots\epsilon_{\dot{\gamma}_{2j}\dot{\beta}_{2j+2j_{1}+1}})\,.
\end{split}
\label{CDC3}
\end{align}
We see that this is of the form of \eqref{ansC00Bar} and \eqref{semigsol} and therefore satisfies \eqref{DcondQ}/\eqref{DcondQ0} and \eqref{C00}.
On the other hand, the function $G(\BZ_2)$ is now evaluated as
\begin{align}
G(\BZ_{2})=\Imu\frac{ \vb{u}_{k n_{1}}^{\dagger}(\BZ_{2})}{\bar{X}^{4}}\!
\left(\bar{X}\bar{\Theta}^{k}\right)_{\delta_{1}}
(\epsilon_{j_{1}n_{2}}\cdots \epsilon_{j_{2R}n_{2R+1}})
(\epsilon_{\alpha_{1}\delta_{2}}\cdots\epsilon_{\alpha_{2j}\delta_{2j+1}})
\tilde{\bar{X}}^{-1}_{\delta_{2j+2}\dot{\delta}_{1}}
\cdots \tilde{\bar{X}}^{-1}_{\delta_{2j+2j_{1}+1}\dot{\delta}_{2j_{1}}}\,.
\end{align}
We see that this expression satisfies the 
%third multiplet 
semi-shortening conditions associated with the third multiplet for any $ j\geq 0$, $
j_{1}\geq 0$ and $R\geq 0$.

\subsubsection{$ \bj_{2}-j_{1}=j+\frac{1}{2} \,, R'=R-\frac{1}{2}$}

The only possible $H(\BZ_{3})$ is of the form
\begin{align}
H(\BZ_{3})&=\frac{1}{X^2}
\Bigg(A 
\epsilon_{\dot{\gamma}_{1}\dot{\beta}_{1}}\tilde{\bar{\Theta}}_{\dot{\beta}_{2} m_{1}}
+\frac{\Theta_{m_{1}}^{~\alpha}\bar{\Theta}^{\dot{\alpha}\dot{\alpha}'}}{X^2}
\Bigg(
B\epsilon_{\dot{\gamma}_{1}\dot{\beta}_{1}}
\epsilon_{\dot{\beta}_{2}\dot{\alpha}'}
X_{\alpha\dot{\alpha}}
+CX_{\alpha\dot{\gamma}_{1}} \epsilon_{\dot{\beta}_{1}\dot{\alpha}}
\epsilon_{\dot{\beta}_{2}\dot{\alpha}'}
+
DX_{\alpha\dot{\beta}_{1}} 
\epsilon_{\dot{\beta}_{2}\dot{\alpha}}
\epsilon_{\dot{\gamma}_{1}\dot{\alpha}'}
\Bigg)\Bigg)
\nonumber\\
&~\times
(\epsilon_{m_{2}k_{1}}\cdots \epsilon_{m_{2R}k_{2R-1}})
(\epsilon_{\dot{\gamma}_{2}\dot{\beta}_{3}}
\cdots\epsilon_{\dot{\gamma}_{2j}\dot{\beta}_{2j+1}})
X_{\beta_{1}\dot{\beta}_{2j+3}}\cdots X_{\beta_{2j_{1}}\dot{\beta}_{2j+2j_{1}+1}}.
\end{align}
In the case of $j>0$, the constraints \eqref{C00} and \eqref{DcondQ} imply that $B=-4 \Imu A +C $
and $D=-C$.
On the other hand, for $ j=0$, the constraints \eqref{C00} and \eqref{DcondQ} imply $B=-4 \Imu A$. 
Note that $C$ and $D$ do not exist for $j=0$.
The function $G(\BZ_2)$ is now evaluated as
\begin{align}
\begin{split}
G(\BZ_{2})&=\frac{1}{\bar{X}^{2}}
\Bigg(
A \epsilon_{\alpha_{1}\delta_{1}}
\left(\tilde{\bar{X}}\bar{\Theta}^{k}\right)_{\delta_{2}}
+\Bigg((-4\Imu A +j C )\epsilon_{\alpha_{1}\delta_{1}}
(\Theta_{j_{1}}\tilde{\bar{X}}^{-1}\bar{\Theta}_{k})
(\tilde{\bar{X}}^{-1}\bar{\Theta}^{k})_{\delta_{2}}
\\
&\quad\qquad+jC 
\left(\tilde{\Theta}_{j_{1}\alpha_{1}}
(\tilde{\bar{X}}^{-1}\bar{\Theta}_{k})_{\delta_{1}}(\tilde{\bar{X}}^{-1}\bar{\Theta}^{k})_{\delta_{2}}
-\tilde{\Theta}_{j_{1}\delta_{1}}
(\tilde{\bar{X}}^{-1}\bar{\Theta}_{k})_{\delta_{2}}(\tilde{\bar{X}}^{-1}\bar{\Theta}^{k})_{\alpha_{1}}\right)
\Bigg)\Bigg)
\\
&~\times 
(\epsilon_{j_{2}n_{1}}\cdots \epsilon_{j_{2R}n_{2R-1}})
(\epsilon_{\alpha_{2}\delta_{3}}\cdots\epsilon_{\alpha_{2j}\delta_{2j+2}})
\tilde{\bar{X}}_{\delta_{1}\dot{\delta}_{2j+3}}\cdots \tilde{\bar{X}}_{\delta_{2j_{1}}\dot{\delta}_{2j+2j_{1}+2}}\,,
\end{split}
\end{align}
which satisfies \eqref{thirdcondC1} and \eqref{thirdcondC2} if and only
if $C=0$.

\subsubsection{$\bj_{2}-j_{1}=j \,, R'=R$}

In this case, the only possible $H(\BZ_3)$ is 
\begin{align}
\begin{split}
H(\BZ_{3})&=
\frac{\left(\epsilon_{m_{1}k_{1}}\cdots \epsilon_{m_{2R}k_{2R}}\right)}{X^2}
\Bigg(
A X_{\beta_{1}\dot{\beta}_{1}}
\tilde{\bar{\Theta}}_{\dot{\gamma}_{1}\dot{\beta}_{2}}
+B X_{\beta_{1}\dot{\alpha}}\epsilon_{\dot{\beta}_{1}\dot{\alpha}'}\bar{\Theta}^{\dot{\alpha}\dot{\alpha}'} \epsilon_{\dot{\gamma}_{1}\dot{\beta}_{2}}
\Bigg)
\\
&~\times 
(\epsilon_{\dot{\gamma}_{2}\dot{\beta}_{3}}\cdots\epsilon_{\dot{\gamma}_{2j}\dot{\beta}_{2j+1}})
X_{\beta_{2}\dot{\beta}_{2j+2}}\cdots X_{\beta_{2j_{1}}\dot{\beta}_{2j+2j_{1}}}\,.
\end{split}
\end{align}
Since $\bar{\Theta}_{(i}\bar{\Theta}_{j)}\bar{\Theta}_{\dot{\gamma}_{1}\dot{\beta}_{2}}
=\bar{\Theta}_{(i}\bar{\Theta}_{j)}\bar{\Theta}^{\dot{\alpha}\dot{\alpha}'}=0$,
the constraint \eqref{C00} is trivially satisfied. 
On the other hand, the constraint \eqref{DcondQ} implies $A=B=0$.

\section{$\langle{\schurC}_{0(0,0)}{\schurC}_{R(j_1,\bj_1)}{\schurC}_{R'({j}_2,\bj_2)}\rangle$}
\label{appCCC}

In this appendix, we show that the only non-vanishing three-point
functions of the form
$\langle{\schurC}_{0(0,0)}{\schurC}_{R(j_{1},\bj_{1})}{\schurC}_{R'(j_{2},\bj_{2})}\rangle$
are those listed in \eqref{CCCR} and \eqref{CCCR-1}. To that end, in sub-section
\ref{subsec:non-zero-r}, we
will show that the three point function vanishes when the
corresponding $H(\BZ_3)$ has non-vanishing $U(1)_r$ charge. In
sub-section \ref{subsec:zero-r}, we show that $\langle
\widehat{\mathcal{C}}_{0(0,0)}\widehat{C}_{R(j+\ell_2,j+\ell_1+\ell_2)}\widehat{\mathcal{C}}_{R+1(j+\ell_1,j)}\rangle$
vanishes for any $j,\ell_1\geq 0$ and $\ell_2>0$.

\subsection{Non-vanishing $U(1)_r$ charge}
\label{subsec:non-zero-r}

We first show that $\langle{\schurC}_{0(0,0)}{\schurC}_{R(j_{1},\bj_{1})}{\schurC}_{R'(j_{2},\bj_{2})}\rangle$
vanishes when $H(\BZ_3)$ has $U(1)_r$ charge $\pm 1$. 
It is straightforward to generalize it to the case of $U(1)_r$ charge $\pm \frac{1}{2}$.
While $U(1)_{r}$ charge vanishes, we can also show $\langle \schurC_{0(0,0)}\schurC_{R(j+\ell_{1}+\ell_{2},j+\ell_{2})}\schurC_{R+1(j,j+\ell_{1})} \rangle$ is zero with same argument for $\ell_{2}>0$.

Since charge conjugation flips the sign of the $U(1)_r$ charge, we focus
on the case of $U(1)_r$ charge $-1$.
Then we see from \eqref{ansC00} that $R'=R$ must hold for a non-trivial solution to exist.
All possible combinations of four spin eigenvalues $(j_{1},\bj_{1},j_{2},\bj_{2})$ are listed in table \ref{jeigen}.
\begin{table}%[!h]
	\centering \renewcommand{\arraystretch}{1.5}
	\begin{tabular}{|p{0.3\textwidth}|p{0.48\textwidth}|}
	\hline
	$(j_{1},\bj_{1},j_{2},\bj_{2})$& $(\Delta,L_{1},L_{2},L_{3},L_{4})$\\
	\hline
	$(j+1,j+\ell_{1},j+\ell_{1}+\ell_{2},j+\ell_{2})$ & $(k+4,2\ell_{2}+k, k+2, 2j-k,2j+2\ell_{1}-k-2)$,\\
	\hline
	$(j+1+\ell_{1}, j, j+\ell_{2},j+\ell_{1}+\ell_{2})$ & $(k+4,2\ell_{2}+k, k+2, 2j+2\ell_{1}-k,2j-k-2)$, \\
	\hline
	$(j+1+\ell_{2},j+\ell_{1}+\ell_{2},j+\ell_{1},j)$ & $(2\ell_{2}+k+4,k,2\ell_{2}+ k+2, 2j-k,2j-k+2\ell_{1}-2)$, \\
	\hline
	$(j+1+\ell_{1}+\ell_{2},j+\ell_{2},j,j+\ell_{1})$ & $(2\ell_{2}+k+4,k,2\ell_{2}+ k+2, 2j+2\ell_{1}-k,2j-k-2)$,\\
	\hline
\end{tabular}
	\caption{{The left column shows possible spin eigenvalues, and the right column shows the corresponding values of parameters appearing in \eqref{apDexpdf}.}
	}
	\label{jeigen}
\end{table}
From \eqref{ansC00}, we see that the general expression for $H(\BZ_{3})$
is of the form
\begin{align}
H(\BZ_{3})=\left(\epsilon_{m_{1}k_{1}}\cdots \epsilon_{m_{2R}k_{2R}}\right)
\TB{\Theta}_{\dot{\gamma}\dot{\gamma}'}f^{(\dot{\gamma}\dot{\gamma}')(\dot\gamma_1\cdots \dot\gamma_{2j_1})(\gamma_1\cdots
	\gamma_{2\bj_1})\II}(X)\,,
\end{align}
where $\II$ stands for the Lorentz indices associated with the third
multiplet, and we write $(X,\Theta,\bar\Theta)$ for
$(X_3,\Theta_3,\bar\Theta_3)$.
Since $ j_{1}>0 $,
the semi-shortening
conditions for the second multiplet contain \eqref{CjjB}.
%It always satisfies \eqref{C00} trivially, because of the Fierz identity $\TB{\Theta}_{\dot{\alpha}\dot{\alpha}'}\bar{\Theta}_{(i}\bar{\Theta}_{i')}=0$. 
%Next, we pick up the first order of $\bar{\Theta}$ in the
%semi-shortening condition \eqref{CjjB}
The condition \eqref{CjjB} particularly implies
\begin{align}
%\bar{\QQ}_{\dot{\gamma}_{1}}^{i}H^{(\dot{\gamma}_{1}\cdots\dot{\gamma}_{2j_{1}})}|_{\bar{\Theta}}=
2\left(\epsilon_{m_{1}k_{1}}\cdots \epsilon_{m_{2R}k_{2R}}\right)
\TB{\Theta}_{\dot{\gamma}'}^{~i}\epsilon_{\dot{\gamma}\dot{\gamma}_{1}}
f^{(\dot{\gamma}\dot{\gamma}')(\dot{\gamma}_{1}\cdots\dot{\gamma}_{2j_{1}})(\gamma_1\cdots \gamma_{2\bj_1})\II}=0\,,
\end{align}
which means that 
%so we can set 
$f^{(\dot{\gamma}\dot{\gamma}')
	(\dot{\gamma}_{1}\cdots\dot{\gamma}_{2j_{1}})(\gamma_1\cdots \gamma_{2\bj_1})\II}
=f^{(\dot{\gamma}\dot{\gamma}'\dot{\gamma}_{1}\cdots\dot{\gamma}_{2j_{1}})(\gamma_1\cdots
	\gamma_{2\bj_1})\II}$.
%Therefore,
Then the remaining constraint arising from \eqref{CjjB} is expressed as
%\note{Takahiro}{Please double check the modifications I made here.}
\begin{align}
\epsilon_{\dot{\alpha}\dot{\gamma}_{1}}\frac{\partial}{\partial
	X_{\alpha\dot{\alpha}}}
f^{(\dot\gamma\dot\gamma'\dot{\gamma}_{1}\cdots\dot{\gamma}_{2j_{1}})(\gamma_1\cdots \gamma_{2\bj_1})\II}(X)=0\,.
\label{CCCredcond}
\end{align}

To show that there is no non-trivial solution to this equation, let us
first write down the most general expression for
$f^{(\dot\gamma\dot\gamma'\dot\gamma_1\cdots
	\dot\gamma_{2j_1})(\gamma_1\cdots \gamma_{2\bj_1})\II}(X_3)$. 
%To reduce clutter, we write $X$
%instead of $X_3$ in what follows. 
Recall first that
$\II$ stands for the Lorentz indices associated with
$\widehat{\mathcal{C}}_{R(j_2,\bj_2)}$. Therefore, the most general
ansatz for $f^{(\dot\gamma\dot\gamma'\dot\gamma_1\cdots
	\dot\gamma_{2j_1})(\gamma_1\cdots \gamma_{2\bj_1})\II}$
is written as
\begin{align}
%&
f(X)&=\sum_{k=0}^{2j-2}\lambda_{k}f_{k}(X)\,,
\\
%&
f_{k}(X)
%:=
&\equiv\frac{1}{(X^2)^{\Delta}}(\tilde{X}^{\dot{\beta}\beta})^{L_{1}}
(\tilde{X}^{\dot{\gamma}\gamma} )^{L_{2}}
( \epsilon^{\dot{\gamma}\dot{\beta}} )^{L_{3}}( \epsilon^{\gamma\beta})^{L_{4}}(\tilde{X}^{\dot{\gamma}\beta} )^{2}\,,
\label{apDexpdf}
\end{align}
where we use the short-hand notation 
$(\tilde{X}^{\dot{\gamma}\gamma} )^{L}\equiv \tilde{X}^{\dot{\gamma}_{1}\gamma_{1}}\cdots
\tilde{X}^{\dot{\gamma}_{L}\gamma_{L}}$ with
$\dot\gamma_{2j_1+1}\equiv \dot\gamma$ and $\dot\gamma_{2j_1+2}\equiv
\dot\gamma'$. 
The parameters $\Delta,L_{1},L_{2},L_{3}$, and $L_{4}$ are
fixed by $j_1,\bj_1,j_2,\bj_2$ and $k$ as in table
\ref{jeigen}.\footnote{Note that $j, \ell_1$, and $\ell_2$ are fixed by
	$j_1,\bj_1,j_2$, and $\bj_2$ as in the left column of the table.}
Substituting this general ansatz into the constraint \eqref{CCCredcond}, we obtain
\begin{align}
0&=\sum_{k=0}^{2j-2}\lambda_{k}\Bigg[
\epsilon^{\alpha\gamma}
\frac{L_{2}(\Delta-L_{2}-L_{3}-3)}{(X^2)^{\Delta}}
(\tilde{X}^{\dot{\beta}\beta} )^{L_{1}}
(\tilde{X}^{\dot{\gamma}\gamma} )^{L_{2}-1}
( \epsilon^{\dot{\gamma}\dot{\beta}} )^{L_{3}}( \epsilon^{\gamma\beta})^{L_{4}}
(\tilde{X}^{\dot{\gamma}\beta} )^{2}
\nonumber\\
&\qquad +\tilde{X}^{\dot{\beta}\alpha}\frac{\Delta L_{3}}{(X^2)^{\Delta+1}}(\tilde{X}^{\dot{\beta}\beta} )^{L_{1}}
(\tilde{X}^{\dot{\gamma}\gamma} )^{L_{2}}
( \epsilon^{\dot{\gamma}\dot{\beta}} )^{L_{3}-1}
( \epsilon^{\gamma\beta})^{L_{4}}(\tilde{X}^{\dot{\gamma}\beta})
\nonumber\\
&\qquad +\text{(linearly independent terms)}\Bigg]\,.
\end{align}
This particularly implies that, for $f(X)$ to be non-vanishing, 
$L_{2}(\Delta-L_{2}-L_{3}-3)=\Delta L_{3}=0$ must hold for some $k$. 
However, this last set of equations has no solution
for $l_{1}\geq 0$, $l_{2}\geq 0$, $j\geq 1$, and $2j-2 \geq k \geq 0$.
Therefore there is no non-trivial solution to the constraint \eqref{CCCredcond}.
In other words, when $H(\BZ_3)$ has $U(1)_{r}$ charge 
$-1$, the correlation function 
$\langle{\schurC}_{0(0,0)}{\schurC}_{R(j_1,\bj_1)}{\schurC}_{R'({j}_2,\bj_2)}\rangle$
vanishes.

\subsection{Vanishing $U(1)_r$ charge}
\label{subsec:zero-r}

In this sub-section we show that, even though the corresponding
$H(\BZ_3)$ is neutral
under $U(1)_r$, the three-point function $\langle
\widehat{\mathcal{C}}_{0(0,0)}\widehat{C}_{R(j+\ell_2,j+\ell_1+\ell_2)}\widehat{\mathcal{C}}_{R+1(j+\ell_1,j)}\rangle$
must vanish
for $j,\ell_1\geq 0$ and $\ell_2>0$.

In the case of $\langle
\widehat{\mathcal{C}}_{0(0,0)}\widehat{C}_{R(j+\ell_2,j+\ell_1+\ell_2)}\widehat{\mathcal{C}}_{R+1(j+\ell_1,j)}\rangle$,
\eqref{semigsol} and the fact that the third multiplet
has two more $SU(2)_R$ indices than the second one imply that
\begin{align}
 H(\BZ_3) &= \bar{\Theta}_{\dot \gamma k_1}\Theta_{\gamma k_2}
 \epsilon_{m_1k_3}\cdots \epsilon_{m_{2R}k_{2R+2}}
 g^{(\gamma\gamma_1\cdots\gamma_{2j+2\ell_1+2\ell_2}) \dot\gamma(
 \dot\gamma_1\cdots\dot\gamma_{2j+2\ell_2})}_{(\beta_1\cdots\beta_{2j+2\ell_1})(\dot\beta_1\cdots\dot\beta_{2j})}(X)
\label{eq:H-last}
\end{align}
is the most general expression for $H(\BZ_3)$, where $g$ is a function of $X$,\, $\{m_i,
\gamma_j,\dot\gamma_j\}$ are indices associated with
$\widehat{\mathcal{C}}_{R(j+\ell_2,j+\ell_1+\ell_2)}$, and
$\{k_i,\beta_j,\dot\beta_j\}$ are indices associated with
$\widehat{\mathcal{C}}_{R+1(j+\ell_1,j)}$.
In the rest of this sub-section, we write $\gamma_0$ and $\dot\gamma_0$
for $\gamma$ and $\dot\gamma$, respectively.
The function $g$ has many possible terms corresponding to different assignments
of the spinor indices to
$\delta^{\gamma_i}_{\beta_j},\,\delta^{\dot\gamma_i}_{\dot\beta_j},\,X^{\gamma_i\dot\gamma_j},\,X_{\beta_i\dot\beta_j},\,X^{\gamma_i}{}_{\dot\beta_j}$
and $X_{\beta_i}{}^{\dot\gamma_j}$. However, since the number of $\gamma_i$ minus
that of $\beta_i$ is at least $2\ell_2 + 1\geq 2$, every term contains at least one of
the following factors
\begin{align}
X^{\gamma_{i_1}\dot\gamma_{j_1}}X^{\gamma_{i_2}\dot\gamma_{j_2}}~,\qquad
 X^{\gamma_{i_1}\dot\gamma_{j_1}}X^{\gamma_{i_2}}{}_{\dot\beta_{j_2}}~,\qquad
 X^{\gamma_{i_1}}{}_{\dot\beta_{j_1}}X^{\gamma_{i_2}}{}_{\dot\beta_{j_2}}~.
\label{eq:XX}
\end{align}
Let us focus on a $k$-th term in $g$, whose coefficient we denote by
$\lambda_k$. Since all indices $\gamma_0,\cdots,\gamma_{2j+2\ell_1+2\ell_2}$ are
symmetric, the $k$-th term in $g$ involves a piece proportional to
$\lambda_k X^{\gamma_0\dot\gamma_{i+1}}$ or $\lambda_k
X^{\gamma_0}{}_{\beta_i}$ for some $i\geq 0$. This means that $H(\BZ_3)$
involves a term proportional to one of the following
\begin{align}
 \lambda_k\bar\Theta_{\dot\gamma_0 k_1}\Theta_{\gamma_0
 k_2}X^{\gamma_0\dot\gamma_{i+1}}~,\qquad \lambda_k\bar\Theta_{\dot\gamma_0
 k_1}\Theta_{\gamma_0 k_2} X^{\gamma_0}{}_{\dot\beta_i}~,
\end{align}
for some $i\geq 0$.
When we change the variables from $(X,\Theta,\bar\Theta)$ to $(\bar X,
\Theta,\bar\Theta)$, it gives rise to a term proportional to one of the following
\begin{align}
\lambda_k \bar\Theta_{\dot\gamma_0 k_1}\bar\Theta^{k\dot\gamma_{i+1}}\Theta_{k_2}\Theta_{k}
~,\qquad \lambda_k \bar\Theta_{\dot\gamma_0
 k_1}\bar\Theta^{k}{}_{\dot\beta_i}\Theta_{k_2}\Theta_{k}~,
\end{align}
which is prohibited by \eqref{C00} since the most general solution to
\eqref{C00} is written as \eqref{ansC00Bar}.\footnote{Recall that
$\Theta_{k}\Theta_{l}$ and $\Theta^{\alpha\beta}$ are linearly independent.} Therefore we must impose
$\lambda_k=0$ for \eqref{eq:H-last} to satisfy the semi-shortening
conditions. Since $k$ is arbitrary, this means that $g=0$ as a function
of $X$. Therefore, there is no non-trivial solution to the
semi-shortening conditions in the case of $\langle
\widehat{\mathcal{C}}_{0(0,0)}\widehat{C}_{R(j+\ell_2,j+\ell_1+\ell_2)}\widehat{\mathcal{C}}_{R+1(j+\ell_1,j)}\rangle$
for $j,\ell_1\geq 0$ and $\ell_2>0$.

\bibliography{hoge}
\bibliographystyle{utphys}

\end{document}